\documentclass[manuscript,screen,prologue,table,xcdraw]{acmart}
\usepackage{listings}
\usepackage{verbatim}
\usepackage{soul}
\usepackage{courier}
\usepackage{graphicx}

\AtBeginDocument{%
  \providecommand\BibTeX{{%
    \normalfont B\kern-0.5em{\scshape i\kern-0.25em b}\kern-0.8em\TeX}}}

\setcopyright{acmcopyright}
\copyrightyear{2023}
\acmYear{2023}
\acmDOI{XXXXXXX.XXXXXXX}

\usepackage{multirow}

\newcommand*\justify{%
  \fontdimen2\font=0.4em
  \fontdimen3\font=0.2em
  \fontdimen4\font=0.1em
  \fontdimen7\font=0.1em
  \hyphenchar\font=`\-
}

\renewcommand{\texttt}[1]{%
  \begingroup
  \ttfamily
  \begingroup\lccode`~=`/\lowercase{\endgroup\def~}{/\discretionary{}{}{}}%
  \begingroup\lccode`~=`[\lowercase{\endgroup\def~}{[\discretionary{}{}{}}%
  \begingroup\lccode`~=`.\lowercase{\endgroup\def~}{.\discretionary{}{}{}}%
  \catcode`/=\active\catcode`[=\active\catcode`.=\active
  \justify\scantokens{#1\noexpand}%
  \endgroup
}

\newcommand{\prim}[2]{
\texttt{#1(#2)}
}

\newcommand{\mesg}[4]{
\texttt{<#1$\rightarrow$#2,#3,[#4]>}
}

\usepackage{color}
\definecolor{baseColour}{rgb}{.90,.90,0.5}
\definecolor{todoColour}{rgb}{.90,.75,0.5}
\definecolor{torefColour}{rgb}{.65,.85,0.95}
\definecolor{daveWriteColor}{rgb}{.95,.55,0.65}

\begin{document}

\title{Unpacking Human-AI interactions: From interaction primitives to a design space}

\author{Konstantinos Tsiakas}
\email{k.tsiakas@tudelft.nl}
\author{Dave Murray-Rust}
\email{d.s.murray-rust@tudelft.nl}
\affiliation{%
  \institution{Delft University of Technology}
  \city{Delft} 
  \country{The Netherlands}
}


\begin{abstract}
This paper aims to develop a semi-formal design space for Human-AI interactions, by building a set of interaction primitives which specify the communication between users and AI systems during their interaction. We show how these primitives can be combined into a set of interaction patterns which can provide an abstract specification for exchanging messages between humans and AI/ML models to carry out purposeful interactions. The motivation behind this is twofold: firstly, to provide a compact generalisation of existing practices, that highlights the similarities and differences between systems in terms of their interaction behaviours; and secondly, to support the creation of new systems, in particular by opening the space of possibilities for interactions with models. We present a short literature review on frameworks, guidelines and taxonomies related to the design and implementation of HAI interactions, including human-in-the-loop, explainable AI, as well as hybrid intelligence and collaborative learning approaches. From the literature review, we define a vocabulary for describing information exchanges in terms of providing and requesting particular model-specific data types. Based on this vocabulary, a message passing model for interactions between humans and models is presented, which we demonstrate can account for existing systems and approaches. Finally, we build this into design patterns as mid-level constructs that capture common interactional structures. We discuss how this approach can be used towards a design space for Human-AI interactions that creates new possibilities for designs as well as keeping track of implementation issues and concerns.
\end{abstract}



\maketitle

\section{Introduction}
Artificial Intelligence and Machine Learning (AI/ML) models are in the midst of a transition from use in back-end data science systems, operated and interacted with by experts, to end-user focused applications that prioritise ease of use and interactional fluidity \cite{xu2019toward}. This transition to human-centered AI approaches is driven by several factors -- the increasing power of the models and systems; the need for more engagement with real-world data; the enlisting of users as participants in model development through annotation or other feedback; the desire by more organisations to make use of the possibilities of AI/ML; and the corresponding need to make sure that the models are operating correctly or adapt them for particular tasks. This has given rise to a growing set of human-AI interaction paradigms, in particular human-in-the-loop (HITL) \cite{chai2020human, nadj2020power}, explainable AI (XAI) \cite{schwalbe2021comprehensive, liao2021human} and hybrid intelligence (HI) and collaborative learning systems \cite{wiethof2021hybrid, dellermann2021future}. This expansion in interactivity, coupled with the need to understand how systems grow and change over time \cite{giaccardi2020technology} and affect diverse stakeholders \cite{birhane2021algorithmic} leads to a need for new ways to design interactions, as we transition from \textit{Human-Computer} to \textit{Human-AI} interactions \cite{xu2022transitioning}. 

Human-AI (HAI) interactions can involve humans in different parts of the algorithmic operation, offering possibilities for designing new types of interactions, considering the different paradigms. XAI methods are used to provide additional information about the underlying AI processes \cite{arrieta2020explainable} through communicating descriptions of model functioning \cite[e.g.][]{balayn2021WhatYou} even if it is not always a silver bullet \cite{edwards2017SlaveAlgorithm, ohara2020ExplainableAI}. HITL approaches aim to enhance system's decisions by incorporating human decision making, when needed, to enable human oversight \cite{gronsund2020augmenting}. Human feedback can be integrated to the learning mechanism of ML models through interactive ML (iML) methods to facilitate model training, as well as to explore new forms of interactivity between humans and ML algorithms \cite{mosqueira2022human}. Hybrid intelligence and collaborative learning methods can combine human capabilities with AI, allowing both parts to efficiently communicate and exchange information in order to mutually make decisions, inform and learn from each other \cite{akata2020research,zhang2021forward}. 

Despite the computational advances and capabilities of AI models, designing HAI interactions remains a challenging task. Designers often have a broad understanding of the possibilities of AI/ML, but lack specifics \cite{dove2017ux}. This is related to two AI attributes: \textit{capability uncertainty}, which indicates the uncertainty of what the possibilities of an ML model are, and \textit{output complexity}, which covers the general difficulty of working with multidimensional, rich outputs from large systems \cite{yang2020re}. Envisioning new AI-based solutions for a given UX problem should consider that HAI interactions need to adjust to different users and evolve over time \cite{giaccardi2020technology}. Going beyond the one-to-one interactions with systems, there is a also the need to investigate the effects of automated algorithmic systems on on the different groups and types of users involved in or affected by the AI decisions \cite{birhane2021algorithmic}, such as in organisational decision making \cite{behymer2016autonomous}. Human oversight is an important aspect which enables users to maintain human agency and accountability, by participating in the decision making. From such a sociotechical perspective, hybrid (or shared) decision making can involve both decision-makers and decision-subjects towards interactions with contestable AI models \cite{alfrink2022contestable}. However, designing for hybrid decision making can be challenging while considering the legal, social, technical and organizational issues \cite{enarsson2022approaching}. Moreover, designing AI-based systems that facilitate \textit{meaningful human control} requires the implementation of moral decision making methods in order to address responsibility gaps related to hybrid decisions \cite{cavalcante2022meaningful}. 

Responses to this complexity and the need for human-centered AI design takes several forms, from high level frameworks through to new interaction paradigms and implementation methods. High level frameworks shape the way that people approach creating HAI systems. Shneiderman's \textit{Human-Centered Artificial Intelligence} (HCAI) framework looks to create reliable, safe and trustworthy HAI interactions through active participation \cite{shneiderman2020human}, with the intent to achieve a high level of human control, while maintaining a high level of computer automation. Xu's HAI framework \cite{xu2019toward} sets out three key components for system design: \textit{technology enhancement}, \textit{ethically aligned design}, and \textit{human factors design}, highlighting the need to design responsible and reliable AI-based solutions. Yurrita et. al's multi-stakeholder framework looks at how human values connect to properties of AI/ML systems \cite{yurrita2022MultistakeholderValuebased}. To address sociotechnical questions, new human-machine configurations use HITL techniques to augment the algorithmic system through model auditing and altering \cite{gronsund2020augmenting}. Such interactions enable the user to be part of the decision making process by querying, evaluating, and editing the underlying model and data. Similarly, a Human-Centered XAI approach (HCXAI) proposes to put the user and human factors in the center of technology design, taking into consideration the interplay between values, interpersonal dynamics, and the socially situated nature of AI systems \cite{ehsan2020human}.  

Design guidelines and frameworks tend to work upwards from an interactional perspective, to support practitioners through a set of design methods, practices and examples for designing with AI, e.g., Google's People + AI Guidebook\footnote{https://pair.withgoogle.com/guidebook}. Other approaches focus on specific contexts and application areas to identify the most important design aspects and challenges for a given context, e.g., design guidelines for AI-based Tutoring Robots \cite{yang2019artificial}. Similarly, design frameworks aim  to address challenges related to \textit{Responsible AI} by integrating ethical analysis into engineering practice \cite{peters2020responsible}, or by addressing ethical challenges for specific AI application contexts, e.g., ethical AI in K-12 education \cite{zhou2020designing}. Research through Design (RtD) has been proposed as an approach to ensure that the role of AI in a system is legible to the end users \cite{lindley2020researching}. Sketching and prototyping are RtD activities which could support the ideation and implementation of HAI interactions. At a higher conceptual level, metaphors can affect expectations of performance \cite{khadpe2020ConceptualMetaphors}, help designers to understand key concepts \cite{dove2020MonstersMetaphors, murray-rust2022MetaphorsDesigners}, or lead to envisioning new kinds of relations between humans and technology \cite{luria2018DesigningRobot}. 

This work attempts to address the gap of ways to specify interactions between humans and models. Our proposed design space aims to be more pragmatic than high level frameworks, as it works from concrete actions at a user-model level, providing a link from guidelines and suggestions for design practices to the actual implementation and prototyping of existing and new types of interaction. It attempts to give designers and developers more direct facility to engage with the potentials for interaction with models by providing a palette of possibilities that map understandable human concepts to specific exchanges of information. We do this through proposing a communication protocol which can describe the interactions that take place around human user(s) and AI/ML model(s). We draw inspiration from agent communication protocols \cite[e.g.][]{o1998fipa,finin1994kqml} which use communicative acts to enable agents to communicate their intent for a specific service \cite{ahmed2009review,ahmad2007intelligence}. Working from this, the intuition behind this paper is that a common set of communicative acts can be defined to describe HAI interaction patterns as information exchanges whether by providing training examples to a model, validating a model's prediction or providing explanations. Such a specification would help to imagine richer interactions with models, and could allow a broader range of people to take part in the design of HAI interactions. 

\begin{figure}[h]
\centering
\includegraphics[width=0.65\columnwidth]{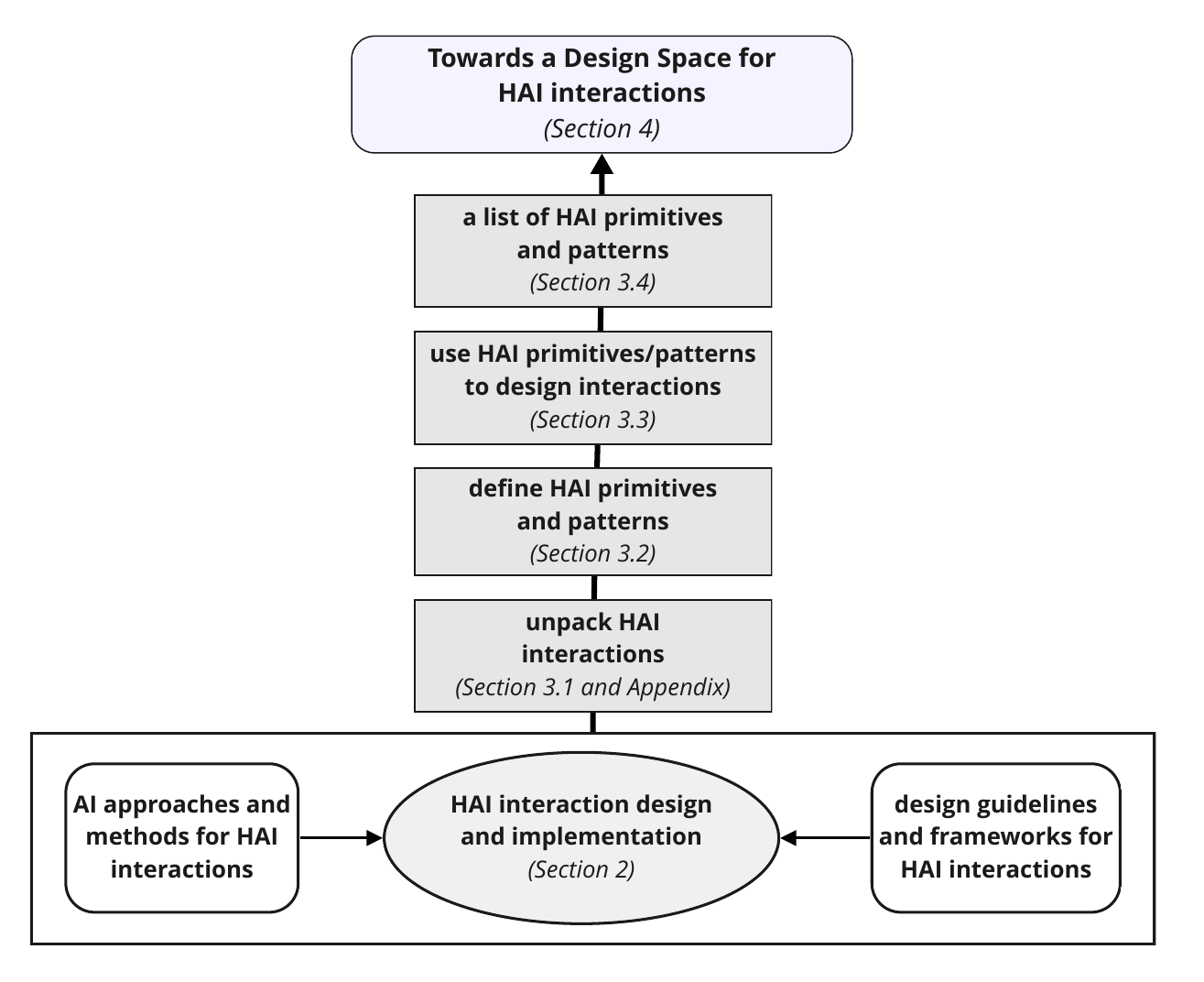}
\caption{Proposed Approach. We review existing guidelines, taxonomies and frameworks related to the design and implementation of human-AI interactions. Our goal is to unpack HAI interactions into a set of interaction primitives and patterns, based on which more complex interactions can be defined. Taking into consideration both technical and design challenges, we describe how the proposed primitives can be used to define a design space which can inform both design and implementation choices for HAI interactions. }
\label{fig:approach}
\end{figure}

Our approach defines interactions between users and models based on the intent and type of the exchanged information. Towards this, we review existing frameworks and guidelines for HAI interactions, as well as application examples, in order to identify a common set of communication channels between users and AI models. Following our approach, we unpack existing HAI interaction into low-level communicative acts. Based on this, we define a set of HAI interaction primitives which consider both design and implementation aspects of an interaction (Figure \ref{fig:approach}). Our motivation for defining a design space based on HAI interaction primitives is to: (a) provide a space in which existing interaction concepts and paradigms can be represented, bridging the gap between human and machine understanding of what an interaction entails, (b) allow for the potential invention of new kinds of interactions through exploring the space of communicative actions and interaction patterns, and (c) provide a specification which carries the required information needed for the prototyping and implementation of HAI interactions. The key contributions of the paper are:
\begin{itemize}
\item A review of HAI interaction paradigms, design frameworks and guidelines for XAI, HITL, and HI, along with a set of use cases, focusing on the different types of HAI communication.    
\item A system of HAI primitives and types that can be used to encode the interactions between users and AI models in the form of actions and message passing, capturing the intent and type of the communicated information. 
\item A collection of interaction patterns as sequences of messages which can be used as design patterns for human-AI interactions to (a) reflect existing practice and (b) create new forms of interactions.   
\end{itemize}

The paper is structured as follows:
\begin{itemize}
\item In order to define the proposed communicative framework, we review existing taxonomies, frameworks, and guidelines for HAI interactions, including HI, XAI, HITL, interactive ML (iML), and collaborative learning systems (Section \ref{related-work}). Considering the range of these interaction paradigms, approaches and concepts, our goal is to specify the interactions between humans and AI models, based on the intent and type of the communication. 
\item From this, we extract a set of communicative acts that can be combined into interaction patterns to represent a range of HAI interactions within these dominant paradigms (Section \ref{approach}). We build these acts and patterns from a combination of simple primitives and a set of types into human-readable verbs that specify an interaction. We demonstrate our approach by describing interactions from existing use cases and frameworks (Section \ref{design-space}).
\item We discuss how a design space using the proposed HAI interaction primitives and patterns can be developed to support both designers and AI practitioners to design and implement HAI interactions considering both design and technical aspects (Section \ref{discussion}).  
\end{itemize}

\section{Background and Related Work} \label{related-work}
The goal of the review is to identify the ways of communication between human users and AI models and characterize them in terms of the intent and type of the exchanged information, exploring existing HAI interaction paradigms. We provide an overview of design guidelines and taxonomies which provide suggestions on how to apply design principles for HAI interaction systems, as well as frameworks for design patterns. Finally, we present a set of HAI interaction systems focusing on the design and implementation aspects, as well as approaches for communicative protocols, towards defining a set of communicative acts and patterns for HAI interactions.  

\subsection{Human-AI Interaction Paradigms}

Human-AI interaction is defined as \textit{“the completion of a user's task with the help of AI support, which may manifest itself in non-intermittent scenarios”} \cite{van2021human}. Following this definition, there are three main HAI interaction paradigms, \textit{intermittent}, and \textit{continuous}, and \textit{proactive}, taking into consideration \textit{“how differences in initiation and control result in diverging user needs”}. Intermittent HAI interactions are user-initiated and turn-taking interactions where the system provides a response for an explicit and predefined cue. Continuous HAI interactions utilize user's implicit input, as part of a continuous input stream, and provide a response which can either be accepted or ignored by the user. Finally, proactive HAI interactions are AI-initiated and triggered by predefined changes in the system. Since such interaction paradigms can exist in parallel, there is an increasing need to consider the open challenges while designing continuous and proactive interactions in terms of how users and AI systems interact with each other.  

One of the key concerns while designing HAI interactions is the black box nature of AI models, where decisions and underlying operations are not visible or explained to humans. Moreover, human intentions are not always clear or they differ from the actual communicated action. This problem is known as the \textit{“two black boxes problem”}, based on which both human cognition (cognitive intelligence) and AI are considered as black boxes. In order to address this problem, a symmetric and collaborative model for HAI interactions has been proposed, highlighting the need for explainability for both communication channels \cite{wenskovitch2020interactive}. Based on this framework, Semantic Interaction (SI) is used as a design philosophy for symmetric and collaborative interactions between humans and AI. In the context of SI, XAI serves as a method to communicate information from AI to human users, while \textit{explainable cognitive intelligence} (XCI) is used to provide information from humans to AI systems. A main challenge while considering both communication channels for the design of HAI is the \textit{semantic gap}. Human high-level decisions and actions need to be translated to a set of model-understandable parameters, values and operations, while AI models must communicate its internal processes and decisions in a human-understandable way. HAI interactions should enable both AI systems and human users to efficiently communicate to each other and make decisions in a collaborative manner to augment both human and artificial intelligence, towards Hybrid Intelligence (HI) interactions. Akata et al. \cite{akata2020research} unpack the research challenge of building HI systems into four research themes: \textit{Collaborative}, \textit{Adaptive}, \textit{Responsible}, and \textit{Explainable HI}. An open challenge towards designing interactions with HI systems is to develop methods for designing negotiation, agreements, planning, and delegation interactions in hybrid teams, considering the needs and role of the team members. \textit{Usable} and \textit{Useful AI} are two terms which describe how interactions with AI models can be designed to provide usable solutions that are easy to understand and apply in order to satisfy user needs \cite{xu2019toward}. Towards this, AI designers and practitioners should consider both the development of AI/ML models and the design of the interactions and behaviors around such models, including user-centric techniques, such as explainability and user control. 

The integration of explainability and user control features creates new types of interactions between users and models. Designing interactions with (X)AI models should follow specific design principles towards the development of \textit{Responsible AI} systems, related to fairness, transparency, and privacy aspects \cite{arrieta2020explainable}. In terms of user control and feedback, HITL and iML techniques enable users to interact with AI/ML models in order to guide, steer and facilitate the learning process. Human users can be involved in the different stages of the ML pipeline: data extraction, integration and cleaning, iterative labeling, as well as model training and inference \cite{chai2020human, wu2022survey}. Related to how human users can intervene with models, model contestation is defined as an interaction with “humans challenging machine predictions” \cite{hirsch2017designing}. A proposed framework for contestable AI by design \cite{alfrink2022contestable} highlights a set of sociotechnical features and practices for model contestation and hybrid decision making, including interactive controls, explanations, and intervention requests. Such approaches highlight the need for designing new types of interactions, considering the different interaction paradigms, leading to the development of design frameworks and guidelines for HAI interactions.   

\subsection{Design Guidelines, Frameworks and Taxonomies for Human-AI interactions}

Design guidelines and frameworks aim to support designers on how to apply specific design principles while designing interactions with AI systems. Microsoft Research has proposed a set of guidelines for designing HAI interactions \cite{amershi2019guidelines}. These guidelines provide design choices and examples for interactions between users and AI systems, considering aspects of system transparency, explainability, user feedback and control. Designing interactions which support such features requires human users to interact with the system at a \textit{user-model} level. That means that designers should take into consideration the different ways that users can efficiently interact with an AI model, as well as the underlying design and technical challenges while prototyping such interactions. A process model for co-creation of AI experiences (AIX) describes how designers can be familiarized with AI models and their functionalities in order to include them in the design process as design materials \cite{subramonyam2021towards}. However, there is a lack of design innovation in envisioning how ML/AI might improve and create UX value to users. Yang et al. \cite{yang2018mapping} highlight the need to support UX practitioners by creating design patterns. They propose four channels based on which AI/ML capabilities can create UX value to users, namely \textit{self}, \textit{context}, \textit{optimal}, and \textit{utility-capability}, considering what users can infer about these aspects through their interaction with models. Such frameworks and guidelines aim to support the ideation process of designers while prototyping and designing interactions between users and an AI system.   

Taxonomies and frameworks for XAI approaches and methods can support the design and implementation of interactions with explainable systems, considering different aspects of XAI-based interactions. A collection of XAI-based questions has been proposed as a design space for XAI interactions  considering diverging user characteristics, e.g., user needs, role, expertise, and experience  \cite{liao2020questioning,liao2021human}. Other taxonomies focus on the different aspects of the use case, e.g., problem definition, explanator properties, and evaluation metrics \cite{schwalbe2021comprehensive} or the relations between the use case, the AI system, and the explanation algorithm \cite{adhikari2022towards}. Design frameworks can provide guidelines on how to design and evaluate XAI-based interactions by mapping design goals to evaluation methods based on the target population \cite{mohseni2021multidisciplinary}. Focusing on user evaluation for XAI interactions, Chromik et al. \cite{chromik2020taxonomy} classify evaluation methods based on task-related, participant-related and study design-related dimensions. Such frameworks can support the selection of appropriate XAI methods for a given interaction concept. Sperrle et al. \cite{sperrle2020should} proposed a dependency model for XAI processes  as a design space for human-XAI interactions, considering the potential bias that can be propagated during the interactions between different stakeholders and end-users. Wang et al. \cite{wang2019designing} followed a theory-driven approach to link explanation features to user reasoning goals, resulting to a conceptual framework which can inform the selection of appropriate explanations towards mitigating possible cognitive biases. Designing XAI interfaces should also consider specific design principles based on the interaction concept and explanatory goals \cite{chromik2021human}. Human users may interact with XAI interfaces for \textit{information transmission}, where the goal of XAI is to help the user understand the underlying AI behavior, while XAI-interaction as \textit{dialogue} refers to an iterative user-driven communication through user queries and AI responses/explanations. For both concepts, the AI behavior does not change in contrast to XAI-interaction as \textit{control}, where the user provides feedback to adjust the (explainable) model until a desired AI behavior is reached, which relates to HITL/iML interactions.   

HITL/iML methods enable human users to participate in the decision making and model training process. Design principles for user control and feedback interfaces include interactivity to promote rich interactions, providing explicit and clear task goals, support user understanding and engagement, and capture user's intent based on their input \cite{dudley2018review, nadj2020power}. Moreover, designing interactions with HITL/iML interfaces requires appropriate learning methods, considering new forms of relationships between humans and ML algorithms \cite{mosqueira2022human, cui2021understanding}.
The design and implementation of iML/HITL interfaces depend on the type of user feedback and can affect both user experience and system/model performance. Michael et al. \cite{michael2020interactive} discuss how the different types of cognitive feedback can be integrated to the iML mechanism, i.e., \textit{self-reporting}, \textit{implicit feedback}, and \textit{modeled feedback}, through different feedback mechanisms. For example, domain-expert feedback can be communicated to the system and translated to model updates, i.e., modify dataset or model parameters \cite{chen2022perspectives}. Such HITL/iML approaches consider the effects of the interaction on human engagement and feedback quality. Since explanations can be used to enhance user's understanding of the model's performance, they can play an important role in ensuring a high quality of user feedback. Explanations can be combined with interactive capabilities enabling users to train ML models from scratch, resulting to a closed loop of XAI- and HITL-based interactions \cite{teso2023}.   

In the domain of hybrid intelligence, a taxonomy for design knowledge organizes the design decisions for HI interactions, including XAI and HITL methods, along four dimensions: task characteristics, learning paradigm, human-AI interaction, and AI-human interaction \cite{dellermann2021future}. 

Focusing on hybrid decision making, a common interaction paradigm is \textit{backward reasoning} design where the system's goal is to provide correct outputs and explain them to the user, following an AI-driven interaction. \textit{Forward reasoning} can provide more agency to human users, involving both human and AI in the decision loop in order to address the issue of system uncertainty \cite{zhang2021forward}. 

Such guidelines, frameworks and taxonomies can help AI designers to select appropriate types and methods for HAI interactions, based on the system requirements and interaction goals. Designing interactions which comply with high-level guidelines can be challenging, especially when multiple requirements need to be met. Design patterns can be used to structure the interaction by defining smaller parts and combining them into more complex interactions.    

\subsection{Design Patterns for Human-AI Interactions} 
Describing an HAI interaction using design patterns is not straightforward but it can simplify the design of complex AI systems and make them transparent in terms of their implementation requirements. Focusing on co-creation applications with Generative Adversarial Networks (GANs), Grabe at al. provide a list of HAI interaction patterns which categorize  interactions based on the AI's co-creativity support level \cite{grabe2022towards}. These interaction patterns are sequences of actions which describe an activity, e.g. initialize, create, adapt model, and can be combined to design interactions with different co-creativity support levels. A co-creative framework for interaction design analyzes the interactions between users, AI models, and the shared product \cite{rezwana2022designing}, resulting to a set of design choices. The design choices can define the behavior of the AI, as a generative, improvisational, or advisor agent. Design patterns in HAI co-learning are used to design interactions which enable humans and AI to share knowledge and experience. Learning Design Patterns (LDP) refer to sequences of interactions that aim to initiate and facilitate the co-learning process \cite{schoonderwoerd2022design}, either by identifying knowledge gaps that team members may have or by enabling team partners to learn from other team members. In the context of AI-based game design, AI can serve different roles based on the selection of the interaction pattern \cite{treanor2015ai}. For example, AI can act as \textit{role-model}, where users need to imitate an AI agent to complete a game, or as an (AI trainee) which enables the user to teach the agent to do something in the game. 

Design patterns have also been proposed considering specific interaction paradigms. Focusing on collaborative learning and hybrid intelligence systems, humans and computers can learn from each other through an iterative process, combining human-in-the-loop with computer-in-the-loop interactions \cite{wiethof2021hybrid}. More specifically, users and computers can communicate through a set of learning process patterns: (a) \textit{decision support}, (b) \textit{exploration}, and (c) \textit{integration}. Each pattern refers to different types of interactions between human users and computers as an exchange of inputs, outputs, and feedback/explanations. iML methods have been used as design materials for movement interaction design \cite{gillies2019understanding}, as an approach to design interactions with models based on human movements. The interaction is defined as a closed loop between designers and software which allows designers to provide information in the form of movement examples and parameters, and receive AI test outputs and visualizations. Through this loop, both AI and designers can learn from each other; designers can reflect on how movement parameters may affect AI behavior and AI can update its internal models based on the feedback received from the designer. However, even prototyping such interactions would require a lower-level specification of how users and models communicate and exchange information during such interactions at a user-model level.  

Design patterns have been used to specify and categorize the interactions with hybrid intelligence and reasoning systems at a user-model level \cite{van2021modular}. A modular approach is used to specify patterns of interactions of hybrid systems, including both data-driven (ML) and knowledge-driven (symbolic AI) approaches. Based on this approach, data (numbers, texts, tensors, and streams) and symbols (labels, relation, traces) are the two types of instances that can be used for AI model operations. The authors present a list of design patterns for hybrid AI systems and demonstrate how such patterns can be used to describe interactions from existing applications. Their proposed approach specifies the interactions in terms of model operations, including training, inference, and transformation, and can be extended to capture the concept of different types of human actors included in interactions with hybrid AI systems.

\subsection{Design and Implementation of Human-AI interactions}

In order to explore how users and AI models communicate and exchange information during HAI interactions, we review applications of HAI systems, focusing on the interaction design and implementation aspects. Advances in computational and learning methods, as well as the plethora of human-generated data has recently led to innovative AI systems which can be used by non-expert users for complex tasks. More specifically, OpenAI has released two types of deep learning models which generate new content based on user's input (promtps); DALL-E generates images based on a textual description \cite{ramesh2021zero} and chatGPT is a Large Language Model (LMM) which generates structured textual content based on user's prompts and questions\footnote{https://openai.com/blog/chatgpt/}. Despite their complex architecture and advanced learning methods, interactions with such models are straightforward; the user provides a prompt and the model returns the generated data. However, the implementation of the training process requires methods to integrate human feedback and expertise in order to facilitate model learning. The training process utilizes an HITL method to integrate human users to the learning process. More specifically, Reinforcement Learning from Human Feedback (RLHF) is used to integrate human expertise to the learning process. Based on this approach, human users provide different types of feedback based on the learning process step. During model initialization, human users demonstrate the desired optimal behavior of the model (response to a prompt). For model optimization, a reward model was trained based on user ranking of possible model outcomes (output) for a given prompt (input). Such an approach demonstrates how different types of users can interact with different types of models, considering both the goal of such interactions, e.g., interface design, and their technical implications, e.g., selection of learning/update mechanisms.  

Interactions can become complex when designing for explainability and user feedback, even with less advanced or pre-trained models. Designing an XAI-based interaction can serve as a channel to communicate additional information about the agent's performance for a given concept. For example, model transparency (e.g., visualizing model's confidence) has shown to improve user's trust during an interaction with a virtual agent with speech recognition capabilities \cite{weitz2021let}. More specifically, human participants interacted with a virtual agent and an underlying speech recognition system. The virtual agent was used as a visualization means for XAI-based feedback to enhance user's understanding and trust of the speech recognition system. During these interactions, the user provides a model input (utterance) and the model communicates both its output and additional feedback (prediction/explanation). HITL-based interactions can allow for user feedback and control within an interaction in various ways \cite{cui2021understanding}. The type of human feedback (as well as amount, frequency, granularity, etc.) should be inline both with the user's characteristics (e.g., role, expertise, preferences, etc.) and the model's characteristics (model architecture, learning/update rules, etc.).   

HAI interactions can also involve multiple users with different roles, expertise and intentions. In such interactions, an AI model should be able to interact with multiple users and behave in a different way based on the user's characteristics. An interactive RL-based framework has been proposed for a personalized robot-based cognitive game which involves two different types of users \cite{tsiakas2018task}. A player (primary user) interacts with a robot (RL) through a game-based interaction (robot sets a game difficulty and user plays the game) and a supervisor (secondary user) who can remotely supervise and control the interactions. In terms of communicated information, the player provides implicit feedback to the model through task performance and engagement, while a supervisor can monitor the model's decisions through a informative UI and modify the robot's decisions, when needed. Both communication channels contribute to the model's learning updates and the decision making process in a different way. Similarly, the Human Experience Transfer Model (HETM) \cite{maettig2020approach} combines two interaction loops; one for a human expert (trainer) who guides the model learning process and one for the human learner (trainee) who can provide feedback during the learning interaction. The trainer loop aims to facilitate the model training while the learner loop aims to personalize the learner's interaction. These loops specify two different interactions with different communication channels and model update methods. Such approaches for designing and implementing HAI interactions highlight the need to specify design patterns considering the underlying implementation requirements. Our work moves towards a design space which takes into consideration the technical aspects while designing such human-model communications.   

\subsection{Communicative Protocols}
Based on the Semantic Interaction paradigm \cite{wenskovitch2020interactive}, both types of interacting agents, humans and models/systems, should exchange information in a mutually understandable way, to ensure meaningful communicative acts from both sides. In order to do so, user actions and intentions should be translated into an understandable model-specific format. On the other hand, model actions should be effectively communicated and understandable to human users. In the domain of communicative protocols, process calculus provides a tool for the high-level description of interactions, communications, and synchronizations between agents or processes. Agent modeling and communication languages make use of communicative acts to enable agents to communicate their intent for a specific service \cite{ahmed2009review,ahmad2007intelligence}. For example, the Knowledge Query and Manipulation Language (KQML) and the Foundation for Intelligent Physical Agents agent communication language (FIPA-ACL) are two major developments in message exchange interaction protocols between agents \cite{o1998fipa,finin1994kqml}. Such protocols aim to encode the communicative acts of an agent, as well as to model the agents' knowledge using semantics. More specifically, the FIPA-ACL defines a set of (primitive and composite) communicative acts, along with a formal definition of the underlying semantic model. The semantic language formalism allows for the specification of the mental model of the agent (e.g., belief, uncertainty) and the effects of the acts on the interacting agents. A set of primitives defines how agents communicate information, including an \textit{assertive inform} act, based on which an agent provides a message (in a proposition form), and a \textit{directive request} act, which describes a request of sender (for an action from the receiver). Taking into consideration multi-agent coordination and social norms, Lightweight Coordination Calculus (LCC) can be used to specify the behaviours required of agents interacting in a given social context \cite{robertson2005lightweight}, as a form of `electronic institution' \cite{dinverno2012CommunicatingOpen}. This formalisation has been extended to model coordination and communication between multiple interacting actors in social context \cite{murray2015softening}. Our approach is inspired by such communicative protocols; however, our aim is to encode the interactions capturing the intent and the type of the exchanged information and not the effects of the act on the underlying semantic models of the interacting agents. 

\section{Proposed Approach: Unpacking Human-AI interactions into interaction primitives} \label{approach}
In this section, we present our approach towards a design space for HAI interactions. Our goal is to identify common interaction patterns from HAI interactions and specify them in terms of the type and intent of the communicated information. We provide a semi-formal definition of interaction primitives and we demonstrate how they can be used to define interactions between users and models. The outcome of the proposed approach is a list of interaction patterns and the actions that define them, along with a description of the interaction in terms of design and implementation aspects. We discuss how the proposed formalization can be used towards a design space. 

\subsection{Unpacking Human-AI interactions to patterns and primitives} \label{use-cases}
\begin{figure}[h]
\centering
\includegraphics[width=\columnwidth]{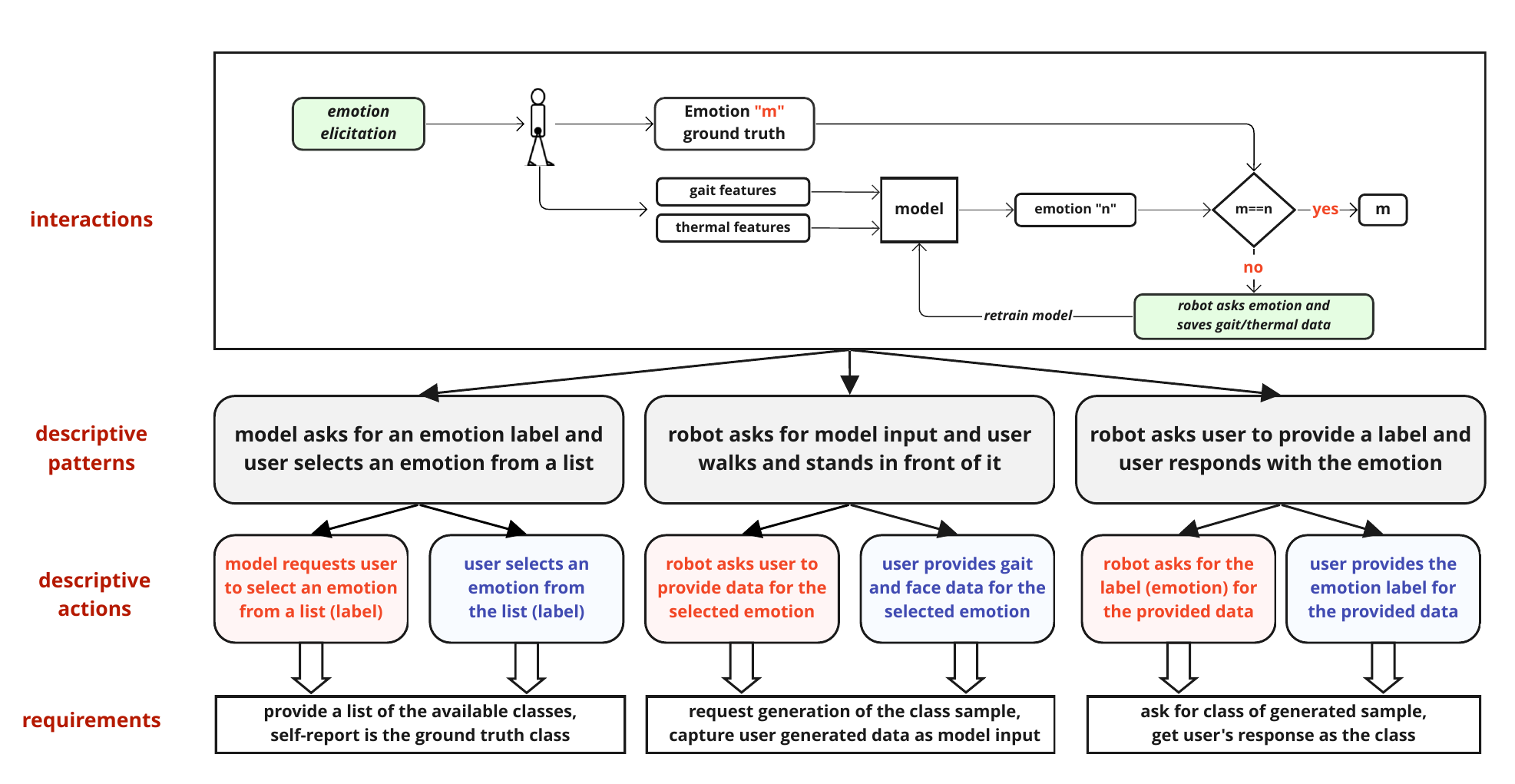}
\caption{Interactive robot learning \cite{yu2019interactive} using descriptions of interaction patterns and primitives. The image shows the original interaction schema as a box and arrow diagram, followed by a high level description of the HAI interactions patterns, the unpacking into actions involving the exchange of information between the human and the system, and a description of the interaction requirements.}
\label{fig:unpacking1}
\end{figure}

Our approach aims to unpack HAI interactions into low-level communicative acts or \textit{interaction primitives} which can specify the type and intent of the communication during a single interaction step. Such interaction steps are defined as user-model interactions, e.g., user provides an input, model provides an output, etc.. Our motivation is that unpacking HAI interactions can provide us with insights about the underlying design and technical challenges for HAI interaction patterns. In order to unpack HAI interactions into interaction primitives, we identify the interaction patterns between users and models and we specify the intent and type of the communicated information. The goal of this approach is to identify a set of communicative acts from different HAI interaction scenarios, towards a formalization for HAI interaction primitives. We demonstrate our approach using an interactive robot learning system for multimodal emotion recognition as a running example \cite{yu2019interactive}. More use cases were selected for unpacking in order to cover different types of interactions, including XAI, HITL and hybrid intelligence interactions (See Appendix section \ref{more-unpacking}). We provide a description of the HAI interactions in the form of patterns and actions, and we highlight the design and implementation aspects of the interactions (Figure \ref{fig:unpacking1}).

\textit{\textbf{Description of the HAI interactions.}} The goal of this system is to collect and annotate human-generated data for an emotion classification model, through the human-robot interaction. In terms of interaction design, the interaction starts with an emotion elicitation session (user watches a clip which invokes a specific emotion). After the session, the user is asked to select their current emotional state from a list of emotions. This emotional state is provided as a model output (class) and it is used by the robot for the interaction as the ground truth emotion. The robot asks the user to walk towards and stand in front of it, demonstrating the selected emotion. The robot uses the gait/thermal data to predict the user's emotional state. If user's response (ground truth) is different from the predicted emotion, the robot asks the user about their current emotional state, annotating the collected gait/thermal data which are used to retrain the model. 

\textbf{\textit{Interaction Patterns.}} We unpack the HAI interactions into three types of interactions between the user and the robot (model): \textit{(a) class selection:} model asks user to select an emotion from a list (class) -- user selects an emotion from the list, \textit{(b) new class sample:} robot asks user to provide an input by demonstrating the selected emotion -- user responds by walking and standing in front of the robot, and \textit{(c) annotate sample:} model makes a prediction and asks the user for labeling, if prediction is different from ground truth -- user provides the correct label by responding to the robot's question. During these interactions, user and robot exchange information in the form of \textit{model input} (gait/thermal data) and \textit{model output} (emotion from list, model prediction, robot question, user response). Model input is communicated as a set of raw data generated and captured during user's activity (walking and standing). Model output is communicated (a) through user's selection from a list, (b) implicitly through the robot's (model) prediction, and (c) as a user response to the robot's request during their interaction.

\textbf{\textit{Design and implementation aspects.}} The robot runs an underlying emotion classification model which uses walking (gait data) and facial expressions (thermal data) as model inputs to predict the user's emotional state. The system utilizes the interaction with the user in order to dynamically improve the classification model by retraining on new (labeled) data. The design of the interactions enables the user to participate in the interactive learning process in an implicit way. The prediction model consists of two models (gait and thermal models) and the final prediction is estimated through a modified confusion matrix. After each interaction with the user, the model is retrained including the new data from the user as an input-output pair. The robot initiates the interactions to request for user's input (emotion label, input data and user response), and uses the responses to make interaction decisions; if model's prediction is different from user's input, the model implicitly requests the correct label from the user without communicating its own prediction. A possible limitation of this implementation is that it highly depends on the quality of user's input. User may provide inaccurate information both during the selection (label) and the demonstration (sample) of the emotion.

\subsection{Defining Interaction Primitives and Patterns for Human-AI interactions} \label{definitions}
\begin{figure}[h]
\centering
\includegraphics[width=0.85\columnwidth]{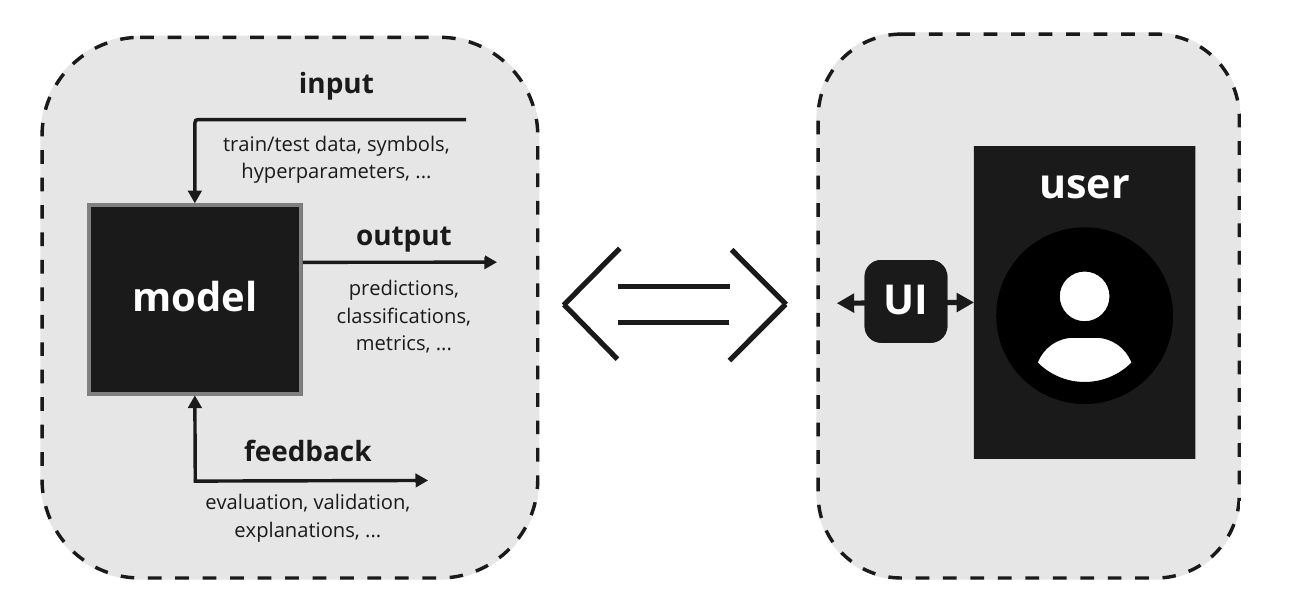}
\caption{Communication types during human-model interactions. AI models provide/receive information in a model-specific format: (a) input (test/train data, relations, hyperparameters, etc.), (c) output (labels, predictions, estimated parameters, projections, etc.) and (c) feedback (explanations, validation, requests, etc.). Users send/receive information through the UI which translates user actions to a model-understandable format and vice versa. Our approach aims to specify this communications  using interaction primitives.}
\label{fig:hmi}
\end{figure}

\begin{figure}[h]
\centering
\includegraphics[width=0.99\columnwidth]{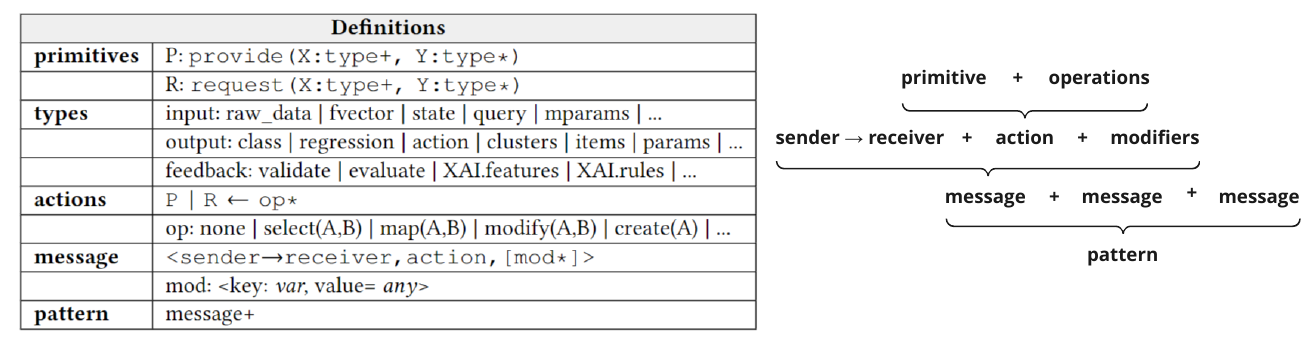}
\caption{Definitions and visual representation for interaction primitives, types and actions. An action is defined as a primitive specified by a set of operations. Actions are communicated through messages, by specifying the interacting agents and the modifiers of the message. Actions are reusable and can be used by different messages. Sequences of messages can form patterns of interaction.}
\label{fig:definitions}
\end{figure}

Our motivation is to characterize HAI interactions in terms of the communicated information. In order to describe such interactions, we need to define how users and models interact with each other at a human-model level to either provide or request information. Based on the Semantic Interaction paradigm \cite{wenskovitch2020interactive}, both interactive agents should be able to communicate in an understandable and meaningful way. Following the taxonomy of instances for (hybrid) AI systems \cite{van2021modular}, model-specific information can be in different formats, including train/test data, learning and estimated parameters, symbols, rules, labels, and others. Moreover, user feedback and explanations can be communicated in a model-specific format. Based on the model specifications, user actions are translated through the user interface into machine-readable information. Considering these different types of communication (Figure \ref{fig:hmi}), we define a set of \textit{interaction primitives} to specify user/model actions in terms of the intent (provide or ask for information) and the type (format) of the communication. We demonstrate how the proposed formalization can be used to design interactions as exchanges of messages between the interacting agents. The descriptive formalization, along with a visual representation is shown in Figure \ref{fig:definitions}. We provide a description of the definitions, as well as examples to demonstrate our approach.  


\subsubsection{Interaction primitives and types}
We define an \textit{interaction primitive} as a low-level communicative act which specifies the type and intent of the communicated information. Considering the intent during an interaction step, the communicating agents can either \textit{provide} or \textit{request} information. We define a \textit{provide primitive} as \prim{P: provide}{X:type+,Y:type*} to describe the act of providing information (provide type \texttt{X} referencing type \texttt{Y}). Similarly, a \textit{request primitive} is defined as \prim{R: request}{X:type+,Y:type*} to describe the act of requesting information (request type \texttt{X} referencing type \texttt{Y}). For both primitives, the optional argument (\texttt{Y:type}) is used to reference an additional type. Considering the format of the information, human users and models can provide and request the following types of information: \texttt{input}, \texttt{output}, and \texttt{feedback}. The definition of the types (and subtypes) can inform the design and implementation of the specific interaction, since they characterize how users and models exchange information. 

We provide an overview of the types (and their possible subtypes) with examples. 

\begin{itemize}
    \item \texttt{input} is information to be fed into a model. The \textit{subtype} depends on the model type and architecture, including numeric vectors, world state information, images, video frames or sequences, user preferences, audio, text and so on. Model-specific parameters (hyperparameters) are also used as model input, e.g., learning rate, number of clusters, model sensitivity, etc.      
    \item \texttt{output} is information coming out of a model --- class labels,  feature estimations, lists of recommendations, selected actions, generated images, projections of the input space, cluster labels, lower dimensional data and so on. This can also include learned parameters such as model weights, confidence values, loss etc.   
    \item \texttt{feedback} refers to the additional information that can be provided both from users and models, including (requests for) evaluation, validation, explanations, etc. Explanations can be provided in different modalities, e.g., salient maps or natural language. Users can provide explanations to justify their decisions, as well as feedback for a model's decision. Similarly with model input/output, feedback is model-specific.  
\end{itemize}

The proposed interaction primitives and types can describe different \textit{actions} between users and AI models. Primitives capture the intent of the communication (provide/request) and types describe the format of the communicated information. For example, actions \prim{u1: provide}{X:input} and \prim{u2: request}{Y:output, X:input} can both describe a user communicating a model input type. However, the first action describes a user providing \texttt{X:input}, while the second one describes a user request for \texttt{Y:output} given the provided input. These actions are similar in terms of the provided input (\texttt{X:input}) but they differ on the type of the action (provide input vs. request output). 

\begin{table}[h]
\centering
\resizebox{0.99\columnwidth}{!}{
\begin{tabular}{|lll|}
\hline
\multicolumn{3}{|c|}{\textbf{Examples of interaction primitives and types}}                                             \\ \hline
\multicolumn{1}{|l|}{\prim{provide}{X:input}} & \multicolumn{1}{l|}{provide an input}                    & upload/capture an image             \\ \hline
\multicolumn{1}{|l|}{\prim{provide}{X:output,Y:input}} & \multicolumn{1}{l|}{provide an output for a given input} & detect a face in the image          \\ \hline
\multicolumn{1}{|l|}{\prim{provide}{[X:input,Y:output]}} & \multicolumn{1}{l|}{provide an input-output pair}        & show an image and the detected face \\ \hline
\multicolumn{1}{|l|}{\prim{request}{X:output,Y:input}} & \multicolumn{1}{l|}{request an output for an input}            & ask if there is a face in a given image           \\ \hline
\multicolumn{1}{|l|}{\prim{request}{[Y:input,X:output]}} & \multicolumn{1}{l|}{request an input-output pair}              & ask for an image with a detected face (if any)    \\ \hline
\multicolumn{1}{|l|}{\prim{provide}{Z:output,[Y:input,X:output]}} & \multicolumn{1}{l|}{provide output for an input-output pair}   & modify an existing bounding box on an image       \\ \hline
\multicolumn{1}{|l|}{\prim{request}{F:feedback,Y:output}} & \multicolumn{1}{l|}{request feedback for the given output} & ask for confirmation about a bounding box\\ \hline
\multicolumn{1}{|l|}{\prim{request}{X:[input]}} & \multicolumn{1}{l|}{request a set of inputs} & ask for a set of input images\\ \hline
\end{tabular}
}
\caption{Examples and descriptions of interaction primitives and types}
\label{tab:primitives}
\end{table}

The selection of the types (and subtypes) depends on the model specifications and can inform design and implementation choices. For example, given that \prim{provide}{X:input.raw\_data} can describe a user who communicates a model input in the form of raw data, the interaction design (e.g., interface) should enable the user to interact with raw data (e.g., images). Primitives and types can also provide information about the requirements related to the goal of the interaction. Let us consider a face recognition model which takes as an input an image (raw data or extracted features) and detects faces (if any) in the form of bounding boxes. For this case, both of the defined actions \prim{request}{Y:output;X:input} and \prim{request}{Y:output, [X:input, Z:output]} can describe a request for an output (detected face image). The first one is an output request for a given input image, and can describe a face detection interaction. The second action is an output request given an an input-output pair (image - bounding box) and can describe a modification request for the model's detection. Table \ref{tab:primitives} provides examples of interaction primitives and types along with a description in the context of face detection.  

\subsubsection{Actions, operations and modifiers} 
In order to specify an interaction between a user and a model for a given interaction context, we define a message as \texttt{msg:}\mesg{sender}{receiver}{action}{mod*} to describe the communication of an action from a sender to a receiver. An action is defined as \texttt{P $\vert$ R $\leftarrow$ operations} and specifies a primitive action by adding a description of how the arguments need to be communicated, through a set of \texttt{operations}. The operations contextualize an action by specifying the preconditions for its communication in a given interaction context. More specifically, the operations define how the argument is being created and describe the relations between multiple arguments. A list of operations includes, but is not limited to: 

\begin{itemize}
    \item \texttt{select(A,\textit{B})}: argument \texttt{A} is selected (from a given set \texttt{B} - optional argument) -- this operation can describe the selection of an item from a list or set of choices, i.e., recommendations, labels, samples, etc.  
    \item \texttt{map(A,B)}: argument \texttt{A} is mapped to argument \texttt{B} --  this operation can be used to describe a model prediction (e.g., classification, regression, clustering), human labeling, evaluation, etc. 
    \item \texttt{modify(A,B)}: modify argument \texttt{A} to argument \texttt{B} -- this operation can describe a modification of a sample (modify input image), an alternate decision (change label), etc.  
    \item \texttt{create(A)}: create new argument \texttt{A} -- this operation can describe data acquisition or generation (e.g., image, sound), a human annotation  (e.g., new label), etc.     
\end{itemize}

Finally, a set of \texttt{modifiers} (mod: \textless{}key: \textit{type}, val=\textit{any}\textgreater{}) can be used to further characterize the message in terms of interaction requirements, as a free-form annotation feature. Modifiers can provide information about the interaction modality, interface elements (e.g., buttons, forms, etc), type of communication (e.g., explicit vs. implicit), etc. Based on the above, the definition of an action includes the primitive (provide/request), the type (input, output, feedback), the operations needed to communicate the types, as well as additional information for the interaction through the modifiers, and describes the communication of an action as a single message from a sender to a receiver. A sequence of actions for a given interaction context is defined as an \textit{interaction pattern}.

The proposed formalization allows for a custom definition of an action and its specification as a message for a given interaction context. For example, we can define the action \texttt{\textbf{req-new\_sample(M)}}$\equiv$\texttt{request(M:input.raw\_data)}$\leftarrow$\texttt{create(M)} to describe a request for a generated input in the form of raw data. Sequences or exchanges of messages can be used to define interactions between users and models for a given context. For interactions with a face recognition model, we can define a message \texttt{msg:} \texttt{<model$\rightarrow$user, \textbf{req-new\_sample(M)},[M:image]>} to describe the model's request for an image from the user. As a response to this request message, the user could respond with either:
\begin{itemize}
    \item[-] \texttt{msg1:} \texttt{<user$\rightarrow$model, \textbf{generate-sample(M)},[M:image]>},\\where: \texttt{generate-sample(M)}$\equiv$\texttt{provide(M:input.raw\_data)}$\leftarrow$\texttt{create(M)}, or 
    \item[-] \texttt{msg2:} \texttt{<user$\rightarrow$model, \textbf{req-gsample\_class(M,L)},[M:image]>},\\
    where: \texttt{req-gsample\_class(M,L)} $\equiv$ \texttt{request(L:output.label,M:input.raw\_data)} \\$\leftarrow$\texttt{[create(M),map(M,L)]}
\end{itemize}

Based on the first message, the user responds by providing the requested input (create/capture image), while with the second message, a request is made to the model for face detection given the generated input (map/assign the captured image to a label). The same action can be communicated by different messages. A message specifies the communication of an action from a sender to a receiver in a given interaction context. For example, a speech recognition model can communicate its request for user-generated input (speech) using \texttt{msg':} \texttt{<model$\rightarrow$user, \textbf{req-new\_sample(M)},[M:speech]>}. From this, a set of actions can be defined as a vocabulary which can be used in different HAI interactions. Such actions and messages can be used to design interaction patterns as sequences of messages between users and models. Given the message definitions above, we can define \texttt{query\_input}$\equiv$\texttt{[msg,msg1]} as a query pattern to describe the request and communication of a model input, while \texttt{query-label\_input}$\equiv$\texttt{[msg,msg2]} could describe an interaction where the user awaits for the model's decision based on a new sample. These patterns serve different interaction goals and also require different implementation. In the next section, we provide a set of action definitions and interaction patterns from existing HAI interactions and frameworks. 

\subsection{Designing interaction patterns using actions and messages} \label{design-use-case} 
In this section, we describe how the proposed primitives can be used to define actions and interaction patterns. Such patterns serve a given goal during the interaction and can be applied for the design of other interactions. For the \textit{interactive robot learning for emotion recognition} (Section \ref{use-cases}), we define the following messages and actions (Table \ref{tab:actions}):

\begin{table}[h]
\resizebox{\columnwidth}{!}{
\begin{tabular}{|ll|l|}
\hline
\multicolumn{2}{|c|}{\textbf{Message}} & \multicolumn{1}{c|}{\textbf{Action definition}} \\ \hline
\multicolumn{1}{|l|}{\texttt{A1}} & \begin{tabular}[c]{@{}l@{}}\texttt{<model$\rightarrow$user,\textbf{req-class\_selection(Y,L)},}\\\texttt{{[}Y:reqSelfReport;L:listEmotions{]}>}\end{tabular} & \begin{tabular}[c]{@{}l@{}}\texttt{req-class\_selection(Y,L) $\equiv$} \\ \texttt{request(Y:output.label, L:{[}output.label{]})}\\$\leftarrow$ \texttt{select(Y,L)}\end{tabular} \\ \hline
\multicolumn{1}{|l|}{\texttt{A2}} & \begin{tabular}[c]{@{}l@{}}\texttt{<user$\rightarrow$model,\textbf{select-class(Y,L)},}\\\texttt{{[}Y:SelfReport;L:listEmotions{]}>}\end{tabular} & \begin{tabular}[c]{@{}l@{}}\texttt{select-class(Y,L) $\equiv$} \\ \texttt{provide(Y:output.label,L:{[}output.label{]})}\\$\leftarrow$ \texttt{select(Y,L)}\end{tabular} \\ \hline
\multicolumn{1}{|l|}{\texttt{A3}} & \begin{tabular}[c]{@{}l@{}}\texttt{<model$\rightarrow$user,\textbf{req-new\_class\_sample(X,Y)},}\\\texttt{{[}X:reqWalkStand,Y:SelfReport{]}>}\end{tabular} & \begin{tabular}[c]{@{}l@{}}\texttt{req-new\_class\_sample(X,Y) $\equiv$} \\\texttt{request(X:input.raw\_data,Y:output.label)}\\$\leftarrow$\texttt{create(X), map(X,Y)}\end{tabular} \\ \hline
\multicolumn{1}{|l|}{\texttt{A4}} & \begin{tabular}[c]{@{}l@{}}\texttt{<user$\rightarrow$model,\textbf{generate-class\_sample(X,Y)},}\\\texttt{{[}X:WalkStand,Y:SelfReport{]}>}\end{tabular} & \begin{tabular}[c]{@{}l@{}}\texttt{generate-class\_sample(X,Y) $\equiv$} \\ \texttt{provide(X:input.raw\_data,Y:output.label)}\\$\leftarrow$\texttt{create(X), map(X,Y)}\end{tabular} \\ \hline
\multicolumn{1}{|l|}{\texttt{A5}} & \begin{tabular}[c]{@{}l@{}}\texttt{<model$\rightarrow$user,\textbf{req-sample\_class(X,Y)},}\\\texttt{{[}X:WalkStand;Y:reqSelfReport{]}>}\end{tabular} & \begin{tabular}[c]{@{}l@{}}\texttt{req-sample\_class(X,Y) $\equiv$}\\\texttt{request(Y:output.label,X:input.raw\_data)}\\$\leftarrow$ \texttt{map(X,Y)}\end{tabular} \\ \hline
\multicolumn{1}{|l|}{\texttt{A6}} & \begin{tabular}[c]{@{}l@{}}\texttt{<model$\rightarrow$user,\textbf{annotate-sample(X,Y)},}\\\texttt{{[}X:WalkStand;Y:SelfReport{]}>}\end{tabular} & \begin{tabular}[c]{@{}l@{}}\texttt{annotate-sample(X,Y)$\equiv$}\\\texttt{provide(Y:output.label, X:input.raw\_data)}\\$\leftarrow$\texttt{map(X,Y)}\end{tabular} \\ \hline
\end{tabular}
}
\caption{Messages and action definitions for the interactive robot learning interactions}
\label{tab:actions}
\end{table}
\begin{figure}[h]
\centering
\includegraphics[width=\columnwidth]{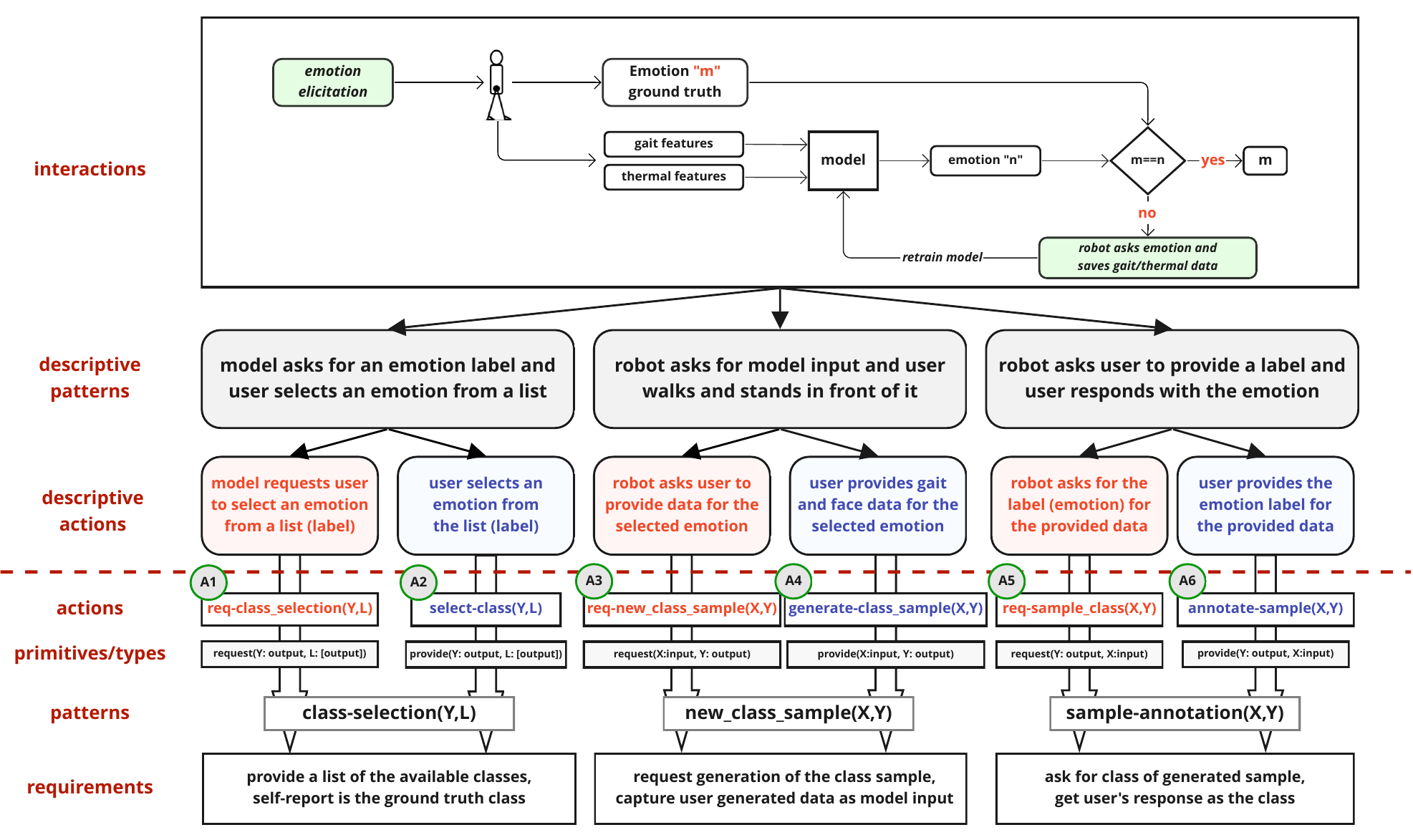}
\caption{Description of the HAI interactions using definitions of interaction primitives and patterns. The diagram illustrates our unpacking approach and the process of describing interaction primitives and patterns in terms of the interaction requirements.}
\label{fig:hai1}
\end{figure}

\textbf{\textit{Interaction Patterns.}} The defined messages and actions can semi-formally specify: (a) the intent of the sender's act (primitive), (b) the type of the communicated information (types and operations), and (c) how the information is communicated (modifiers). Based on the unpacking of the interactions, we define the following interaction patterns: \texttt{class\_selection$ \equiv$ [A1,A2]} , \texttt{new\_class\_sample $\equiv$ [A3,A4]} and \texttt{sample\_annotation $\equiv$ [A5,A6]} (Table \ref{tab:interaction1}). All patterns describe a query-response interaction initiated by the robot (model). The goal of the first pattern is to set the target class, without requiring any information about the model input. The class represents the user's emotional state selected from a predefined list of emotions (after an emotion elicitation activity). The goal of the second pattern is to receive a model input for the target class (emotion), resulting to a training example (input-output pair). The robot asks the user to demonstrate the selected emotion (model input - waling and standing). For the third pattern, the model captures the gait/thermal data and makes a prediction for the user's emotion. This prediction is not communicated to the user but it is used for the design of the interaction; if the prediction is inaccurate, the robot interacts with the user and uses the response to retrain its model. These patterns can be used to design the interactions between the human user and the model (Figure \ref{fig:hai1}). 

\textbf{\textit{Design and implementation aspects.}} The selection of the patterns and the actions that define them can provide insights about the design and implementation aspects. For example, designing the \texttt{class-selection} interaction requires an interface for the user to choose an emotion so it can be communicated to the model (robot). For the \texttt{new\_class\_sample} pattern, the model needs to capture the gait/thermal data and process them for a model prediction (feature extraction). In terms of technical aspects, the implementation does not consider user's input uncertainty, making the assumption that the user provides the correct data (model input and output). Such choices can affect both the performance of the model and its interaction with the user. Different actions or patterns can lead to different types of interactions between users and models, e.g., negotiation, which may require different methods for implementation, e.g., shared decision making. 

\begin{table}[h]
\resizebox{0.9\columnwidth}{!}{
\begin{tabular}{|c|c|c|}
\hline
\textbf{pattern} & \textbf{actions} & \textbf{message description} \\ \hline
\multirow{2}{*}{class-selection} & \texttt{req-class\_selection(Y,L)} & model asks user for a class from list \\ \cline{2-3} 
 & \texttt{select-class(Y,L)} & user selects a class from a list \\ \hline
\multirow{2}{*}{new\_class\_sample} & \texttt{req-new\_class\_sample(X,Y)} & model asks user to provide a sample for the given class \\ \cline{2-3}
 & \texttt{generate-class\_sample(X,Y)} & user provides a sample for the given class \\ \hline
\multirow{2}{*}{sample-annotation} & \texttt{req-sample\_class(X,Y)} & model asks user to annotate sample \\ \cline{2-3} 
 & \texttt{annotate-sample(S,M)} & user provides a label for the sample \\ \hline
\end{tabular}
}
\caption{Interaction patterns for the interactive robot learning interactions}
\label{tab:interaction1}
\end{table}
Based on the unpacking process of HAI interactions to primitives and patterns on this running example, we compiled a list of three patterns, and the actions that define them, which can describe the interactions between the user and the robot. Following the proposed formalization for actions and patterns, this approach aims to (a) characterize existing interactions, and (b) modify or design new interactions. Our motivation is to compile a list of commonly used actions and patterns, towards a design space for HAI interactions.     

\subsection{A list of interaction patterns and actions}
Based on the unpacking of HAI interaction use cases (Appendix \ref{more-unpacking}) and review on frameworks (Section \ref{related-work}), we compiled a list of actions, messages and interaction patterns. We provide a set of action definitions with their description (Table \ref{tab:action-defs}). Such actions can be further specified as messages and used for a given interaction context. For example, \texttt{annotate-sample(X,Y)} describes the annotation of a sample during a classification task, as the mapping of input sample X to output label Y. A similar action, \texttt{show-policy(S,A)}, describes an RL policy as the mapping of input state S to output action A. This list is not exhaustive since it does not cover all possible actions and patterns, but rather provides examples from a range of HAI interaction scenarios. More actions can be defined following the proposed formalization, resulting to an extendable library of actions. While there are different ways to define these actions and messages for the selected use cases, the goal of this approach is to specify the intent and type of the communicated information considering existing interaction concepts and HAI interaction paradigms. 

\begin{table}[h]
\resizebox{0.935\columnwidth}{!}{
\begin{tabular}{|l|l|}
\hline
\multicolumn{1}{|c|}{\textbf{Action Definition}} & \multicolumn{1}{|c|}{\textbf{Description}} \\ \hline
\multirow{3}{*}{\begin{tabular}[c]{@{}l@{}}\textbf{\texttt{req-class\_selection(Y,L)}} $\equiv$ \\ \texttt{request(Y:output.label,L{[}output.label{]})} \\ $\rightarrow$  \texttt{select(Y,L)}\end{tabular}} &  \multirow{3}{*}{} \\
 &  request the selection of a class Y from list L \\
 &   \\ \hline
\begin{tabular}[c]{@{}l@{}}\textbf{\texttt{select-class(Y,L)}} $\equiv$ \\ \texttt{provide(Y:output.label, L:{[}output.label{]})} \\ $\rightarrow$ \texttt{select(Y,L)}\end{tabular} & select a class Y from a list L \\ \hline
\begin{tabular}[c]{@{}l@{}}\textbf{\texttt{req-new\_class\_sample(X,Y)}} $\equiv$\\ \texttt{request(X:input.raw\_data|fvector, Y:output.label)} \\$\rightarrow$  \texttt{create(X),map(X,Y)}\end{tabular} & ask for a new sample X for a given class Y  \\ \hline
\begin{tabular}[c]{@{}l@{}}\textbf{\texttt{req-class\_sample(X,Y)}} $\equiv$\\ \texttt{provide(X:input.raw\_data|fvector, Y:output.label)} \\$\rightarrow$  \texttt{select(X), map(X,Y)}\end{tabular} & select a sample X of a given class Y \\ \hline
\begin{tabular}[c]{@{}l@{}}\textbf{\texttt{req-sample\_class(X,Y)}} $\equiv$\\ \texttt{request(Y:output.label, X:input.raw\_data|fvector)} \\   $\rightarrow$ \texttt{map(X,Y)}\end{tabular} & ask for the class Y of a given sample X \\ \hline
\begin{tabular}[c]{@{}l@{}}\textbf{\texttt{req-gsample\_class(X,Y)}} $\equiv$\\ \texttt{request(Y:output.label, X:input.raw\_data|fvector)} \\ $\rightarrow$ \texttt{create(X),map(X,Y)}\end{tabular} & request the annotation of a generated sample  \\ \hline
\begin{tabular}[c]{@{}l@{}}\textbf{\texttt{req-sel\_sample\_class(X,Y)}} $\equiv$\\ \texttt{request(Y:output.label, X:input.raw\_data|fvector)} \\ $\rightarrow$ \texttt{select(X),map(X,Y)}\end{tabular} & request the annotation of a selected sample  \\ \hline
\begin{tabular}[c]{@{}l@{}}\textbf{\texttt{annotate-sample(X,Y)}} $\equiv$ \\ \texttt{provide(Y:output.label, X:input.raw\_data|fvector)} \\   $\rightarrow$ \texttt{map(X,Y)}\end{tabular} & annotate a given sample  \\ \hline
\begin{tabular}[c]{@{}l@{}}\textbf{\texttt{show-policy(S,A)}} $\equiv$ \\ \texttt{provide(Y:output.action, X:input.state)} \\   $\rightarrow$ \texttt{map(S,A)}\end{tabular} & show selected action A for state S   \\ \hline
\begin{tabular}[c]{@{}l@{}}\textbf{\texttt{give-evaluative\_advice(S,A,R)}} $\equiv$ \\ \texttt{provide(R:feedback.eval, {[}X:input.state,Y:output.action{]})} \\   $\rightarrow$ \texttt{select(R), map(S,A,R)}\end{tabular} & select a reward R for the mapping from S to A   \\ \hline
\begin{tabular}[c]{@{}l@{}}\textbf{\texttt{modify-prediction(X,Y,Z)}} $\equiv$\\ \texttt{provide(Z:output.label,{[}X:input.raw\_data|fvector, Y:label{]})} \\ $\rightarrow$  \texttt{modify(Y,Z), map(X,Z)}\end{tabular} & modify prediction of X from Y to Z \\ \hline
\begin{tabular}[c]{@{}l@{}}\textbf{\texttt{show-candidate\_samples(CS,S)}} $\equiv$\\ \texttt{provide(CS:{[}input.raw\_data{]}, S:{[}input.raw\_data{]}{]})} \\ $\rightarrow$  \texttt{select(CS,S)}\end{tabular} &  show a list of candidate samples CS based on sample S \\ \hline
\begin{tabular}[c]{@{}l@{}}\textbf{\texttt{select-sample(X,CS)}} $\equiv$ \\ \texttt{provide(X: input.raw\_data, CS:{[}input.raw\_data{]}{]})} \\ $\rightarrow$ \texttt{select(X,CS)}\end{tabular} & select sample X from a list of samples CS  \\ \hline
\begin{tabular}[c]{@{}l@{}}\textbf{\texttt{modify-sample(X,M)}} $\equiv$ \\ \texttt{provide(M:input.raw\_data,X:input.raw\_data)}\\ $\rightarrow$ \texttt{modify(X,M)}\end{tabular} & modify a sample from X to M  \\ \hline
\begin{tabular}[c]{@{}l@{}}\textbf{\texttt{generate-sample(X)}} $\equiv$ \\ \texttt{provide(X:input.raw\_data)}\\ $\rightarrow$ \texttt{create(X)}\end{tabular} & modify a sample from X to M  \\ \hline
\begin{tabular}[c]{@{}l@{}}\textbf{\texttt{modify-mparams(P,M)}} $\equiv$ \\ \texttt{provide(M:input.mparams,X:input.mparams)}\\ $\rightarrow$ \texttt{modify(X,M)}\end{tabular} & modify model parameter from P to M  \\ \hline
\begin{tabular}[c]{@{}l@{}}\textbf{\texttt{modify-features(X,M)}} $\equiv$ \\ \texttt{provide(M:input.fvector,X:input.fvector)}\\ $\rightarrow$ \texttt{modify(X,M)}\end{tabular} & modify a feature vector from X to M  \\ \hline
\begin{tabular}[c]{@{}l@{}}\textbf{\texttt{req-prediction\_evaluation(X,Y,F)}} $\equiv$ \\ \texttt{request(F:feedback.eval,{[}X:input.raw\_data,}\\ \texttt{Y:output.label{]})} $\rightarrow$ \texttt{select(F), map(X,Y,F)}\end{tabular} & ask for feedback F to evaluate the mapping of X to Y   \\ \hline
\begin{tabular}[c]{@{}l@{}}\textbf{\texttt{evaluate-prediction(X,Y,F)}} $\equiv$ \\ \texttt{provide(F:feedback.eval,{[}X:input.raw\_data,}\\ \texttt{Y:output.label{]})} $\rightarrow$ \texttt{select(F), map(X,Y,F)}\end{tabular} & select feedback F to evaluate the mapping of X to Y   \\ \hline
\begin{tabular}[c]{@{}l@{}}\textbf{\texttt{show-prediction\_XAI(X,Y,F)}} $\equiv$ \\ \texttt{provide(F:feedback.XAI,{[}X:input.raw\_data,}\\ \texttt{Y:output.label{]})} $\rightarrow$ \texttt{map(F,X,Y)}\end{tabular} & provide explanation for the mapping of X to Y   \\ \hline
\end{tabular}
}
\caption{A list of action definitions and their description, extracted by unpacking existing HAI interactions into interaction primitives.}
\label{tab:action-defs}
\end{table}

Following the proposed definitions, we provide a collection of interaction patterns (Table \ref{tab:summary-patterns}). Each pattern is described as a sequence of messages based on the action definitions and the unpacking process of the selected HAI interactions. The selection of the messages and actions play an important role to the definition of an interaction pattern. For example, \texttt{req-class\_selection} and \texttt{req-sample\_class} can both describe a model's request for an output. However, the first action specifies the selection of a class from a list, while the second action requires a class given an input sample. The selection of an action for a given pattern depends on the goal of the interaction, as well as the design and technical requirements. Combining different actions can lead to patterns with different interaction goals. For example, model queries for (informative/evaluative) advice aim to improve model's performance, while XAI-based interaction patterns can be used to support the user by providing justifications about model predictions. This collection captures a range of interaction patterns in terms of the interaction concept and requirements. Combinations of interaction patterns can serve multiple goals and concepts. For example, providing explanations to users (XAI) can enhance the quality of human feedback (HITL) resulting to collaborative learning systems. The long-term goal of this research is the formalization of a new design space for HAI interactions which will support designers to explore, modify, and apply interaction patterns by providing design and implementation choices towards the prototyping of new types of interactions between human users and AI models. 

\begin{table}[h]
\resizebox{\columnwidth}{!}{
\begin{tabular}{|c|c|c|}
\hline
\textbf{Patterns} & \textbf{Actions} & \textbf{Description} \\ \hline
\multirow{2}{*}{class-selection} & \texttt{req-class\_selection(Y,L)} & request for a class from a list \\ \cline{2-3} 
 & \texttt{select-class(Y,L)} & select a class from a list \\ \hline
 \multirow{2}{*}{new\_sample} & \texttt{req-new\_sample(X)} & ask for a new input sample \\ \cline{2-3} 
 & \texttt{generate-sample(X)} & generate an input sample \\ \hline
\multirow{2}{*}{new\_class\_sample} & \texttt{req-new\_class\_sample(X,Y)} & request a new sample for a given class \\ \cline{2-3} 
 & \texttt{generate-class\_sample(X,Y)} & generate a sample for a given class \\ \hline
\multirow{2}{*}{sample-annotation} & \texttt{req-sample\_class(X,Y)} & ask for the class of a given sample \\ \cline{2-3} 
 & \texttt{annotate-sample(X,Y)} & select a class for a given sample \\ \hline
\multirow{2}{*}{new\_sample-annotation} & \texttt{req-new\_sample(X)} & ask for a new input sample \\ \cline{2-3} 
 & \texttt{req-gsample-class(X,Y)} & ask for the class of a new sample \\ \hline
\multirow{2}{*}{candidate\_samples} & \texttt{req-candidate\_samples(CS,S)} & ask for a set of candidate input samples \\ \cline{2-3} 
 & \texttt{show-candidate\_samples(CS,S)} & show a set of candidate input samples \\ \hline
\multirow{2}{*}{sample-modification} & \texttt{req-modified\_sample(X,M)} & ask for a modified input sample \\ \cline{2-3} 
 & \texttt{modify-sample(X,M)} & modify an input sample \\ \hline
 \multirow{2}{*}{feature-modification} & \texttt{req-modified\_feature(X,M)} & ask for a modified feature vector \\ \cline{2-3} 
 & \texttt{modify-feature(X,M)} & provide a modified feature vector \\ \hline
 \multirow{2}{*}{parameter-modification} & \texttt{req-mparam-modification(P,M)} & ask for a modified model parameter \\ \cline{2-3} 
 & \texttt{modify-mparam(X,M)} & modify a model input parameter \\ \hline
 \multirow{2}{*}{prediction-modification} & \texttt{annotate-sample(X,Y)} & provide a label for a sample \\ \cline{2-3} 
 & \texttt{modify-prediction(X,Y,M)} & modify a prediction of a given sample \\ \hline
 policy-visualization & \texttt{show-policy(S,A)} & show selected action for current state \\ \hline
\multirow{2}{*}{informative\_advice} 
 & \texttt{req-informative\_advice(S,A,B)} & ask for informative advice based on state-action \\ \cline{2-3} & \texttt{give-informative\_advice(S,A,B)} & modify (or not) the selected action \\ \hline
\multirow{2}{*}{evaluative\_advice} & \texttt{req-evaluative\_advice(S,A,B)} & ask for evaluative feedback for state-action \\ \cline{2-3} 
 & \texttt{give-evaluative\_advice(S,A,B)} & evaluate the state-action pair \\ \hline
\multirow{2}{*}{prediction-based\_XAI} & \texttt{req-prediction\_XAI(X,Y,F)} & ask for explanations for a given input-output \\ \cline{2-3} 
 & \texttt{show-prediction\_XAI(X,Y,F)} & show explanation for a given input-output pair \\ \hline
 \multirow{2}{*}{outcome-evaluation} & \texttt{req-outcome\_evaluation(Y,F)} & request evaluative feedback for a given outcome \\ \cline{2-3} 
 & \texttt{evaluate-outcome(Y,F)} & provide evaluative feedback for a given outcome \\ \hline
\multirow{2}{*}{prediction\_parameters} & \texttt{req-prediction\_params(X,Y,P)} & request predictions with model output parameters \\ \cline{2-3} 
 & \texttt{show-prediction\_params(X,Y,P)} & show predictions with model output parameters \\ \hline
\multirow{3}{*}{turn\_taking-evaluation} & \texttt{generate-and-turn(X,Y)} & request a modified sample based on a generated input \\ \cline{2-3} 
 & \texttt{capture-and-generate(X,Y)} & provide a modified sample based on input \\ \cline{2-3} 
 & \texttt{evaluate-outcome(Y)} & provide evaluative feedback based on outcome \\ \hline
\multirow{3}{*}{prediction-with-XAI} & \texttt{select-sample(X)} & select a sample \\ \cline{2-3} 
 & \texttt{show-prediction\_XAI(F,X,Y)} & show explanation for the sample prediction \\ \cline{2-3} 
 & \texttt{modify-annotation(X,Y,M)} & modify the outcome based on the input-output pair \\ \hline
\multirow{3}{*}{recommendations} & \texttt{req-recommendations(M,R)} & modify a model input parameter \\ \cline{2-3} 
 & \texttt{show-recommendations(M,R)} & show recommended items for a given model input \\ \cline{2-3} 
 & \texttt{evaluate-recommendation(S,V)} & evaluate a selected recommended item (accept/reject) \\ \hline
\end{tabular}
}
\caption{A list of interaction patterns as sequences of the defined actions/messages.}
\label{tab:summary-patterns}
\end{table}

\section{From interaction primitives to a design space for HAI interactions} \label{design-space}
In this section, we discuss how the proposed formalization can be used towards the definition of a design space for HAI interactions. We demonstrate how the proposed interaction primitives and patterns can be used as design materials to prototype HAI interactions. We highlight how differences in patterns can serve different interaction goals and concepts. The goal of the proposed design space for HAI interactions is to support AI designers and practitioners by providing appropriate design and implementation choices for a given interaction concept. Based on the literature review and the unpacking of existing HAI interactions, we provide an overview of how interaction patterns can be designed for given interaction paradigms, aiming to bridge the gap between high-level guidelines and implementation requirements.   

\subsection{Interaction primitives and actions as design materials}
\begin{figure}[h]
\centering
\includegraphics[width=0.85\columnwidth]{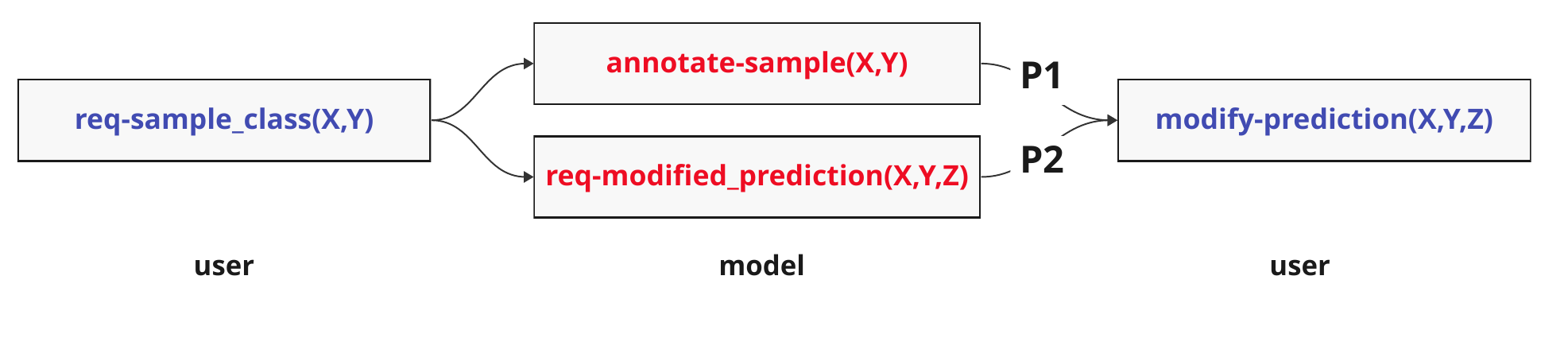}
\caption{Query Patterns. The user asks for a model prediction for a given input. Based on pattern P1, the model provides the prediction and the user chooses to modify the prediction. Based on pattern P2, the model makes an explicit request for a modified prediction.}
\label{fig:query}
\end{figure}

Human-centered AI approaches, including explainability, transparency, interactivity, and human control, have provided opportunities to design new types of interactions, e.g., model auditing and contestation, negotiation, shared decision making, and others. HAI interactions can utilize the available information provided by the models, apart from decisions and predictions. For example, model uncertainty can be measured, communicated and used as a design material through transparency \cite{bhatt2021uncertainty}. AI designers and practitioners need to consider interactions with AI as a design material, based on its capabilities, limitations, and the challenges that may arise while designing for transparency, unpredictability, learning, and shared control \cite{holmquist2017intelligence}. Our proposed formalization for interaction primitives and patterns aims to enable designers and AI practitioners to design HAI interactions by exploring appropriate patterns for a given set of design and technical requirements. The following example illustrates two alternatives of a query pattern, where the user queries the model for a prediction using a model input - \texttt{req-sample\_class(X,Y)}, receives the model's prediction, and provides a modified prediction - \texttt{modify-prediction(X,Y,Z)}. Using the same user actions, we can define two query patterns based on two different model actions (Figure \ref{fig:query}): 

\begin{itemize}
    \item[-] \texttt{\textbf{annotate-sample(X,Y)$\equiv$}} \texttt{provide(Y:output,X:input)$\leftarrow$...}, and
    \item[-] \texttt{\textbf{req-modified\_prediction(X,Y,Z)$\equiv$}}
    \texttt{request(Z:output,[X:input,Y:output])$\leftarrow$...}
\end{itemize} 

In the first case (P1), the model provides the user with a prediction on user's input and the user contests the prediction by providing an alternate output. The second case (P2) describes a model which queries the user to provide an alternative output given its prediction on the user's input. In terms of design and implementation choices, the first query alternative may require a mechanism to decide who makes the decision based on the quality of user/model predictions and how to update the model based on user's prediction (hybrid decision making). For the second query, the model is designed to explicitly ask the user for an alternative output and could be an example of active learning, where the model requests human annotation for (uncertain) predictions. In that case, human's decision should have a larger effect on the model updates, compared to the first version. Through these two versions of a query, we demonstrate how primitives can be used as design materials to generate different patterns given an interaction goal. Both patterns could describe the interactive control feature for the design of \textit{contestable AI} interactions \cite{alfrink2022contestable}, which enable the users to intervene and modify a decision made by the system.

\subsection{Prototyping interactions} 
\begin{figure}[h]
\centering
\includegraphics[width=0.85\columnwidth]{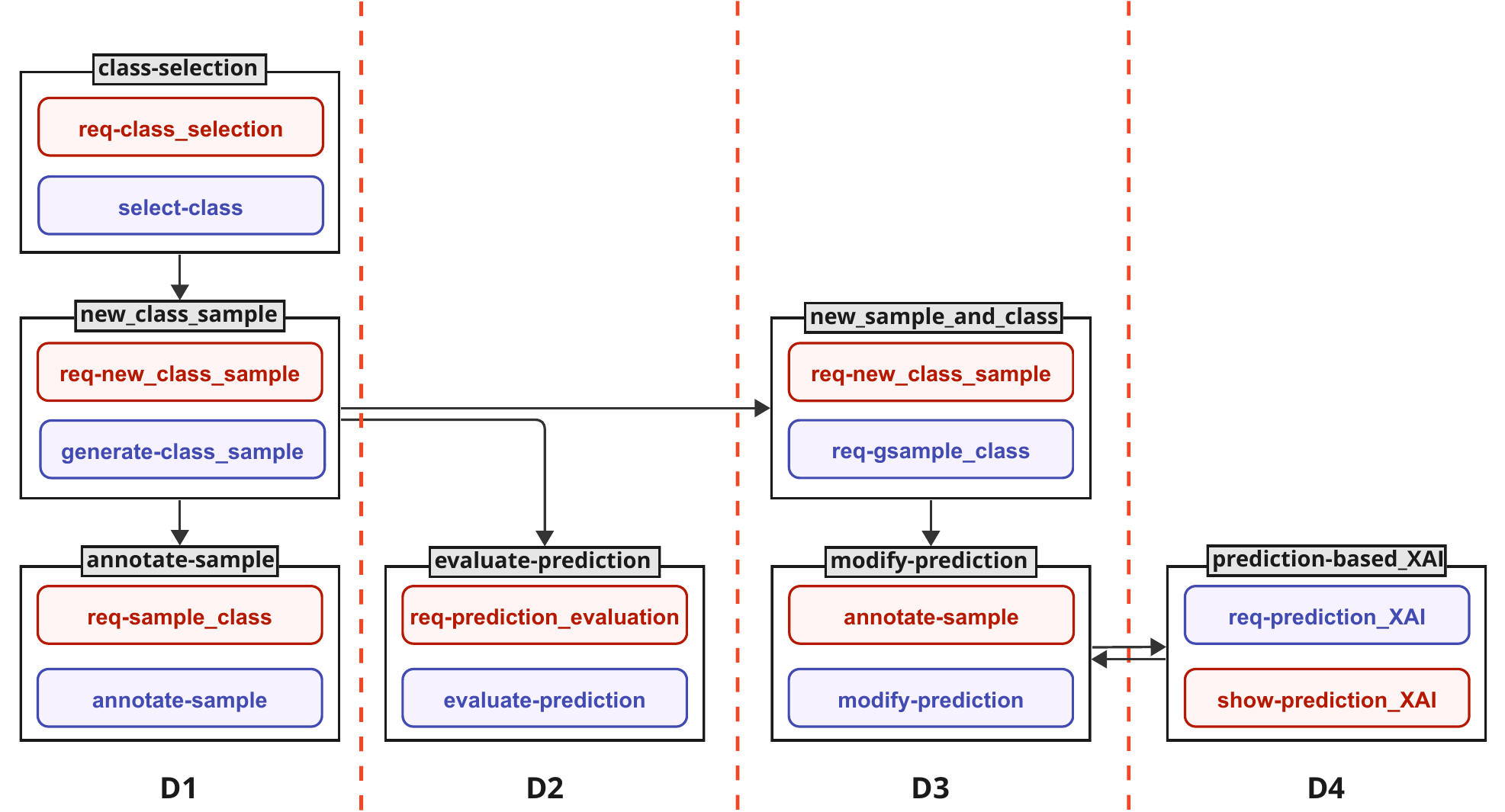}
\caption{Design alternatives for the interactive robot learning interactions \cite{yu2019interactive}. Different sequences of patterns result to a set of design alternatives for different interaction concepts. Each alternative is related to specific design/implementation aspects and requirements. D1 describes the original design. D2 describes a robot-initiated interaction for user feedback. Based on D3, user asks the robot for its prediction based on the generated sample which can be modified. D4 adds an XAI-based interaction for model explanations for the robot's prediction.}
\label{fig:choices}
\end{figure}

A key aspect of the proposed design space is the ability to create and explore alternatives of interactions by selecting different patterns or actions. We consider a set of alternative designs for the interactive robot learning scenario \cite{yu2019interactive}, using the defined interaction primitives and patterns as design materials. We demonstrate how different sequences of patterns can result to different types of interactions. For each design alternative, we provide a description of the design and technical aspects, as well as the interaction concept (Figure \ref{fig:choices}). 
The first design (D1) represents the original interaction design (Figure \ref{fig:unpacking1}); a robot-initiated interaction where the user provides an annotated sample in a query-response manner. The second design (D2) describes an interaction where the robot communicates its prediction for the user's emotion, followed by a request for user's validation. This is achieved by replacing the last pattern \texttt{(annotate-sample)} with a pattern for evaluating the model's prediction \texttt{(evaluate-prediction)}. The third design (D3) can describe a system where the user asks the robot to make a prediction which they can modify. This design introduces two patterns to allow the user to request model's prediction for the generated sample \texttt{(new\_sample\_and\_class)} and modify it \texttt{(modify-prediction)}. The fourth design (D4) includes an additional XAI-based interaction pattern, where the model provides explanations about its prediction to the user \texttt{(prediction-based\_XAI)}. 

These alternatives can characterize different types of interactions with respect to the goal of the interaction. The original design is proposed as an interactive robot learning approach where the goal is to evaluate and improve the model's predictions through its robot-initiated interactions with the user. The second design requires the user to be more active in the interaction through an interaction pattern for model's prediction and user's feedback (evaluation). Based on the third design, the user initiates the interaction for the model's prediction. Finally, adding the XAI interaction pattern enables the user to further interact with the robot for explanations.       
These designs are also related to specific design and implementation aspects. The original design assumes that user's provided sample and self-reported emotion are accurate. The second one requires an interaction where the robot communicates the prediction and asks for user's evaluation. Such an interaction requires an appropriate learning mechanism which can integrate user's feedback to the learning process. The XAI-based interaction requires the implementation of a specific explanation method. These alternatives can also be related to different roles of users involved in the interaction. The first design (D1) has been proposed for an end-user who implicitly participates in the model training process. The fourth alternative (D4) could be a design for an interaction between the model and the developer for model evaluation, where XAI is used to enhance the user's understanding about the system capabilities, e.g., identify "hard-to-predict" emotions.     

\begin{figure}[h]
\centering
\includegraphics[width=0.9\columnwidth]{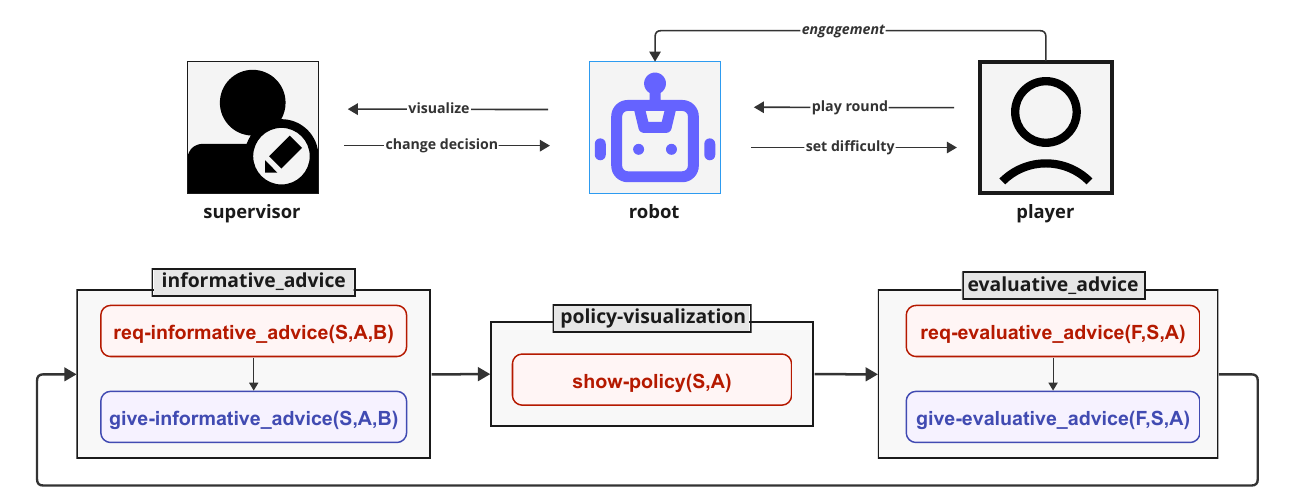}
\caption{Interaction Patterns for the robot-based multi-user interaction \cite{tsiakas2018task}. The robot visualizes its policy to the supervisor (\texttt{policy-visualization}) who can alter it (\texttt{(informative\_advice)}). The robot receives implicit feedback from user's engagement \texttt{(evaluative\_advice)}. Both types of feedback are integrated to model learning updates.}
\label{fig:irl}
\end{figure}

Considering multi-user HAI interactions, different types of users can participate in the interaction for different goals. For example, the robot-based game interactions \cite{tsiakas2018task} are divided into two interaction loops; player-AI and supervisor-AI interaction. Each loop serves a specific interaction goal. The player-AI interaction goal is to elicit feedback from the player implicitly in order to personalize the model's decisions, while the supervisor-AI loop aims to enhance safety through the interventions of a supervisor using a transparent interface. Considering these, different interaction patterns should be used to serve each goal (Figure \ref{fig:irl}).

The patterns describe the interactions between the robot, the player and the supervisor. The robot  adjusts its policy (difficulty selection) based on user's performance and engagement, and the supervisor's interventions. Two main design aspects of the proposed system are (a) transparency and explainability and (b) user feedback and control. Considering the player-AI interactions, the model communicates the selected difficulty through the robot's announcement and the player provides implicit feedback through task performance and engagement. For the supervisor-AI interaction, model's transparency through the UI aims to enhance user's decision making by providing appropriate interventions. Considering both types of interactions and their patterns, a learning mechanism is required to learn from both types of users in an online way so the robot can dynamically improve its policy and select the appropriate levels of difficulty. The goal of the interaction is to include both types of users in the personalization process.   

Designing multi-user interactions may also require combining interaction patterns for different goals. Considering the framework for contestable AI \cite{alfrink2022contestable}, we provide a description of possible interactions during model contestation, highlighting the different interaction concepts between users and the AI model. The framework follows the paradigm of mixed-initiative interaction, based on which human controllers and decision subjects can both participate in the decision making process. Decision subjects can interact with the system to negotiate a decision which affects them. Such decisions may be the outcome of decision support interactions between the model and the human controller (semi-automated decisions). In order to support such types of interactions, the proposed framework provides the following features: \textit{interactive controls}, \textit{explanations}, and \textit{intervention requests}. Interactive controls enable both types of users to provide feedback to the system for different purposes. Explanations are used to provide justification for the model's decisions and aim to support the semi-automated decision making process. Intervention requests enable the decision subject to initiate a model auditing process. XAI-based interaction patterns can be used to: (a) make a user aware of a decision made by the system, (b) inform the user about how to contest a model decision and (c) provide an explanation to justify the decision of the system. HITL-based interaction patterns can provide both types of users with the ability to negotiate and even override AI decisions (interactive controls). Moreover, collaborative learning interactions can enable both users and system to learn from their interactions and augment their decision making towards hybrid intelligence. 

\begin{figure}[h]
\centering
\includegraphics[width=0.99\columnwidth]{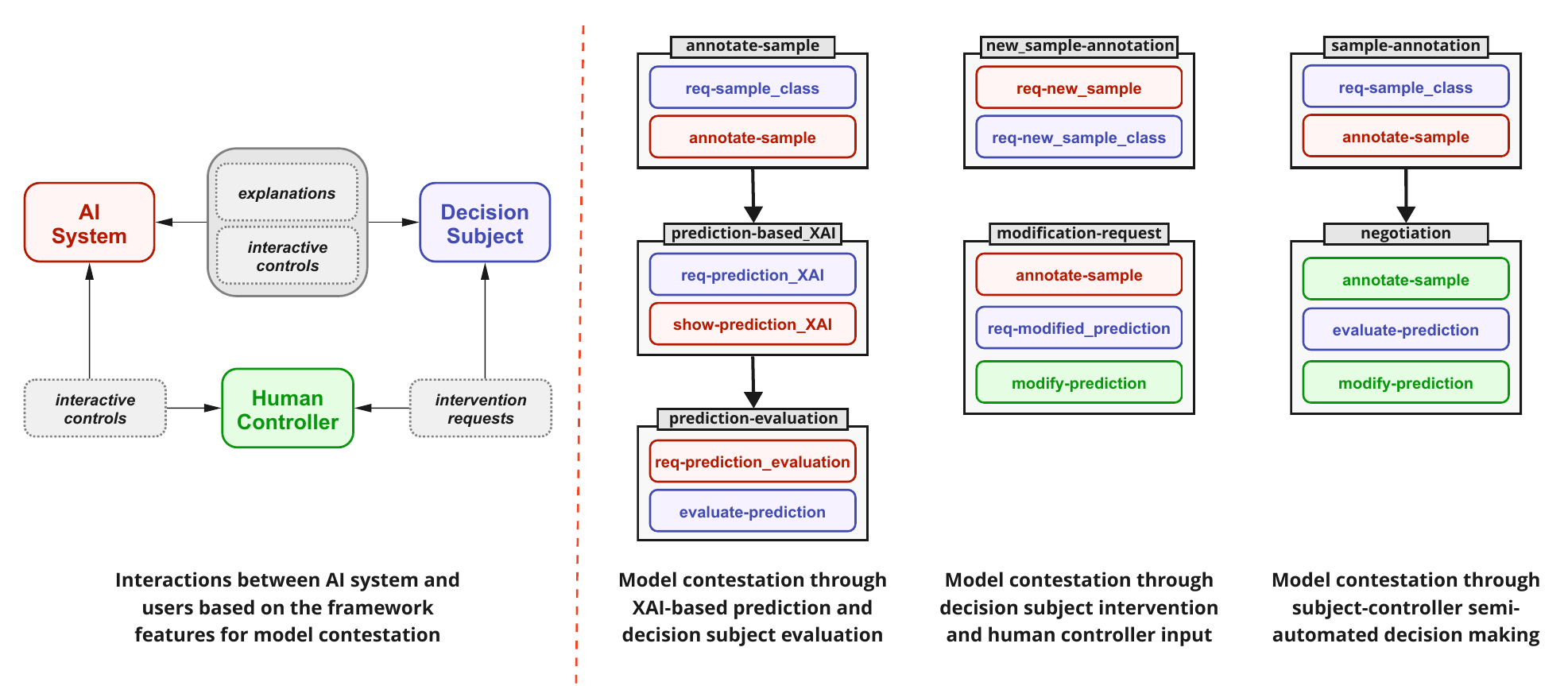}
\caption{Examples of Interaction Patterns for contestable AI interactions \cite{alfrink2022contestable}. We describe three examples of interactions based on the proposed framework for mixed-initiative interactions between the AI system, the decision subject and the human controller. Each interaction describes a different contestation aspect using interactive controls, explanations and intervention requests.}
\label{fig:contest}
\end{figure}

Considering the above, we provide examples of interaction based on our defined actions and patterns (Figure \ref{fig:contest}). The first example is a combination of three interaction patterns and describes the communication between AI and the decision subject. During the interaction, the user asks for the model's prediction for a sample (\texttt{sample-annotation}) and for explanations for this prediction (\texttt{prediction-based\_XAI}). The subject can utilize the explanations to evaluate the outcome. if needed (\texttt{prediction-evaluation}). The second example is a multi-user interaction, where the decision subject needs to generate a new sample (e.g., submit a form) and ask for a decision (\texttt{new\_sample-annotation}). The AI provides its decision to subject who makes a request to the human controller for the modification of the decision (\texttt{modification-request}). Finally, the third example describes a semi-automated decision making process, where the human controller can provide a decision (considering the AI's output) and modify it based on the decision subject's evaluative feedback for the model's prediction (\texttt{negotiation}).

\subsection{Interaction Patterns and Requirements} \label{requirements}
We demonstrated how the proposed formalization can be used towards prototyping HAI interactions using patterns. In order to implement such interactions, we need to consider the interaction goals that each pattern can serve. Towards this, based on our literature review (Section \ref{related-work}) and the extracted patterns and actions (See Tables \ref{tab:action-defs} and \ref{tab:summary-patterns}) from the unpacking process of HAI interaction use cases (See Appendix \ref{more-unpacking}), we provide an overview of how interaction patterns can be used in the context of different concepts for XAI-based, HITL-based, and HI-based interactions. The goal of the overview is to identify possible relations between interaction patterns and requirements. 

\subsubsection{XAI-based interactions}
Human-XAI interactions can be designed for several interaction concepts and goals (e.g., debugging, persuasion, decision support, etc.). Designing explainable and transparent models is not trivial, especially while considering the various parameters that can affect the interaction, e.g., user's expertise, perception and understanding, cognitive load, preferences, etc. From the unpacking process, we identified the following XAI-based patterns and interaction goals: 
\begin{itemize}
\item \textit{XAI-based interactions can manage user's expectations about the AI model's behavior.} A Meeting Scheduling Assistant \cite{kocielnik2019will} (See Appendix \ref{more-unpacking} - Figure \ref{fig:unpacking7} for description) uses explanations to calibrate user's trust and expectations about its model predictions. The model predicts if an email is a meeting request to help the user make a decision. The \texttt{prediction-with-XAI} pattern enables the model to be transparent by communicating its accuracy rate along with the prediction. The system visualizes the model's accuracy rate, as well as a set of prediction/explanation examples with different levels of uncertainty to enhance user's understanding.
\item \textit{XAI-based interactions can enhance user's perception about the model's performance.} In the context of active learning for emotion recognition \cite{heimerl2019nova} (See Appendix \ref{more-unpacking} - Figure \ref{fig:unpacking4} for description), XAI-based interactions aim to facilitate the selection of appropriate samples for model refinement. Based on the unpacking process of the proposed system, we identify two patterns with different goals: the \texttt{prediction-parameters} pattern aims to support the selection of appropriate samples for annotation by visualizing the confidence of predictions, and the \texttt{prediction-based\_XAI} pattern is used to support the annotation process by providing additional information about the model's prediction for a given sample through visual explanations.
\item \textit{Model transparency can support human trainers while providing feedback to iML models.} Policy visualization \cite{bignold2022human} (See Appendix \ref{more-unpacking} - Figure \ref{fig:unpacking3} for description) serves as a transparency method to engage the user to provide feedback to the model during task performance. The model utilizes user feedback to facilitate its learning process, i.e., faster convergence to the optimal policy. Based on our unpacking approach, the \texttt{policy-visualization} pattern is used to communicate the model's current policy, by visualizing the current state (input) and the selected action (output).  
\item \textit{Explanations can justify model's prediction based on user preferences.} For an explainable Music Recommendation system \cite{martijn2022knowing} (see Appendix \ref{more-unpacking} - Figure \ref{fig:unpacking6} for description), explainable user (preference) models are used to enhance user's perception about their own preferences and how these affect model's recommendations. Considering this, the \texttt{prediction-based\_XAI} pattern is used to provide predictions and justify them through explanations based on user's preferences. In terms of design aspects, different visualizations (and explanation methods) are required considering the individual characteristics of users, e.g., need for cognition.
\end{itemize}

\subsubsection{HITL-based interactions} The goal of HITL methods is to efficiently integrate the human user to the learning and decision making process of an AI system. Human users can participate in the model's development and deployment phases. The selection of appropriate iML/HITL methods and approaches depends on several aspects of the interaction, including user role and expertise. We present a set of HITL-based patterns extracted from the unpacking process, considering the different interaction goals.
\begin{itemize}
\item \textit{HITL methods can be used for interactive data collection and labeling.} For the interactive robot learning scenario for emotion recognition (Figure \ref{fig:unpacking1}), the model utilizes human-robot interaction data in order to evaluate its predictions and collect training data. Model re-trains without making the user aware of their participation in the data collection, annotation and training processes. Based on our unpacking, \texttt{sample-annotation} defines a human labeling action given a generated sample.     
\item \textit{HITL methods can be used for model improvement by relabeling uncertain predictions.} In the context of active learning, the NOVA system (See Appendix \ref{more-unpacking} - Figure \ref{fig:unpacking4} for description) asks the user to select uncertain samples and modify their predictions. In terms of the interaction design, the \texttt{prediction-modification} pattern enables the user to validate the model's prediction or provide a new label. Manual corrections are used to update the model. Since the user makes the final decision (label), XAI methods are used to enhance user's perception and, thus,  the quality of feedback. 
\item \textit{Feedback interfaces should be user-friendly and intuitive in order to ensure human feedback quality.} Training an RL agent through human advice (Appendix \ref{more-unpacking} - Figure \ref{fig:unpacking3}) requires an appropriate learning methods to integrate human feedback. The interaction supports two types of feedback. The \texttt{informative-advice} pattern is used to receive human advice in the form of a corrective action and the \texttt{evaluative-advice} pattern describes the evaluation of the model's decision in the form of binary feedback. These types of feedback are integrated through different feedback interfaces and learning methods (policy/reward shaping).
\item \textit{Users can control model predictions and  parameters.} For the Meeting Scheduling Assistant (Appendix \ref{more-unpacking} - Figure \ref{fig:unpacking7}), HITL methods enable the user to (a) provide feedback for a prediction by accepting or rejecting it, and (b) control the model's sensitivity parameters through a UI slider, affecting the model's predictions. The \texttt{modify-prediction} action enables the user to validate (or not) a prediction and the \texttt{modify-mparams} action allows the user to adjust the model's sensitivity parameters until they are satisfied with the model predictions. 
\end{itemize} 

\subsubsection{Collaborative Learning and Hybrid Intelligence} The goal of collaborative learning and hybrid intelligence interactions is to enable both humans and machines (AI systems) to learn from each other in a collaborative manner. Collaborative learning interactions can be designed by combining XAI-based and HITL-based interactions, enabling both users and AI models to exchange information while solving a task. Based on our unpacking process, we identify and discuss interaction patterns used in collaborative learning and hybrid intelligence interactions.
\begin{itemize}
\item \textit{User can control and evaluate the collaboration with a model.} In the context of a robot-based collaborative sketching interaction \cite{lin2020your} (See Appendix \ref{more-unpacking} - Figure \ref{fig:unpacking5} for description), both user and robot work together during the co-ideation process in a turn-taking interaction. Both agents generate ideas building on their partner's generated sketch. The model uses the captured image to generate a variation of the user's sketch - to provide alternative ideas. During the interaction, the user can control what the robot will capture as an input by moving the robot and also provide feedback for the generated outcome. The \texttt{turn\_taking-evaluation} pattern is repeated until the user is satisfied. The model uses image classification for the user's input and generates a variation of this. The goal of such interactions is for the user to explore and identify new insights, rather than to identify a correct solution. 
\item \textit{Model can support the user through semi-automated decision making.} An interactive sound segmentation system \cite{kim2018human} (Appendix \ref{more-unpacking} - Figure \ref{fig:unpacking2}) enables the user and the model to work together in order to complete an audio segmentation and annotation task. The model supports the user by providing a list of candidate samples which can be selected, edited and annotated by the user, towards a collaborative interaction. The \texttt{candidate-samples} pattern requires a model mechanism to identify the candidate samples and a proper visualization to highlight the segments. Based on this visualization, users provide feedback to adjust the model's predictions.
\item \textit{Explanations can be provided both by users and models to justify their predictions.} In a game-based scenario \cite{guo2022building} (See Appendix \ref{more-unpacking} - Figure \ref{fig:unpacking8} for description), XAI and iML methods are used to enhance user's trust about the model's decisions and allow for corrections. The \texttt{turn-taking\_XAI} pattern describes an interaction where both user and model communicate their prediction justification in the form of rule-based XAI. In terms of the design and implementation of this interaction, an appropriate visualization of the rules is needed to enable the user to understand the reasoning of the model in order to provide appropriate modifications and justification. 
\end{itemize}

Based on this overview, we can observe that each pattern can serve a specific goal within the interaction. For example, considering the XAI-based patterns, the \texttt{prediction-with-XAI} pattern is used to calibrate end-user's trust, while \texttt{prediction\_XAI} is used to enhance user's understanding about a prediction. For HITL-based patterns, \texttt{annotate-sample} enables a user to improve a model by providing annotations, while \texttt{modify-mparams} is used as a control pattern which enables the user to alter the model's predictions until the user is satisfied with the decision. Considering possible commonalities and differences between patterns and goals, we envision a design space which can provide suggestions for interaction patterns based on a given interaction concept. This would allow for fast prototyping of interactions using patterns, considering the interaction goals and requirements. 

\section{Discussion} \label{discussion}
\subsection{Limitations}
A basic limitation of our proposed formalization is that it describes interactions without providing information about how the interacting agents are affected by the communication of a message, or their underlying processes (e.g., predict or fit/update) and the dataflow during the interaction. Moreover, the current formalization allows for multiple ways to define a given interaction. In order to support the design and prototyping of HAI interactions following our approach, we need to introduce a formalization of the primitives and actions as design materials, including instances of objects and agents (user profiles, model cards, data sheets) and their operations (model operations, human decisions, etc.). For example, the \texttt{new\_sample-annotation} pattern (Table \ref{tab:summary-patterns}) includes an action for the generation of the new sample and an action for the sample annotation. These actions are related to specific model operations; generating a sample needs a \texttt{preprocessing} step so it can be used by the model for a \texttt{model.predict} operation. Moreover, retraining a model by altering its predictions, e.g., \texttt{modify-prediction}, can be implemented as a \texttt{model.fit} operation using updated training data.  
Such formalization will allow for the design of interactions between multiple agents and objects (users with different roles, models/data with different levels of access, etc.). Another limitation of the current approach is that the XAI-based interactions can be further described considering the explanation method. Our current approach considers XAI as a separate type for the defined primitives. For example, LIME-based visualization \cite{heimerl2019nova} (Figure \ref{fig:unpacking4}) communicates the important features/pixels of an image which affect the prediction. With our current approach, this interaction is described using the \texttt{feedback.XAI} type and the LIME method is defined as a modifier. In order to support designing with XAI and HITL methods, we will define an in-depth description of XAI/HITL-based interaction patterns based on existing taxonomies \cite{arrieta2020explainable, mohseni2021multidisciplinary,mosqueira2022classification, dudley2018review, teso2023}. A mapping from interaction patterns to a set of implementation techniques will help us explore possible commonalities between design and implementation aspects. In order to address both limitations, our proposed design space will be informed by existing guidelines and frameworks related to the design and implementation of HAI interactions. 

\subsection{Envisioned Applications}
We propose a design space which can support designers and AI practitioners to design and implement HAI interactions based on interaction primitives and patterns. More specifically, we envision a design space which can enable users to explore and choose between existing patterns and modify them towards new types of interactions. The design space will be developed as a prototyping tool by providing suggestions about the design and implementation aspects of existing and new patterns. Building on the existing list of patterns and actions (Tables \ref{tab:action-defs}, \ref{tab:summary-patterns}), an extendable collection of interaction patterns and their design/implementation choices will be used as design materials for more complex HAI interactions. A mapping from interaction patterns to common implementation aspects could support fast prototyping of HAI interactions, in the form of auto-generated code for basic model operations during HAI interactions, e.g., data operations, model predict, fit, etc.. A key aspect of the proposed space is to link design aspects with implementation choices for a given interaction concept. Towards this, we provide a short description of the implementation aspects of extracted patterns (Section \ref{requirements}), aiming to identify possible commonalities between patterns and implementation issues. Finally, we discuss how the proposed space can be informed by existing frameworks and guidelines. 

\subsubsection{Manifesting implementation concerns}
A main motivation for the proposed design space is the need to bridge the gap between design and implementation choices. This can be achieved by characterizing the defined interaction patterns in terms of interface and implementation requirements. We envisage that developing a collection of interaction patterns as well-defined, named entities, can enable us to extract a list of common implementation issues attached to a given interaction pattern. Towards this, we provide an overview of the implementation aspects for the defined patterns, considering the different HAI interaction paradigms.

\textit{XAI-based interaction patterns.} Based on the literature review and the unpacking process, we identify the following implementation aspects and challenges regarding XAI-based interaction patterns: (a) the selection of appropriate explanation strategies and methods considering the interaction goals \cite{arrieta2020explainable,adhikari2022towards,nunes2017systematic}, (b) the adaptation of explanations based on user's roles and characteristics (e.g., expertise, perception, etc.) \cite{liao2020questioning,chromik2021human}, and (c) the evaluation of XAI methods in terms of the interaction goal and user's behavior \cite{schwalbe2021comprehensive, mohseni2021multidisciplinary, chromik2020taxonomy}. These can be related to the patterns found above:
\begin{itemize}
    \item The \texttt{prediction-with-XAI} is used to help a non-technical user to understand if they can trust the model's prediction or not. This requires the model to communicate its accuracy rate along with a prediction, in an intuitive way (e.g., chart) in order to manage user's expectations about the model decisions.  
    \item The \texttt{prediction\_based-XAI} requires a proper explanation method considering the role of the user. For a domain expert user, it needs to provide appropriate information towards altering the model's decisions, while for a non-technical user, it considers the user's characteristics, e.g., cognitive load or preferences. The accuracy of the explanations is a key factor towards an effective interaction.   
    \item The \texttt{prediction-parameters} is used to help the user identify the weaknesses of the model, e.g., prediction with low confidence. This requires the model to communicate its confidence values along with its predictions for a set of input samples, towards a scrutable model. The user is able to explore and alter the model decisions, providing feedback for model updates.  
    \item The \texttt{policy visualization} is used to communicate the model's policy in an online fashion. In terms of implementation, the model communicates and updates its policy (input-output) during the interactions. This requires an online policy update mechanism to enable the user get an understanding of how the model's performance changes over time. 
\end{itemize}

\textit{HITL-based interaction patterns.} The goal of HITL methods and approaches is to efficiently integrate the human user to the learning and decision making process of an AI system. Human users can participate in the model's development and deployment phases. The selection of appropriate iML/HITL methods and approaches depends on several aspects of the interaction, e.g., goal, user role and expertise, etc. For a given context, HITL-based patterns can be used to design interactions where the user is part of the decision making and learning process. We identify the following implementation challenges for HITL/iML-based interactions: (a) the selection of teaching strategies considering user roles and expertise \cite{chen2022perspectives, cui2021understanding}, (b) the design of intuitive and user-friendly feedback/control interfaces to ensure high-quality feedback \cite{endert2012semantic, dudley2018review}, and (c) the integration of feedback (models) to model updates and decision making \cite{michael2020interactive, mosqueira2022human}. We can relate these to the patterns above as:
\begin{itemize}
    \item \texttt{sample-annotation} is used to enable the user provide a label for a sample. A key decision relates to the quality of the user-provided data. For the interactive robot learning, the model considers the human label as the ground truth for the generated sample, based on which it performs the learning update (online training). User is able to provide training data implicitly during the interaction. 
    \item \texttt{evaluative-advice} is used to enable a user to evaluate the model's performance. Evaluation feedback requires an interface for numerical feedback, as well as a reward shaping mechanism to integrate human feedback into model updates. User needs to be able to perceive and evaluate the model's predictions (policy) in an online manner. 
    \item \texttt{informative-advice} is used to enable provide an alternate decision. Informative advice requires an interface for action selection, as well as a policy shaping mechanism to integrate human feedback into model updates. User needs to be able to perceive and modify the model's predictions (policy) in an online manner. 
    \item \texttt{modify-mparams} is used as a control pattern which enables the user to alter the model's decisions by changing a model's parameter. In terms of implementation, the model should be able to dynamically alter its decisions based on the modified input. The model uses feedback only to alter its predictions for a new parameter and not to update its learning weights.  
\end{itemize}

\textit{Hybrid Intelligence and Collaborative learning interaction patterns}
The goal of collaborative learning and hybrid intelligence interactions is to enable both humans and machines (AI systems) to learn from each other in a collaborative manner. Collaborative learning interactions can be designed by combining XAI-based and HITL-based interactions, enabling both users and AI models to exchange information while solving a task. Based on our unpacking process, we identify and discuss interaction patterns used in collaborative learning and hybrid intelligence interactions. Implementation challenges for hybrid intelligence and collaborative learning systems include, but not limited to: (a) identify the appropriate levels of human control and AI automation both for model learning and decision making \cite{shneiderman2020human,zhang2021forward}, (b) design systems to facilitate the interaction between explainable artificial and cognitive intelligence \cite{wenskovitch2020interactive}, and (c) develop methods for the adaptation of HI systems considering user needs and capabilities to enhance user's perception towards improving the model's learning process \cite{akata2020research}. Considering these, we identify the following challenges for the extracted patterns: 

\begin{itemize}
    \item The \texttt{turn\_taking-evaluation} pattern is used to enable the user collaborate with a model in a turn-taking interaction. In terms of implementation, the model needs to capture the user's input during the interaction and generate a modified sketch. The user can control what the robot will capture, as well as when to provide feedback and terminate the interaction. 
    \item The \texttt{candidate-samples} pattern is used to support the human labeling process in a semi-automated way. The user makes the final decisions based on the model's initial decisions. The model is being updated based on user's feedback. The model needs to make online updates to improve the selection of the candidate samples, which can affect the user's and thus model's performance. 
    \item The \texttt{turn-taking\_XAI} pattern enables both model and user to justify their decisions using explanations. The implementation of this pattern requires an interface where the user can modify the model's prediction and explanations. The type of explanations (visualization) depends on user characteristics, e.g., preferences, cognitive load. 
\end{itemize}

Focusing on the implementation aspects of these patterns, we observe that each pattern can be linked to a set of technical aspects that need to be considered. We envision a library of patterns which can characterize a pattern in terms of the possible implementation choices to manifest common concerns and issues. In order to support both design and implementation choices, our proposed design space will be informed by existing frameworks and guidelines for HAI interactions.    

\subsubsection{Relation to existing frameworks and guidelines.}
\begin{figure}[h]
\centering
\includegraphics[width=0.9\columnwidth]{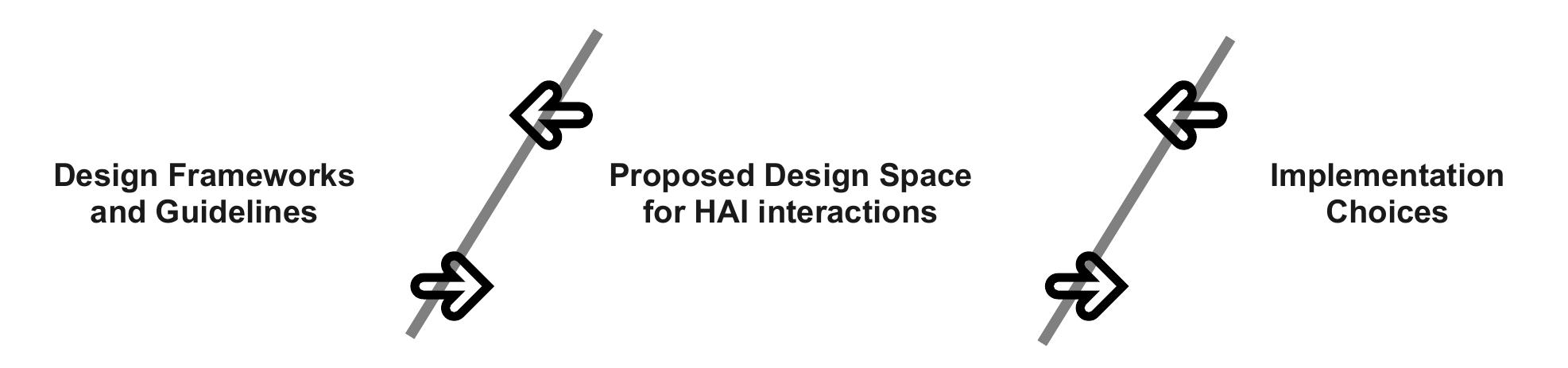}
\caption{Our proposed design space as a link between design guidelines and implementation choices.}
\label{fig:relation}
\end{figure}

In order to get insights about the formalization of the proposed design space, we briefly discuss how existing design guidelines and frameworks are related to our proposed approach. Our vision is a design space which can support AI designers to explore this space between design guidelines and implementation practices by enabling the collaboration of such frameworks and guidelines (Figure \ref{fig:relation}).

For example, the Microsoft guidelines for Human-AI interactions \cite{amershi2019guidelines} provide a set of pattern examples along with descriptions of interactions. These design guidelines can be used as a set of suggestions and pattern examples for the ideation phase of an HAI interaction system. For example, a guideline suggests helping the user understands what the system can do\footnote{https://www.microsoft.com/en-us/haxtoolkit/guideline/make-clear-what-the-system-can-do/}. One of the suggested pattern examples for this given guideline is to use explanation patterns in order to enable users to gain insights into system capabilities \textit{(XAI-based interaction patterns)}. Based on another guideline, the design should encourage granular feedback and enable the user to indicate their preferences\footnote{https://www.microsoft.com/en-us/haxtoolkit/guideline/encourage-granular-feedback/}. A suggested pattern for this guideline is to request explicit feedback on selected system outputs in order to assess the system and help it improve over time (HITL-based interactions). Our proposed design space could be informed by the provided guidelines and pattern examples in order to develop a collection of design patterns considering these guidelines. Working at a higher-level, the Assessment List for Trustworthy AI (ALTAI)\footnote{https://altai.insight-centre.org/} was developed for the assessment of AI systems in terms of seven requirements specified in the Ethics Guidelines for Trustworthy AI \cite{ala2020assessment}. \textit{Human Agency and Oversight} is one of the requirements and refers to the ability of human users to make informed decisions and to monitor and supervise the system. HITL-based interaction patterns can be used to integrate the user to the decision making and model learning process. Another ALTAI guideline is \textit{Transparency} and refers to the ability of the data, system and AI models to be transparent and explainable to the user. XAI-based interaction patterns can describe different approaches to provide explanations to the user based on the role and intent in the interaction. Considering the assessment list and requirements, our proposed design space can provide support towards selecting appropriate patterns and interactions to comply with specific assessment items.   

Apart from high-level guidelines, our proposed design space will be informed by technical and implementation frameworks for AI/ML models. Considering interactions with ML systems, implementation methods are required to enable (a) the communication of information between users and models and (b) the integration of user-provided data to the model learning (and decision making) mechanism. A classification of methods and approaches for interactive ML systems \cite{mosqueira2022classification} considers the development lifecycle of ML systems and provides a list of implementation choices based on a given category of methods, including interactive learning and explainability. However, designing efficient HITL-based interaction patterns requires both the selection of appropriate design choices and learning methods for feedback integration. Focusing on detecting and mitigating bias in ML pipelines, FaiPrep is an open library which extracts dataflow representations to support fairness during model development \cite{yang2020fairness}. Such representations can be used in our context to further characterize HAI interactions in terms of dataflows and model/data operations, e.g., fit, predict, update data. Finally, our proposed design space will be informed by existing approaches for design patterns for AI systems. Design patterns have been proposed for hybrid AI and reasoning systems, combining both data-driven and knowledge-driven AI models \cite{van2021modular}. These hybrid AI patterns can specify the operations that take place during the interactions (e.g., fit, update, predict) and can provide guidelines for the selection of appropriate learning methods.

\section{Conclusion}

The design space developed here is aimed at the complex space between concepts and practice, between formality and accessibility. The interaction patterns developed here act as intermediate level knowledge for the design and understanding of HAI systems, giving a formal representation of the configurations used in existing work.  Starting from a small set of interaction primitives and types to specify the communicated information between the interacting agents, we showed that the proposed primitives can be used to describe patterns of interactions from a range of systems, resulting in a collection of crisply defined actions and interaction patterns. We demonstrated that these patterns and actions are consistent with key HAI paradigms of HITL, XAI and hybrid intelligence, and that they can be used to explore and prototype a range of alternate interactions for a given situation.  

This space has been built  from theoretical ideas about communication between humans and models, based on ideas from agent communication languages and the semantic interaction framework for understanding human-model communication. It has then been developed and tested with examples of systems and interaction paradigms from the literature, demonstrating that it can meaningfully describe existing work. The representation language starts with data types and primitive communicative actions of providing and requesting information, and works up to high level conceptual activities --- interaction patterns --- that both capture common structures and describe the intent of the computational architectures. Extracting these patterns gives people of varying technicalities a common language to talk about what a particular system is doing, by building re-usable descriptions of the interactions taking place. This level of description allows for alternative design choices to be explored, while highlighting concerns that might arise and giving a framework for implementing the interactions. This provides a way to document existing practices, re-use well tested solutions and also speculate about new interaction possibilities through an exploration of the design space. Finally, through focussing on the interactive and communicative possibilities around models, the design space helps to shift thinking from a single user, single model, single purpose viewpoint to one where various stakeholders can carry out different kinds of interaction with a single model, creating a more ecosystemic view of human-model interactions.

\bibliographystyle{ACM-Reference-Format}
\bibliography{sample-manuscript}


\begin{thebibliography}{86}


\ifx \showCODEN    \undefined \def \showCODEN     #1{\unskip}     \fi
\ifx \showDOI      \undefined \def \showDOI       #1{#1}\fi
\ifx \showISBNx    \undefined \def \showISBNx     #1{\unskip}     \fi
\ifx \showISBNxiii \undefined \def \showISBNxiii  #1{\unskip}     \fi
\ifx \showISSN     \undefined \def \showISSN      #1{\unskip}     \fi
\ifx \showLCCN     \undefined \def \showLCCN      #1{\unskip}     \fi
\ifx \shownote     \undefined \def \shownote      #1{#1}          \fi
\ifx \showarticletitle \undefined \def \showarticletitle #1{#1}   \fi
\ifx \showURL      \undefined \def \showURL       {\relax}        \fi
\providecommand\bibfield[2]{#2}
\providecommand\bibinfo[2]{#2}
\providecommand\natexlab[1]{#1}
\providecommand\showeprint[2][]{arXiv:#2}

\bibitem[Adhikari et~al\mbox{.}(2022)]%
        {adhikari2022towards}
\bibfield{author}{\bibinfo{person}{Ajaya Adhikari}, \bibinfo{person}{Edwin
  Wenink}, \bibinfo{person}{Jasper van~der Waa}, \bibinfo{person}{Cornelis
  Bouter}, \bibinfo{person}{Ioannis Tolios}, {and} \bibinfo{person}{Stephan
  Raaijmakers}.} \bibinfo{year}{2022}\natexlab{}.
\newblock \showarticletitle{Towards FAIR Explainable AI: a standardized
  ontology for mapping XAI solutions to use cases, explanations, and AI
  systems}. In \bibinfo{booktitle}{\emph{Proceedings of the 15th International
  Conference on PErvasive Technologies Related to Assistive Environments}}.
  \bibinfo{pages}{562--568}.
\newblock


\bibitem[Ahmad et~al\mbox{.}(2007)]%
        {ahmad2007intelligence}
\bibfield{author}{\bibinfo{person}{Raheel Ahmad}, \bibinfo{person}{Shahram
  Rahimi}, {and} \bibinfo{person}{Bidyut Gupta}.}
  \bibinfo{year}{2007}\natexlab{}.
\newblock \showarticletitle{An Intelligence-Aware Process Calculus for
  Multi-Agent System Modeling}. In \bibinfo{booktitle}{\emph{2007 International
  Conference on Integration of Knowledge Intensive Multi-Agent Systems}}. IEEE,
  \bibinfo{pages}{210--215}.
\newblock


\bibitem[Ahmed et~al\mbox{.}(2009)]%
        {ahmed2009review}
\bibfield{author}{\bibinfo{person}{Moamin Ahmed},
  \bibinfo{person}{Mohd~Sharifuddin Ahmad}, {and} \bibinfo{person}{Mohd
  Zaliman~Mohd Yusoff}.} \bibinfo{year}{2009}\natexlab{}.
\newblock \showarticletitle{A review and development of agent communication
  language}.
\newblock \bibinfo{journal}{\emph{Electronic Journal of Computer Science and
  Information Technology}} \bibinfo{volume}{1}, \bibinfo{number}{1}
  (\bibinfo{year}{2009}).
\newblock


\bibitem[Akata et~al\mbox{.}(2020)]%
        {akata2020research}
\bibfield{author}{\bibinfo{person}{Zeynep Akata}, \bibinfo{person}{Dan
  Balliet}, \bibinfo{person}{Maarten De~Rijke}, \bibinfo{person}{Frank Dignum},
  \bibinfo{person}{Virginia Dignum}, \bibinfo{person}{Guszti Eiben},
  \bibinfo{person}{Antske Fokkens}, \bibinfo{person}{Davide Grossi},
  \bibinfo{person}{Koen Hindriks}, \bibinfo{person}{Holger Hoos},
  {et~al\mbox{.}}} \bibinfo{year}{2020}\natexlab{}.
\newblock \showarticletitle{A research agenda for hybrid intelligence:
  augmenting human intellect with collaborative, adaptive, responsible, and
  explainable artificial intelligence}.
\newblock \bibinfo{journal}{\emph{Computer}} \bibinfo{volume}{53},
  \bibinfo{number}{08} (\bibinfo{year}{2020}), \bibinfo{pages}{18--28}.
\newblock


\bibitem[Ala-Pietil{\"a} et~al\mbox{.}(2020)]%
        {ala2020assessment}
\bibfield{author}{\bibinfo{person}{Pekka Ala-Pietil{\"a}},
  \bibinfo{person}{Yann Bonnet}, \bibinfo{person}{Urs Bergmann},
  \bibinfo{person}{Maria Bielikova}, \bibinfo{person}{Cecilia Bonefeld-Dahl},
  \bibinfo{person}{Wilhelm Bauer}, \bibinfo{person}{Loubna Bouarfa},
  \bibinfo{person}{Raja Chatila}, \bibinfo{person}{Mark Coeckelbergh},
  \bibinfo{person}{Virginia Dignum}, {et~al\mbox{.}}}
  \bibinfo{year}{2020}\natexlab{}.
\newblock \bibinfo{booktitle}{\emph{The assessment list for trustworthy
  artificial intelligence (ALTAI)}}.
\newblock \bibinfo{publisher}{European Commission}.
\newblock


\bibitem[Alfrink et~al\mbox{.}(2022)]%
        {alfrink2022contestable}
\bibfield{author}{\bibinfo{person}{Kars Alfrink}, \bibinfo{person}{Ianus
  Keller}, \bibinfo{person}{Gerd Kortuem}, {and} \bibinfo{person}{Neelke
  Doorn}.} \bibinfo{year}{2022}\natexlab{}.
\newblock \showarticletitle{Contestable AI by Design: Towards a Framework}.
\newblock \bibinfo{journal}{\emph{Minds and Machines}} (\bibinfo{year}{2022}),
  \bibinfo{pages}{1--27}.
\newblock


\bibitem[Amershi et~al\mbox{.}(2019)]%
        {amershi2019guidelines}
\bibfield{author}{\bibinfo{person}{Saleema Amershi}, \bibinfo{person}{Dan
  Weld}, \bibinfo{person}{Mihaela Vorvoreanu}, \bibinfo{person}{Adam Fourney},
  \bibinfo{person}{Besmira Nushi}, \bibinfo{person}{Penny Collisson},
  \bibinfo{person}{Jina Suh}, \bibinfo{person}{Shamsi Iqbal},
  \bibinfo{person}{Paul~N Bennett}, \bibinfo{person}{Kori Inkpen},
  {et~al\mbox{.}}} \bibinfo{year}{2019}\natexlab{}.
\newblock \showarticletitle{Guidelines for human-AI interaction}. In
  \bibinfo{booktitle}{\emph{Proceedings of the 2019 chi conference on human
  factors in computing systems}}. \bibinfo{pages}{1--13}.
\newblock


\bibitem[Arrieta et~al\mbox{.}(2020)]%
        {arrieta2020explainable}
\bibfield{author}{\bibinfo{person}{Alejandro~Barredo Arrieta},
  \bibinfo{person}{Natalia D{\'\i}az-Rodr{\'\i}guez}, \bibinfo{person}{Javier
  Del~Ser}, \bibinfo{person}{Adrien Bennetot}, \bibinfo{person}{Siham Tabik},
  \bibinfo{person}{Alberto Barbado}, \bibinfo{person}{Salvador Garc{\'\i}a},
  \bibinfo{person}{Sergio Gil-L{\'o}pez}, \bibinfo{person}{Daniel Molina},
  \bibinfo{person}{Richard Benjamins}, {et~al\mbox{.}}}
  \bibinfo{year}{2020}\natexlab{}.
\newblock \showarticletitle{Explainable Artificial Intelligence (XAI):
  Concepts, taxonomies, opportunities and challenges toward responsible AI}.
\newblock \bibinfo{journal}{\emph{Information fusion}}  \bibinfo{volume}{58}
  (\bibinfo{year}{2020}), \bibinfo{pages}{82--115}.
\newblock


\bibitem[Balayn et~al\mbox{.}(2021)]%
        {balayn2021WhatYou}
\bibfield{author}{\bibinfo{person}{Agathe Balayn}, \bibinfo{person}{Panagiotis
  Soilis}, \bibinfo{person}{Christoph Lofi}, \bibinfo{person}{Jie Yang}, {and}
  \bibinfo{person}{Alessandro Bozzon}.} \bibinfo{year}{2021}\natexlab{}.
\newblock \showarticletitle{What Do {{You Mean}}? {{Interpreting Image
  Classification}} with {{Crowdsourced Concept Extraction}} and {{Analysis}}}.
  In \bibinfo{booktitle}{\emph{Proceedings of the {{Web Conference}} 2021}}.
  \bibinfo{publisher}{{ACM}}, \bibinfo{address}{{Ljubljana Slovenia}},
  \bibinfo{pages}{1937--1948}.
\newblock
\showISBNx{978-1-4503-8312-7}
\urldef\tempurl%
\url{https://doi.org/10.1145/3442381.3450069}
\showDOI{\tempurl}


\bibitem[Behymer and Flach(2016)]%
        {behymer2016autonomous}
\bibfield{author}{\bibinfo{person}{Kyle~J Behymer} {and}
  \bibinfo{person}{John~M Flach}.} \bibinfo{year}{2016}\natexlab{}.
\newblock \showarticletitle{From autonomous systems to sociotechnical systems:
  Designing effective collaborations}.
\newblock \bibinfo{journal}{\emph{She Ji: The Journal of Design, Economics, and
  Innovation}} \bibinfo{volume}{2}, \bibinfo{number}{2} (\bibinfo{year}{2016}),
  \bibinfo{pages}{105--114}.
\newblock


\bibitem[Bhatt et~al\mbox{.}(2021)]%
        {bhatt2021uncertainty}
\bibfield{author}{\bibinfo{person}{Umang Bhatt}, \bibinfo{person}{Javier
  Antor{\'a}n}, \bibinfo{person}{Yunfeng Zhang}, \bibinfo{person}{Q~Vera Liao},
  \bibinfo{person}{Prasanna Sattigeri}, \bibinfo{person}{Riccardo Fogliato},
  \bibinfo{person}{Gabrielle Melan{\c{c}}on}, \bibinfo{person}{Ranganath
  Krishnan}, \bibinfo{person}{Jason Stanley}, \bibinfo{person}{Omesh Tickoo},
  {et~al\mbox{.}}} \bibinfo{year}{2021}\natexlab{}.
\newblock \showarticletitle{Uncertainty as a form of transparency: Measuring,
  communicating, and using uncertainty}. In
  \bibinfo{booktitle}{\emph{Proceedings of the 2021 AAAI/ACM Conference on AI,
  Ethics, and Society}}. \bibinfo{pages}{401--413}.
\newblock


\bibitem[Bignold et~al\mbox{.}(2022)]%
        {bignold2022human}
\bibfield{author}{\bibinfo{person}{Adam Bignold}, \bibinfo{person}{Francisco
  Cruz}, \bibinfo{person}{Richard Dazeley}, \bibinfo{person}{Peter Vamplew},
  {and} \bibinfo{person}{Cameron Foale}.} \bibinfo{year}{2022}\natexlab{}.
\newblock \showarticletitle{Human engagement providing evaluative and
  informative advice for interactive reinforcement learning}.
\newblock \bibinfo{journal}{\emph{Neural Computing and Applications}}
  (\bibinfo{year}{2022}), \bibinfo{pages}{1--16}.
\newblock


\bibitem[Birhane(2021)]%
        {birhane2021algorithmic}
\bibfield{author}{\bibinfo{person}{Abeba Birhane}.}
  \bibinfo{year}{2021}\natexlab{}.
\newblock \showarticletitle{Algorithmic injustice: a relational ethics
  approach}.
\newblock \bibinfo{journal}{\emph{Patterns}} \bibinfo{volume}{2},
  \bibinfo{number}{2} (\bibinfo{year}{2021}), \bibinfo{pages}{100205}.
\newblock


\bibitem[Cavalcante~Siebert et~al\mbox{.}(2022)]%
        {cavalcante2022meaningful}
\bibfield{author}{\bibinfo{person}{Luciano Cavalcante~Siebert},
  \bibinfo{person}{Maria~Luce Lupetti}, \bibinfo{person}{Evgeni Aizenberg},
  \bibinfo{person}{Niek Beckers}, \bibinfo{person}{Arkady Zgonnikov},
  \bibinfo{person}{Herman Veluwenkamp}, \bibinfo{person}{David Abbink},
  \bibinfo{person}{Elisa Giaccardi}, \bibinfo{person}{Geert-Jan Houben},
  \bibinfo{person}{Catholijn~M Jonker}, {et~al\mbox{.}}}
  \bibinfo{year}{2022}\natexlab{}.
\newblock \showarticletitle{Meaningful human control: Actionable properties for
  AI system development}.
\newblock \bibinfo{journal}{\emph{AI and Ethics}} (\bibinfo{year}{2022}),
  \bibinfo{pages}{1--15}.
\newblock


\bibitem[Chai and Li(2020)]%
        {chai2020human}
\bibfield{author}{\bibinfo{person}{Chengliang Chai} {and}
  \bibinfo{person}{Guoliang Li}.} \bibinfo{year}{2020}\natexlab{}.
\newblock \showarticletitle{Human-in-the-loop Techniques in Machine Learning.}
\newblock \bibinfo{journal}{\emph{IEEE Data Eng. Bull.}} \bibinfo{volume}{43},
  \bibinfo{number}{3} (\bibinfo{year}{2020}), \bibinfo{pages}{37--52}.
\newblock


\bibitem[Chen et~al\mbox{.}(2022)]%
        {chen2022perspectives}
\bibfield{author}{\bibinfo{person}{Valerie Chen}, \bibinfo{person}{Umang
  Bhatt}, \bibinfo{person}{Hoda Heidari}, \bibinfo{person}{Adrian Weller},
  {and} \bibinfo{person}{Ameet Talwalkar}.} \bibinfo{year}{2022}\natexlab{}.
\newblock \showarticletitle{Perspectives on Incorporating Expert Feedback into
  Model Updates}.
\newblock \bibinfo{journal}{\emph{arXiv preprint arXiv:2205.06905}}
  (\bibinfo{year}{2022}).
\newblock


\bibitem[Chromik and Butz(2021)]%
        {chromik2021human}
\bibfield{author}{\bibinfo{person}{Michael Chromik} {and}
  \bibinfo{person}{Andreas Butz}.} \bibinfo{year}{2021}\natexlab{}.
\newblock \showarticletitle{Human-XAI interaction: a review and design
  principles for explanation user interfaces}. In
  \bibinfo{booktitle}{\emph{IFIP Conference on Human-Computer Interaction}}.
  Springer, \bibinfo{pages}{619--640}.
\newblock


\bibitem[Chromik and Schuessler(2020)]%
        {chromik2020taxonomy}
\bibfield{author}{\bibinfo{person}{Michael Chromik} {and}
  \bibinfo{person}{Martin Schuessler}.} \bibinfo{year}{2020}\natexlab{}.
\newblock \showarticletitle{A Taxonomy for Human Subject Evaluation of
  Black-Box Explanations in XAI.}
\newblock \bibinfo{journal}{\emph{Exss-atec@ iui}}  \bibinfo{volume}{94}
  (\bibinfo{year}{2020}).
\newblock


\bibitem[Cui et~al\mbox{.}(2021)]%
        {cui2021understanding}
\bibfield{author}{\bibinfo{person}{Yuchen Cui}, \bibinfo{person}{Pallavi
  Koppol}, \bibinfo{person}{Henny Admoni}, \bibinfo{person}{Scott Niekum},
  \bibinfo{person}{Reid~G Simmons}, \bibinfo{person}{Aaron Steinfeld}, {and}
  \bibinfo{person}{Tesca Fitzgerald}.} \bibinfo{year}{2021}\natexlab{}.
\newblock \showarticletitle{Understanding the Relationship between Interactions
  and Outcomes in Human-in-the-Loop Machine Learning.}. In
  \bibinfo{booktitle}{\emph{IJCAI}}. \bibinfo{pages}{4382--4391}.
\newblock


\bibitem[Dellermann et~al\mbox{.}(2021)]%
        {dellermann2021future}
\bibfield{author}{\bibinfo{person}{Dominik Dellermann}, \bibinfo{person}{Adrian
  Calma}, \bibinfo{person}{Nikolaus Lipusch}, \bibinfo{person}{Thorsten Weber},
  \bibinfo{person}{Sascha Weigel}, {and} \bibinfo{person}{Philipp Ebel}.}
  \bibinfo{year}{2021}\natexlab{}.
\newblock \showarticletitle{The future of human-AI collaboration: a taxonomy of
  design knowledge for hybrid intelligence systems}.
\newblock \bibinfo{journal}{\emph{arXiv preprint arXiv:2105.03354}}
  (\bibinfo{year}{2021}).
\newblock


\bibitem[{d'Inverno} et~al\mbox{.}(2012)]%
        {dinverno2012CommunicatingOpen}
\bibfield{author}{\bibinfo{person}{Mark {d'Inverno}}, \bibinfo{person}{Michael
  Luck}, \bibinfo{person}{Pablo Noriega}, \bibinfo{person}{Juan~A.
  {Rodriguez-Aguilar}}, {and} \bibinfo{person}{Carles Sierra}.}
  \bibinfo{year}{2012}\natexlab{}.
\newblock \showarticletitle{Communicating Open Systems}.
\newblock \bibinfo{journal}{\emph{Artificial Intelligence}}
  \bibinfo{volume}{186} (\bibinfo{date}{July} \bibinfo{year}{2012}),
  \bibinfo{pages}{38--94}.
\newblock
\showISSN{0004-3702}
\urldef\tempurl%
\url{https://doi.org/10.1016/j.artint.2012.03.004}
\showDOI{\tempurl}


\bibitem[Dove and Fayard(2020)]%
        {dove2020MonstersMetaphors}
\bibfield{author}{\bibinfo{person}{Graham Dove} {and}
  \bibinfo{person}{Anne-Laure Fayard}.} \bibinfo{year}{2020}\natexlab{}.
\newblock \showarticletitle{Monsters, {{Metaphors}}, and {{Machine Learning}}}.
  In \bibinfo{booktitle}{\emph{Proceedings of the 2020 {{CHI Conference}} on
  {{Human Factors}} in {{Computing Systems}}}} \emph{(\bibinfo{series}{{{CHI}}
  '20})}. \bibinfo{publisher}{{Association for Computing Machinery}},
  \bibinfo{address}{{New York, NY, USA}}, \bibinfo{pages}{1--17}.
\newblock
\showISBNx{978-1-4503-6708-0}
\urldef\tempurl%
\url{https://doi.org/10.1145/3313831.3376275}
\showDOI{\tempurl}


\bibitem[Dove et~al\mbox{.}(2017)]%
        {dove2017ux}
\bibfield{author}{\bibinfo{person}{Graham Dove}, \bibinfo{person}{Kim Halskov},
  \bibinfo{person}{Jodi Forlizzi}, {and} \bibinfo{person}{John Zimmerman}.}
  \bibinfo{year}{2017}\natexlab{}.
\newblock \showarticletitle{UX design innovation: Challenges for working with
  machine learning as a design material}. In
  \bibinfo{booktitle}{\emph{Proceedings of the 2017 chi conference on human
  factors in computing systems}}. \bibinfo{pages}{278--288}.
\newblock


\bibitem[Dudley and Kristensson(2018)]%
        {dudley2018review}
\bibfield{author}{\bibinfo{person}{John~J Dudley} {and}
  \bibinfo{person}{Per~Ola Kristensson}.} \bibinfo{year}{2018}\natexlab{}.
\newblock \showarticletitle{A review of user interface design for interactive
  machine learning}.
\newblock \bibinfo{journal}{\emph{ACM Transactions on Interactive Intelligent
  Systems (TiiS)}} \bibinfo{volume}{8}, \bibinfo{number}{2}
  (\bibinfo{year}{2018}), \bibinfo{pages}{1--37}.
\newblock


\bibitem[Edwards and Veale(2017)]%
        {edwards2017SlaveAlgorithm}
\bibfield{author}{\bibinfo{person}{Lilian Edwards} {and}
  \bibinfo{person}{Michael Veale}.} \bibinfo{year}{2017}\natexlab{}.
\newblock \bibinfo{booktitle}{\emph{Slave to the {{Algorithm}}? {{Why}} a
  'right to an Explanation' Is Probably Not the Remedy You Are Looking For}}.
\newblock \bibinfo{type}{Preprint}. \bibinfo{institution}{{LawArXiv}}.
\newblock
\urldef\tempurl%
\url{https://doi.org/10.31228/osf.io/97upg}
\showDOI{\tempurl}


\bibitem[Ehsan and Riedl(2020)]%
        {ehsan2020human}
\bibfield{author}{\bibinfo{person}{Upol Ehsan} {and} \bibinfo{person}{Mark~O
  Riedl}.} \bibinfo{year}{2020}\natexlab{}.
\newblock \showarticletitle{Human-centered explainable ai: Towards a reflective
  sociotechnical approach}. In \bibinfo{booktitle}{\emph{International
  Conference on Human-Computer Interaction}}. Springer,
  \bibinfo{pages}{449--466}.
\newblock


\bibitem[Enarsson et~al\mbox{.}(2022)]%
        {enarsson2022approaching}
\bibfield{author}{\bibinfo{person}{Therese Enarsson}, \bibinfo{person}{Lena
  Enqvist}, {and} \bibinfo{person}{Markus Naarttij{\"a}rvi}.}
  \bibinfo{year}{2022}\natexlab{}.
\newblock \showarticletitle{Approaching the human in the loop--legal
  perspectives on hybrid human/algorithmic decision-making in three contexts}.
\newblock \bibinfo{journal}{\emph{Information \& Communications Technology
  Law}} \bibinfo{volume}{31}, \bibinfo{number}{1} (\bibinfo{year}{2022}),
  \bibinfo{pages}{123--153}.
\newblock


\bibitem[Endert et~al\mbox{.}(2012)]%
        {endert2012semantic}
\bibfield{author}{\bibinfo{person}{Alex Endert}, \bibinfo{person}{Patrick
  Fiaux}, {and} \bibinfo{person}{Chris North}.}
  \bibinfo{year}{2012}\natexlab{}.
\newblock \showarticletitle{Semantic interaction for sensemaking: inferring
  analytical reasoning for model steering}.
\newblock \bibinfo{journal}{\emph{IEEE Transactions on Visualization and
  Computer Graphics}} \bibinfo{volume}{18}, \bibinfo{number}{12}
  (\bibinfo{year}{2012}), \bibinfo{pages}{2879--2888}.
\newblock


\bibitem[Finin et~al\mbox{.}(1994)]%
        {finin1994kqml}
\bibfield{author}{\bibinfo{person}{Tim Finin}, \bibinfo{person}{Richard
  Fritzson}, \bibinfo{person}{Don McKay}, {and} \bibinfo{person}{Robin
  McEntire}.} \bibinfo{year}{1994}\natexlab{}.
\newblock \showarticletitle{KQML as an agent communication language}. In
  \bibinfo{booktitle}{\emph{Proceedings of the third international conference
  on Information and knowledge management}}. \bibinfo{pages}{456--463}.
\newblock


\bibitem[Giaccardi and Redstr{\"o}m(2020)]%
        {giaccardi2020technology}
\bibfield{author}{\bibinfo{person}{Elisa Giaccardi} {and}
  \bibinfo{person}{Johan Redstr{\"o}m}.} \bibinfo{year}{2020}\natexlab{}.
\newblock \showarticletitle{Technology and more-than-human design}.
\newblock \bibinfo{journal}{\emph{Design Issues}} \bibinfo{volume}{36},
  \bibinfo{number}{4} (\bibinfo{year}{2020}), \bibinfo{pages}{33--44}.
\newblock


\bibitem[Gillies(2019)]%
        {gillies2019understanding}
\bibfield{author}{\bibinfo{person}{Marco Gillies}.}
  \bibinfo{year}{2019}\natexlab{}.
\newblock \showarticletitle{Understanding the role of interactive machine
  learning in movement interaction design}.
\newblock \bibinfo{journal}{\emph{ACM Transactions on Computer-Human
  Interaction (TOCHI)}} \bibinfo{volume}{26}, \bibinfo{number}{1}
  (\bibinfo{year}{2019}), \bibinfo{pages}{1--34}.
\newblock


\bibitem[Grabe et~al\mbox{.}(2022)]%
        {grabe2022towards}
\bibfield{author}{\bibinfo{person}{Imke Grabe}, \bibinfo{person}{Miguel
  Gonz{\'a}lez-Duque}, \bibinfo{person}{Sebastian Risi}, {and}
  \bibinfo{person}{Jichen Zhu}.} \bibinfo{year}{2022}\natexlab{}.
\newblock \showarticletitle{Towards a Framework for Human-AI Interaction
  Patterns in Co-Creative GAN Applications}.
\newblock  (\bibinfo{year}{2022}).
\newblock


\bibitem[Gr{\o}nsund and Aanestad(2020)]%
        {gronsund2020augmenting}
\bibfield{author}{\bibinfo{person}{Tor Gr{\o}nsund} {and}
  \bibinfo{person}{Margunn Aanestad}.} \bibinfo{year}{2020}\natexlab{}.
\newblock \showarticletitle{Augmenting the algorithm: Emerging
  human-in-the-loop work configurations}.
\newblock \bibinfo{journal}{\emph{The Journal of Strategic Information
  Systems}} \bibinfo{volume}{29}, \bibinfo{number}{2} (\bibinfo{year}{2020}),
  \bibinfo{pages}{101614}.
\newblock


\bibitem[Guo et~al\mbox{.}(2022)]%
        {guo2022building}
\bibfield{author}{\bibinfo{person}{Lijie Guo}, \bibinfo{person}{Elizabeth~M
  Daly}, \bibinfo{person}{Oznur Alkan}, \bibinfo{person}{Massimiliano
  Mattetti}, \bibinfo{person}{Owen Cornec}, {and} \bibinfo{person}{Bart
  Knijnenburg}.} \bibinfo{year}{2022}\natexlab{}.
\newblock \showarticletitle{Building Trust in Interactive Machine Learning via
  User Contributed Interpretable Rules}. In \bibinfo{booktitle}{\emph{27th
  International Conference on Intelligent User Interfaces}}.
  \bibinfo{pages}{537--548}.
\newblock


\bibitem[Heimerl et~al\mbox{.}(2019)]%
        {heimerl2019nova}
\bibfield{author}{\bibinfo{person}{Alexander Heimerl}, \bibinfo{person}{Tobias
  Baur}, \bibinfo{person}{Florian Lingenfelser}, \bibinfo{person}{Johannes
  Wagner}, {and} \bibinfo{person}{Elisabeth Andr{\'e}}.}
  \bibinfo{year}{2019}\natexlab{}.
\newblock \showarticletitle{NOVA-a tool for eXplainable Cooperative Machine
  Learning}. In \bibinfo{booktitle}{\emph{2019 8th International Conference on
  Affective Computing and Intelligent Interaction (ACII)}}. IEEE,
  \bibinfo{pages}{109--115}.
\newblock


\bibitem[Hirsch et~al\mbox{.}(2017)]%
        {hirsch2017designing}
\bibfield{author}{\bibinfo{person}{Tad Hirsch}, \bibinfo{person}{Kritzia
  Merced}, \bibinfo{person}{Shrikanth Narayanan}, \bibinfo{person}{Zac~E Imel},
  {and} \bibinfo{person}{David~C Atkins}.} \bibinfo{year}{2017}\natexlab{}.
\newblock \showarticletitle{Designing contestability: Interaction design,
  machine learning, and mental health}. In
  \bibinfo{booktitle}{\emph{Proceedings of the 2017 Conference on Designing
  Interactive Systems}}. \bibinfo{pages}{95--99}.
\newblock


\bibitem[Holmquist(2017)]%
        {holmquist2017intelligence}
\bibfield{author}{\bibinfo{person}{Lars~Erik Holmquist}.}
  \bibinfo{year}{2017}\natexlab{}.
\newblock \showarticletitle{Intelligence on tap: artificial intelligence as a
  new design material}.
\newblock \bibinfo{journal}{\emph{interactions}} \bibinfo{volume}{24},
  \bibinfo{number}{4} (\bibinfo{year}{2017}), \bibinfo{pages}{28--33}.
\newblock


\bibitem[Khadpe et~al\mbox{.}(2020)]%
        {khadpe2020ConceptualMetaphors}
\bibfield{author}{\bibinfo{person}{Pranav Khadpe}, \bibinfo{person}{Ranjay
  Krishna}, \bibinfo{person}{Li {Fei-Fei}}, \bibinfo{person}{Jeffrey~T.
  Hancock}, {and} \bibinfo{person}{Michael~S. Bernstein}.}
  \bibinfo{year}{2020}\natexlab{}.
\newblock \showarticletitle{Conceptual Metaphors Impact Perceptions of
  Human-{{AI}} Collaboration}.
\newblock \bibinfo{journal}{\emph{Proceedings of the ACM on Human-Computer
  Interaction}} \bibinfo{volume}{4}, \bibinfo{number}{CSCW2}
  (\bibinfo{year}{2020}), \bibinfo{pages}{1--26}.
\newblock


\bibitem[Kim and Pardo(2018)]%
        {kim2018human}
\bibfield{author}{\bibinfo{person}{Bongjun Kim} {and} \bibinfo{person}{Bryan
  Pardo}.} \bibinfo{year}{2018}\natexlab{}.
\newblock \showarticletitle{A human-in-the-loop system for sound event
  detection and annotation}.
\newblock \bibinfo{journal}{\emph{ACM Transactions on Interactive Intelligent
  Systems (TiiS)}} \bibinfo{volume}{8}, \bibinfo{number}{2}
  (\bibinfo{year}{2018}), \bibinfo{pages}{1--23}.
\newblock


\bibitem[Kocielnik et~al\mbox{.}(2019)]%
        {kocielnik2019will}
\bibfield{author}{\bibinfo{person}{Rafal Kocielnik}, \bibinfo{person}{Saleema
  Amershi}, {and} \bibinfo{person}{Paul~N Bennett}.}
  \bibinfo{year}{2019}\natexlab{}.
\newblock \showarticletitle{Will you accept an imperfect ai? exploring designs
  for adjusting end-user expectations of ai systems}. In
  \bibinfo{booktitle}{\emph{Proceedings of the 2019 CHI Conference on Human
  Factors in Computing Systems}}. \bibinfo{pages}{1--14}.
\newblock


\bibitem[Liao et~al\mbox{.}(2020)]%
        {liao2020questioning}
\bibfield{author}{\bibinfo{person}{Q~Vera Liao}, \bibinfo{person}{Daniel
  Gruen}, {and} \bibinfo{person}{Sarah Miller}.}
  \bibinfo{year}{2020}\natexlab{}.
\newblock \showarticletitle{Questioning the AI: informing design practices for
  explainable AI user experiences}. In \bibinfo{booktitle}{\emph{Proceedings of
  the 2020 CHI Conference on Human Factors in Computing Systems}}.
  \bibinfo{pages}{1--15}.
\newblock


\bibitem[Liao and Varshney(2021)]%
        {liao2021human}
\bibfield{author}{\bibinfo{person}{Q~Vera Liao} {and} \bibinfo{person}{Kush~R
  Varshney}.} \bibinfo{year}{2021}\natexlab{}.
\newblock \showarticletitle{Human-centered explainable ai (xai): From
  algorithms to user experiences}.
\newblock \bibinfo{journal}{\emph{arXiv preprint arXiv:2110.10790}}
  (\bibinfo{year}{2021}).
\newblock


\bibitem[Lin et~al\mbox{.}(2020)]%
        {lin2020your}
\bibfield{author}{\bibinfo{person}{Yuyu Lin}, \bibinfo{person}{Jiahao Guo},
  \bibinfo{person}{Yang Chen}, \bibinfo{person}{Cheng Yao}, {and}
  \bibinfo{person}{Fangtian Ying}.} \bibinfo{year}{2020}\natexlab{}.
\newblock \showarticletitle{It is your turn: collaborative ideation with a
  co-creative robot through sketch}. In \bibinfo{booktitle}{\emph{Proceedings
  of the 2020 CHI conference on human factors in computing systems}}.
  \bibinfo{pages}{1--14}.
\newblock


\bibitem[Lindley et~al\mbox{.}(2020)]%
        {lindley2020researching}
\bibfield{author}{\bibinfo{person}{Joseph Lindley}, \bibinfo{person}{Haider~Ali
  Akmal}, \bibinfo{person}{Franziska Pilling}, {and} \bibinfo{person}{Paul
  Coulton}.} \bibinfo{year}{2020}\natexlab{}.
\newblock \showarticletitle{Researching AI legibility through design}. In
  \bibinfo{booktitle}{\emph{Proceedings of the 2020 CHI Conference on Human
  Factors in Computing Systems}}. \bibinfo{pages}{1--13}.
\newblock


\bibitem[Luria(2018)]%
        {luria2018DesigningRobot}
\bibfield{author}{\bibinfo{person}{Michal Luria}.}
  \bibinfo{year}{2018}\natexlab{}.
\newblock \showarticletitle{Designing {{Robot Personality Based}} on
  {{Fictional Sidekick Characters}}}. In \bibinfo{booktitle}{\emph{Companion of
  the 2018 {{ACM}}/{{IEEE International Conference}} on {{Human-Robot
  Interaction}}}} \emph{(\bibinfo{series}{{{HRI}} '18})}.
  \bibinfo{publisher}{{Association for Computing Machinery}},
  \bibinfo{address}{{New York, NY, USA}}, \bibinfo{pages}{307--308}.
\newblock
\showISBNx{978-1-4503-5615-2}
\urldef\tempurl%
\url{https://doi.org/10.1145/3173386.3176912}
\showDOI{\tempurl}


\bibitem[Maettig and Foot(2020)]%
        {maettig2020approach}
\bibfield{author}{\bibinfo{person}{Benedikt Maettig} {and}
  \bibinfo{person}{Hermann Foot}.} \bibinfo{year}{2020}\natexlab{}.
\newblock \showarticletitle{Approach to improving training of human workers in
  industrial applications through the use of intelligence augmentation and
  human-in-the-loop}. In \bibinfo{booktitle}{\emph{2020 15th International
  Conference on Computer Science \& Education (ICCSE)}}. IEEE,
  \bibinfo{pages}{283--288}.
\newblock


\bibitem[Martijn et~al\mbox{.}(2022)]%
        {martijn2022knowing}
\bibfield{author}{\bibinfo{person}{Millecamp Martijn},
  \bibinfo{person}{Cristina Conati}, {and} \bibinfo{person}{Katrien Verbert}.}
  \bibinfo{year}{2022}\natexlab{}.
\newblock \showarticletitle{“Knowing me, knowing you”: personalized
  explanations for a music recommender system}.
\newblock \bibinfo{journal}{\emph{User Modeling and User-Adapted Interaction}}
  \bibinfo{volume}{32}, \bibinfo{number}{1} (\bibinfo{year}{2022}),
  \bibinfo{pages}{215--252}.
\newblock


\bibitem[Michael et~al\mbox{.}(2020)]%
        {michael2020interactive}
\bibfield{author}{\bibinfo{person}{Chris~J Michael}, \bibinfo{person}{Dina
  Acklin}, {and} \bibinfo{person}{Jaelle Scheuerman}.}
  \bibinfo{year}{2020}\natexlab{}.
\newblock \showarticletitle{On interactive machine learning and the potential
  of cognitive feedback}.
\newblock \bibinfo{journal}{\emph{arXiv preprint arXiv:2003.10365}}
  (\bibinfo{year}{2020}).
\newblock


\bibitem[Mohseni et~al\mbox{.}(2021)]%
        {mohseni2021multidisciplinary}
\bibfield{author}{\bibinfo{person}{Sina Mohseni}, \bibinfo{person}{Niloofar
  Zarei}, {and} \bibinfo{person}{Eric~D Ragan}.}
  \bibinfo{year}{2021}\natexlab{}.
\newblock \showarticletitle{A multidisciplinary survey and framework for design
  and evaluation of explainable AI systems}.
\newblock \bibinfo{journal}{\emph{ACM Transactions on Interactive Intelligent
  Systems (TiiS)}} \bibinfo{volume}{11}, \bibinfo{number}{3-4}
  (\bibinfo{year}{2021}), \bibinfo{pages}{1--45}.
\newblock


\bibitem[Mosqueira-Rey et~al\mbox{.}(2022a)]%
        {mosqueira2022human}
\bibfield{author}{\bibinfo{person}{Eduardo Mosqueira-Rey},
  \bibinfo{person}{Elena Hern{\'a}ndez-Pereira}, \bibinfo{person}{David
  Alonso-R{\'\i}os}, \bibinfo{person}{Jos{\'e} Bobes-Bascar{\'a}n}, {and}
  \bibinfo{person}{{\'A}ngel Fern{\'a}ndez-Leal}.}
  \bibinfo{year}{2022}\natexlab{a}.
\newblock \showarticletitle{Human-in-the-loop machine learning: a state of the
  art}.
\newblock \bibinfo{journal}{\emph{Artificial Intelligence Review}}
  (\bibinfo{year}{2022}), \bibinfo{pages}{1--50}.
\newblock


\bibitem[Mosqueira-Rey et~al\mbox{.}(2022b)]%
        {mosqueira2022classification}
\bibfield{author}{\bibinfo{person}{Eduardo Mosqueira-Rey},
  \bibinfo{person}{Elena~Hern{\'a}ndez Pereira}, \bibinfo{person}{David
  Alonso-R{\'\i}os}, {and} \bibinfo{person}{Jos{\'e} Bobes-Bascar{\'a}n}.}
  \bibinfo{year}{2022}\natexlab{b}.
\newblock \showarticletitle{A classification and review of tools for developing
  and interacting with machine learning systems}. In
  \bibinfo{booktitle}{\emph{Proceedings of the 37th ACM/SIGAPP Symposium on
  Applied Computing}}. \bibinfo{pages}{1092--1101}.
\newblock


\bibitem[{Murray-Rust} et~al\mbox{.}(2022)]%
        {murray-rust2022MetaphorsDesigners}
\bibfield{author}{\bibinfo{person}{Dave {Murray-Rust}},
  \bibinfo{person}{Iohanna Nicenboim}, {and} \bibinfo{person}{Dan Lockton}.}
  \bibinfo{year}{2022}\natexlab{}.
\newblock \showarticletitle{Metaphors for Designers Working with {{AI}}}. In
  \bibinfo{booktitle}{\emph{{{DRS Biennial Conference Series}}}}.
\newblock
\urldef\tempurl%
\url{https://doi.org/10.21606/drs.2022.667}
\showDOI{\tempurl}


\bibitem[Murray-Rust et~al\mbox{.}(2015)]%
        {murray2015softening}
\bibfield{author}{\bibinfo{person}{Dave Murray-Rust}, \bibinfo{person}{Petros
  Papapanagiotou}, {and} \bibinfo{person}{Dave Robertson}.}
  \bibinfo{year}{2015}\natexlab{}.
\newblock \showarticletitle{Softening electronic institutions to support
  natural interaction}.
\newblock \bibinfo{journal}{\emph{Human Computation}} \bibinfo{volume}{2},
  \bibinfo{number}{2} (\bibinfo{year}{2015}).
\newblock


\bibitem[Nadj et~al\mbox{.}(2020)]%
        {nadj2020power}
\bibfield{author}{\bibinfo{person}{Mario Nadj}, \bibinfo{person}{Merlin
  Knaeble}, \bibinfo{person}{Maximilian~Xiling Li}, {and}
  \bibinfo{person}{Alexander Maedche}.} \bibinfo{year}{2020}\natexlab{}.
\newblock \showarticletitle{Power to the oracle? design principles for
  interactive labeling systems in machine learning}.
\newblock \bibinfo{journal}{\emph{KI-K{\"u}nstliche Intelligenz}}
  \bibinfo{volume}{34}, \bibinfo{number}{2} (\bibinfo{year}{2020}),
  \bibinfo{pages}{131--142}.
\newblock


\bibitem[Nunes and Jannach(2017)]%
        {nunes2017systematic}
\bibfield{author}{\bibinfo{person}{Ingrid Nunes} {and} \bibinfo{person}{Dietmar
  Jannach}.} \bibinfo{year}{2017}\natexlab{}.
\newblock \showarticletitle{A systematic review and taxonomy of explanations in
  decision support and recommender systems}.
\newblock \bibinfo{journal}{\emph{User Modeling and User-Adapted Interaction}}
  \bibinfo{volume}{27}, \bibinfo{number}{3} (\bibinfo{year}{2017}),
  \bibinfo{pages}{393--444}.
\newblock


\bibitem[O'Brien and Nicol(1998)]%
        {o1998fipa}
\bibfield{author}{\bibinfo{person}{Paul~D O'Brien} {and}
  \bibinfo{person}{Richard~C Nicol}.} \bibinfo{year}{1998}\natexlab{}.
\newblock \showarticletitle{FIPA—towards a standard for software agents}.
\newblock \bibinfo{journal}{\emph{BT Technology Journal}} \bibinfo{volume}{16},
  \bibinfo{number}{3} (\bibinfo{year}{1998}), \bibinfo{pages}{51--59}.
\newblock


\bibitem[O'Hara(2020)]%
        {ohara2020ExplainableAI}
\bibfield{author}{\bibinfo{person}{Kieron O'Hara}.}
  \bibinfo{year}{2020}\natexlab{}.
\newblock \showarticletitle{Explainable {{AI}} and the Philosophy and Practice
  of Explanation}.
\newblock \bibinfo{journal}{\emph{Computer Law \& Security Review}}
  \bibinfo{volume}{39} (\bibinfo{date}{Nov.} \bibinfo{year}{2020}),
  \bibinfo{pages}{105474}.
\newblock
\showISSN{02673649}
\urldef\tempurl%
\url{https://doi.org/10.1016/j.clsr.2020.105474}
\showDOI{\tempurl}


\bibitem[Peters et~al\mbox{.}(2020)]%
        {peters2020responsible}
\bibfield{author}{\bibinfo{person}{Dorian Peters}, \bibinfo{person}{Karina
  Vold}, \bibinfo{person}{Diana Robinson}, {and} \bibinfo{person}{Rafael~A
  Calvo}.} \bibinfo{year}{2020}\natexlab{}.
\newblock \showarticletitle{Responsible AI—two frameworks for ethical design
  practice}.
\newblock \bibinfo{journal}{\emph{IEEE Transactions on Technology and Society}}
  \bibinfo{volume}{1}, \bibinfo{number}{1} (\bibinfo{year}{2020}),
  \bibinfo{pages}{34--47}.
\newblock


\bibitem[Ramesh et~al\mbox{.}(2021)]%
        {ramesh2021zero}
\bibfield{author}{\bibinfo{person}{Aditya Ramesh}, \bibinfo{person}{Mikhail
  Pavlov}, \bibinfo{person}{Gabriel Goh}, \bibinfo{person}{Scott Gray},
  \bibinfo{person}{Chelsea Voss}, \bibinfo{person}{Alec Radford},
  \bibinfo{person}{Mark Chen}, {and} \bibinfo{person}{Ilya Sutskever}.}
  \bibinfo{year}{2021}\natexlab{}.
\newblock \showarticletitle{Zero-shot text-to-image generation}. In
  \bibinfo{booktitle}{\emph{International Conference on Machine Learning}}.
  PMLR, \bibinfo{pages}{8821--8831}.
\newblock


\bibitem[Rezwana and Maher(2022)]%
        {rezwana2022designing}
\bibfield{author}{\bibinfo{person}{Jeba Rezwana} {and}
  \bibinfo{person}{Mary~Lou Maher}.} \bibinfo{year}{2022}\natexlab{}.
\newblock \showarticletitle{Designing Creative AI Partners with COFI: A
  Framework for Modeling Interaction in Human-AI Co-Creative Systems}.
\newblock \bibinfo{journal}{\emph{ACM Transactions on Computer-Human
  Interaction}} (\bibinfo{year}{2022}).
\newblock


\bibitem[Robertson(2005)]%
        {robertson2005lightweight}
\bibfield{author}{\bibinfo{person}{David Robertson}.}
  \bibinfo{year}{2005}\natexlab{}.
\newblock \showarticletitle{A lightweight coordination calculus for agent
  systems}. In \bibinfo{booktitle}{\emph{Declarative Agent Languages and
  Technologies II: Second International Workshop, DALT 2004, New York, NY, USA,
  July 19, 2004, Revised Selected Papers 2}}. Springer,
  \bibinfo{pages}{183--197}.
\newblock


\bibitem[Schoonderwoerd et~al\mbox{.}(2022)]%
        {schoonderwoerd2022design}
\bibfield{author}{\bibinfo{person}{Tjeerd~AJ Schoonderwoerd},
  \bibinfo{person}{Emma~M van Zoelen}, \bibinfo{person}{Karel van~den Bosch},
  {and} \bibinfo{person}{Mark~A Neerincx}.} \bibinfo{year}{2022}\natexlab{}.
\newblock \showarticletitle{Design patterns for human-AI co-learning: A
  wizard-of-Oz evaluation in an urban-search-and-rescue task}.
\newblock \bibinfo{journal}{\emph{International Journal of Human-Computer
  Studies}}  \bibinfo{volume}{164} (\bibinfo{year}{2022}),
  \bibinfo{pages}{102831}.
\newblock


\bibitem[Schwalbe and Finzel(2021)]%
        {schwalbe2021comprehensive}
\bibfield{author}{\bibinfo{person}{Gesina Schwalbe} {and}
  \bibinfo{person}{Bettina Finzel}.} \bibinfo{year}{2021}\natexlab{}.
\newblock \showarticletitle{A Comprehensive Taxonomy for Explainable Artificial
  Intelligence: A Systematic Survey of Surveys on Methods and Concepts}.
\newblock \bibinfo{journal}{\emph{arXiv e-prints}} (\bibinfo{year}{2021}),
  \bibinfo{pages}{arXiv--2105}.
\newblock


\bibitem[Shneiderman(2020)]%
        {shneiderman2020human}
\bibfield{author}{\bibinfo{person}{Ben Shneiderman}.}
  \bibinfo{year}{2020}\natexlab{}.
\newblock \showarticletitle{Human-centered artificial intelligence: Reliable,
  safe \& trustworthy}.
\newblock \bibinfo{journal}{\emph{International Journal of Human--Computer
  Interaction}} \bibinfo{volume}{36}, \bibinfo{number}{6}
  (\bibinfo{year}{2020}), \bibinfo{pages}{495--504}.
\newblock


\bibitem[Sperrle et~al\mbox{.}(2020)]%
        {sperrle2020should}
\bibfield{author}{\bibinfo{person}{Fabian Sperrle},
  \bibinfo{person}{Mennatallah El-Assady}, \bibinfo{person}{Grace Guo},
  \bibinfo{person}{Duen~Horng Chau}, \bibinfo{person}{Alex Endert}, {and}
  \bibinfo{person}{Daniel Keim}.} \bibinfo{year}{2020}\natexlab{}.
\newblock \showarticletitle{Should we trust (x) AI? Design dimensions for
  structured experimental evaluations}.
\newblock \bibinfo{journal}{\emph{arXiv preprint arXiv:2009.06433}}
  (\bibinfo{year}{2020}).
\newblock


\bibitem[Subramonyam et~al\mbox{.}(2021)]%
        {subramonyam2021towards}
\bibfield{author}{\bibinfo{person}{Hariharan Subramonyam},
  \bibinfo{person}{Colleen Seifert}, {and} \bibinfo{person}{Eytan Adar}.}
  \bibinfo{year}{2021}\natexlab{}.
\newblock \showarticletitle{Towards a process model for co-creating AI
  experiences}. In \bibinfo{booktitle}{\emph{Designing Interactive Systems
  Conference 2021}}. \bibinfo{pages}{1529--1543}.
\newblock


\bibitem[Teso et~al\mbox{.}(2023)]%
        {teso2023}
\bibfield{author}{\bibinfo{person}{Stefano Teso}, \bibinfo{person}{Öznur
  Alkan}, \bibinfo{person}{Wolfgang Stammer}, {and} \bibinfo{person}{Elizabeth
  Daly}.} \bibinfo{year}{2023}\natexlab{}.
\newblock \showarticletitle{Leveraging explanations in interactive machine
  learning: An overview}.
\newblock \bibinfo{journal}{\emph{Frontiers in Artificial Intelligence}}
  \bibinfo{volume}{6} (\bibinfo{year}{2023}).
\newblock
\showISSN{2624-8212}
\urldef\tempurl%
\url{https://doi.org/10.3389/frai.2023.1066049}
\showDOI{\tempurl}


\bibitem[Treanor et~al\mbox{.}(2015)]%
        {treanor2015ai}
\bibfield{author}{\bibinfo{person}{Mike Treanor}, \bibinfo{person}{Alexander
  Zook}, \bibinfo{person}{Mirjam~P Eladhari}, \bibinfo{person}{Julian
  Togelius}, \bibinfo{person}{Gillian Smith}, \bibinfo{person}{Michael Cook},
  \bibinfo{person}{Tommy Thompson}, \bibinfo{person}{Brian Magerko},
  \bibinfo{person}{John Levine}, {and} \bibinfo{person}{Adam Smith}.}
  \bibinfo{year}{2015}\natexlab{}.
\newblock \showarticletitle{AI-based game design patterns}.
\newblock  (\bibinfo{year}{2015}).
\newblock


\bibitem[Tsiakas et~al\mbox{.}(2018)]%
        {tsiakas2018task}
\bibfield{author}{\bibinfo{person}{Konstantinos Tsiakas},
  \bibinfo{person}{Maher Abujelala}, {and} \bibinfo{person}{Fillia Makedon}.}
  \bibinfo{year}{2018}\natexlab{}.
\newblock \showarticletitle{Task engagement as personalization feedback for
  socially-assistive robots and cognitive training}.
\newblock \bibinfo{journal}{\emph{Technologies}} \bibinfo{volume}{6},
  \bibinfo{number}{2} (\bibinfo{year}{2018}), \bibinfo{pages}{49}.
\newblock


\bibitem[van Bekkum et~al\mbox{.}(2021)]%
        {van2021modular}
\bibfield{author}{\bibinfo{person}{Michael van Bekkum}, \bibinfo{person}{Maaike
  de Boer}, \bibinfo{person}{Frank van Harmelen}, \bibinfo{person}{Andr{\'e}
  Meyer-Vitali}, {and} \bibinfo{person}{Annette~ten Teije}.}
  \bibinfo{year}{2021}\natexlab{}.
\newblock \showarticletitle{Modular design patterns for hybrid learning and
  reasoning systems}.
\newblock \bibinfo{journal}{\emph{Applied Intelligence}} \bibinfo{volume}{51},
  \bibinfo{number}{9} (\bibinfo{year}{2021}), \bibinfo{pages}{6528--6546}.
\newblock


\bibitem[van Berkel et~al\mbox{.}(2021)]%
        {van2021human}
\bibfield{author}{\bibinfo{person}{Niels van Berkel}, \bibinfo{person}{Mikael~B
  Skov}, {and} \bibinfo{person}{Jesper Kjeldskov}.}
  \bibinfo{year}{2021}\natexlab{}.
\newblock \showarticletitle{Human-AI interaction: intermittent, continuous, and
  proactive}.
\newblock \bibinfo{journal}{\emph{Interactions}} \bibinfo{volume}{28},
  \bibinfo{number}{6} (\bibinfo{year}{2021}), \bibinfo{pages}{67--71}.
\newblock


\bibitem[Wang et~al\mbox{.}(2019)]%
        {wang2019designing}
\bibfield{author}{\bibinfo{person}{Danding Wang}, \bibinfo{person}{Qian Yang},
  \bibinfo{person}{Ashraf Abdul}, {and} \bibinfo{person}{Brian~Y Lim}.}
  \bibinfo{year}{2019}\natexlab{}.
\newblock \showarticletitle{Designing theory-driven user-centric explainable
  AI}. In \bibinfo{booktitle}{\emph{Proceedings of the 2019 CHI conference on
  human factors in computing systems}}. \bibinfo{pages}{1--15}.
\newblock


\bibitem[Weitz et~al\mbox{.}(2021)]%
        {weitz2021let}
\bibfield{author}{\bibinfo{person}{Katharina Weitz}, \bibinfo{person}{Dominik
  Schiller}, \bibinfo{person}{Ruben Schlagowski}, \bibinfo{person}{Tobias
  Huber}, {and} \bibinfo{person}{Elisabeth Andr{\'e}}.}
  \bibinfo{year}{2021}\natexlab{}.
\newblock \showarticletitle{“Let me explain!”: exploring the potential of
  virtual agents in explainable AI interaction design}.
\newblock \bibinfo{journal}{\emph{Journal on Multimodal User Interfaces}}
  \bibinfo{volume}{15}, \bibinfo{number}{2} (\bibinfo{year}{2021}),
  \bibinfo{pages}{87--98}.
\newblock


\bibitem[Wenskovitch and North(2020)]%
        {wenskovitch2020interactive}
\bibfield{author}{\bibinfo{person}{John Wenskovitch} {and}
  \bibinfo{person}{Chris North}.} \bibinfo{year}{2020}\natexlab{}.
\newblock \showarticletitle{Interactive Artificial Intelligence: Designing for
  the" Two Black Boxes" Problem}.
\newblock \bibinfo{journal}{\emph{Computer}} \bibinfo{volume}{53},
  \bibinfo{number}{8} (\bibinfo{year}{2020}), \bibinfo{pages}{29--39}.
\newblock


\bibitem[Wiethof and Bittner(2021)]%
        {wiethof2021hybrid}
\bibfield{author}{\bibinfo{person}{Christina Wiethof} {and} \bibinfo{person}{E
  Bittner}.} \bibinfo{year}{2021}\natexlab{}.
\newblock \showarticletitle{Hybrid intelligence-combining the human in the loop
  with the computer in the loop: a systematic literature review}. In
  \bibinfo{booktitle}{\emph{Forty-Second International Conference on
  Information Systems, Austin}}.
\newblock


\bibitem[Wu et~al\mbox{.}(2022)]%
        {wu2022survey}
\bibfield{author}{\bibinfo{person}{Xingjiao Wu}, \bibinfo{person}{Luwei Xiao},
  \bibinfo{person}{Yixuan Sun}, \bibinfo{person}{Junhang Zhang},
  \bibinfo{person}{Tianlong Ma}, {and} \bibinfo{person}{Liang He}.}
  \bibinfo{year}{2022}\natexlab{}.
\newblock \showarticletitle{A survey of human-in-the-loop for machine
  learning}.
\newblock \bibinfo{journal}{\emph{Future Generation Computer Systems}}
  (\bibinfo{year}{2022}).
\newblock


\bibitem[Xu(2019)]%
        {xu2019toward}
\bibfield{author}{\bibinfo{person}{Wei Xu}.} \bibinfo{year}{2019}\natexlab{}.
\newblock \showarticletitle{Toward human-centered AI: a perspective from
  human-computer interaction}.
\newblock \bibinfo{journal}{\emph{interactions}} \bibinfo{volume}{26},
  \bibinfo{number}{4} (\bibinfo{year}{2019}), \bibinfo{pages}{42--46}.
\newblock


\bibitem[Xu et~al\mbox{.}(2022)]%
        {xu2022transitioning}
\bibfield{author}{\bibinfo{person}{Wei Xu}, \bibinfo{person}{Marvin~J Dainoff},
  \bibinfo{person}{Liezhong Ge}, {and} \bibinfo{person}{Zaifeng Gao}.}
  \bibinfo{year}{2022}\natexlab{}.
\newblock \showarticletitle{Transitioning to human interaction with AI systems:
  New challenges and opportunities for HCI professionals to enable
  human-centered AI}.
\newblock \bibinfo{journal}{\emph{International Journal of Human--Computer
  Interaction}} (\bibinfo{year}{2022}), \bibinfo{pages}{1--25}.
\newblock


\bibitem[Yang and Zhang(2019)]%
        {yang2019artificial}
\bibfield{author}{\bibinfo{person}{Jinyu Yang} {and} \bibinfo{person}{Bo
  Zhang}.} \bibinfo{year}{2019}\natexlab{}.
\newblock \showarticletitle{Artificial intelligence in intelligent tutoring
  robots: A systematic review and design guidelines}.
\newblock \bibinfo{journal}{\emph{Applied Sciences}} \bibinfo{volume}{9},
  \bibinfo{number}{10} (\bibinfo{year}{2019}), \bibinfo{pages}{2078}.
\newblock


\bibitem[Yang et~al\mbox{.}(2020a)]%
        {yang2020fairness}
\bibfield{author}{\bibinfo{person}{Ke Yang}, \bibinfo{person}{Biao Huang},
  \bibinfo{person}{Julia Stoyanovich}, {and} \bibinfo{person}{Sebastian
  Schelter}.} \bibinfo{year}{2020}\natexlab{a}.
\newblock \showarticletitle{Fairness-Aware Instrumentation of Preprocessing\~{}
  Pipelines for Machine Learning}. In \bibinfo{booktitle}{\emph{Workshop on
  Human-In-the-Loop Data Analytics (HILDA'20)}}.
\newblock


\bibitem[Yang et~al\mbox{.}(2018)]%
        {yang2018mapping}
\bibfield{author}{\bibinfo{person}{Qian Yang}, \bibinfo{person}{Nikola
  Banovic}, {and} \bibinfo{person}{John Zimmerman}.}
  \bibinfo{year}{2018}\natexlab{}.
\newblock \showarticletitle{Mapping machine learning advances from hci research
  to reveal starting places for design innovation}. In
  \bibinfo{booktitle}{\emph{Proceedings of the 2018 CHI conference on human
  factors in computing systems}}. \bibinfo{pages}{1--11}.
\newblock


\bibitem[Yang et~al\mbox{.}(2020b)]%
        {yang2020re}
\bibfield{author}{\bibinfo{person}{Qian Yang}, \bibinfo{person}{Aaron
  Steinfeld}, \bibinfo{person}{Carolyn Ros{\'e}}, {and} \bibinfo{person}{John
  Zimmerman}.} \bibinfo{year}{2020}\natexlab{b}.
\newblock \showarticletitle{Re-examining whether, why, and how human-AI
  interaction is uniquely difficult to design}. In
  \bibinfo{booktitle}{\emph{Proceedings of the 2020 chi conference on human
  factors in computing systems}}. \bibinfo{pages}{1--13}.
\newblock


\bibitem[Yu and Tapus(2019)]%
        {yu2019interactive}
\bibfield{author}{\bibinfo{person}{Chuang Yu} {and} \bibinfo{person}{Adriana
  Tapus}.} \bibinfo{year}{2019}\natexlab{}.
\newblock \showarticletitle{Interactive robot learning for multimodal emotion
  recognition}. In \bibinfo{booktitle}{\emph{International Conference on Social
  Robotics}}. Springer, \bibinfo{pages}{633--642}.
\newblock


\bibitem[Yurrita et~al\mbox{.}(2022)]%
        {yurrita2022MultistakeholderValuebased}
\bibfield{author}{\bibinfo{person}{Mireia Yurrita}, \bibinfo{person}{Dave
  {Murray-Rust}}, \bibinfo{person}{Agathe Balayn}, {and}
  \bibinfo{person}{Alessandro Bozzon}.} \bibinfo{year}{2022}\natexlab{}.
\newblock \showarticletitle{Towards a Multi-Stakeholder Value-Based Assessment
  Framework for Algorithmic Systems}. In \bibinfo{booktitle}{\emph{2022 {{ACM
  Conference}} on {{Fairness}}, {{Accountability}}, and {{Transparency}}}}
  \emph{(\bibinfo{series}{{{FAccT}} '22})}. \bibinfo{publisher}{{Association
  for Computing Machinery}}, \bibinfo{address}{{New York, NY, USA}},
  \bibinfo{pages}{535--563}.
\newblock
\showISBNx{978-1-4503-9352-2}
\urldef\tempurl%
\url{https://doi.org/10.1145/3531146.3533118}
\showDOI{\tempurl}


\bibitem[Zhang et~al\mbox{.}(2021)]%
        {zhang2021forward}
\bibfield{author}{\bibinfo{person}{Zelun~Tony Zhang}, \bibinfo{person}{Yuanting
  Liu}, {and} \bibinfo{person}{Heinrich Hussmann}.}
  \bibinfo{year}{2021}\natexlab{}.
\newblock \showarticletitle{Forward reasoning decision support: toward a more
  complete view of the human-AI interaction design space}. In
  \bibinfo{booktitle}{\emph{CHItaly 2021: 14th Biannual Conference of the
  Italian SIGCHI Chapter}}. \bibinfo{pages}{1--5}.
\newblock


\bibitem[Zhou et~al\mbox{.}(2020)]%
        {zhou2020designing}
\bibfield{author}{\bibinfo{person}{Xiaofei Zhou}, \bibinfo{person}{Jessica
  Van~Brummelen}, {and} \bibinfo{person}{Phoebe Lin}.}
  \bibinfo{year}{2020}\natexlab{}.
\newblock \showarticletitle{Designing AI learning experiences for K-12:
  emerging works, future opportunities and a design framework}.
\newblock \bibinfo{journal}{\emph{arXiv preprint arXiv:2009.10228}}
  (\bibinfo{year}{2020}).
\newblock


\end{thebibliography}
\newpage

\appendix 
\section{APPENDIX}
\subsection{Unpacking of human-AI interactions: use cases} \label{more-unpacking}
This section includes a list of uses cases to demonstrate our proposed unpacking approach. Based on the selected use cases, we extract actions and patterns for different interaction concepts, including XAI-based and HITL interactions. More specifically, the list includes an interactive sound annotation system \cite{kim2018human}, an explainable active learning tool for image classification \cite{heimerl2019nova}, a human-robot interaction for collaborative sketching \cite{lin2020your}, an  music recommendation system using personalized explanations \cite{martijn2022knowing}, an interactive meeting scheduling assistant \cite{kocielnik2019will} and an interactive ML approach for game-based collaboration\cite{guo2022building}. For each use case, we describe the interactions (unpacking) and we define the actions and patterns based on the defined primitives. 

\begin{figure}[h]
\centering
\includegraphics[width=\columnwidth]{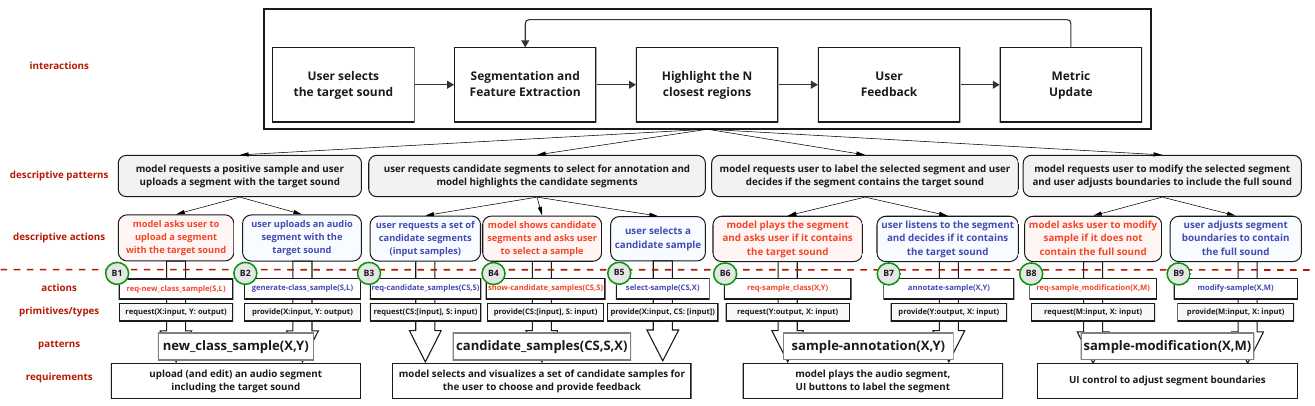}
\caption{Unpacking an interactive sound annotation interaction \cite{kim2018human} into interaction primitives and patterns.}
\label{fig:unpacking2}
\end{figure}

\subsubsection{Human-in-the-loop sound event detection and annotation (Figure \ref{fig:unpacking2}).} The proposed system integrates a user interface for interactive sound event detection and annotation. The user sets a target sound event by selecting or uploading a sound segment which includes the target sound (e.g., door knocking). With this interaction, the user sets the possible model output classes: positive when the segment includes the sound target and negative otherwise. The model selects and highlights segments similar to the positive sample (user's input) and asks the user to classify the segments. The user can either provide a label for a segment (target or not) or adjust the segment boundaries, if the segment does not include the full sound. The model utilizes user's feedback (re-labeling/re-segmentation) to update its model parameters in an iterative and interactive manner.
    
We identify three types of interactions: \textit{(a) positive sample:} the model asks the user for a positive example (segment that includes the target sound) - user provides a positive sample, \textit{(b) select candidate sample:} user requests a set of candidate segments -- model highlights the candidate samples -- user selects a candidate segment, \textit{(c) label sample:} model plays segment and asks user if it contains the target sound -- user responds by labeling the segment, and \textit{(d) modify sample:} model plays segment and asks user if it contains the full sound -- user adjusts the boundaries of the segment until full sound is included. During these interactions, user and model exchange information through positive samples, visualization and selection of candidate segments and their current labels (highlighted), playing/listening to segments, and re-labeling or re-segmenting selected samples. Considering the design and implementation approach, the proposed approach focuses on two aspects: (a) the design of a human-friendly interface which can provide interactivity and feedback control, and (b) the iterative labeling and model training approach, based on which the model dynamically recalculates the model parameters (weights) based on user feedback. In terms of evaluation metrics, both user-based (interaction overhead) and algorithm-based (model accuracy) measurements were considered. Their analysis indicates that minimizing the interaction overhead (through design) can maximize machine performance and speed. Table \ref{tab:actions2} shows the messages and action definitions. 

\begin{table}[h]
\resizebox{\columnwidth}{!}{
\begin{tabular}{|ll|l|}
\hline
\multicolumn{2}{|c|}{\textbf{Message}} & \multicolumn{1}{c|}{\textbf{Action definition}} \\ \hline

\multicolumn{1}{|l|}{\texttt{B1}} & \begin{tabular}[c]{@{}l@{}}\texttt{<model$\rightarrow$user,\textbf{req-new\_class\_sample(S,L)},}\\\texttt{{[}X:uploadBtn,targetSound;Y:positiveLabel{]}>}\end{tabular} & \begin{tabular}[c]{@{}l@{}}\texttt{req-new\_class\_sample(S,L)$\equiv$}\\\texttt{request(X:input.raw\_data,Y:output.label)}\\$\leftarrow$ \texttt{create(S),map(S,L)}\end{tabular} \\ \hline

\multicolumn{1}{|l|}{\texttt{B2}} & \begin{tabular}[c]{@{}l@{}}\texttt{<user$\rightarrow$model,\textbf{generate-class\_sample(S,L)},}\\\texttt{{[}X:uploadTargetSound;Y:positiveLabel{]}>}\end{tabular} & \begin{tabular}[c]{@{}l@{}}\texttt{generate-class\_sample(S,L) $\equiv$} \\ \texttt{provide(S:input.raw\_data,L:output.label)}\\$\leftarrow$\texttt{create(S),map(S,L)}\end{tabular} \\ \hline

\multicolumn{1}{|l|}{\texttt{B3}} & \begin{tabular}[c]{@{}l@{}}\texttt{<user$\rightarrow$model,\textbf{req-candidate\_samples(CS,S)},}\\\texttt{{[}CS:similarSegs,highlightSeg;S:posSample{]}>}\end{tabular} & \begin{tabular}[c]{@{}l@{}}\texttt{req-candidate\_samples(CS,S)$\equiv$}\\
\texttt{request(CS:{[}input.raw\_data{]},S:input.raw\_data)}\\$\leftarrow$\texttt{select(CS),map(CS,S) }\end{tabular} \\ \hline

\multicolumn{1}{|l|}{\texttt{B4}} & \begin{tabular}[c]{@{}l@{}}\texttt{<model$\rightarrow$user,\textbf{show-candidate\_samples(CS,S)},}\\\texttt{{[}CS:similarSegs,highlightSeg;S:posSample{]}>}\end{tabular} & \begin{tabular}[c]{@{}l@{}}\texttt{show-candidate\_samples(CS,S)$\equiv$}\\
\texttt{provide(CS:{[}input.raw\_data{]},S:input.raw\_data)}\\$\leftarrow$\texttt{select(CS),map(CS,S) }\end{tabular} \\ \hline

\multicolumn{1}{|l|}{\texttt{B5}} & \begin{tabular}[c]{@{}l@{}}\texttt{<user$\rightarrow$model,\textbf{select-sample(X,CS)},}\\\texttt{{[}X:selSegment;CS:highlightSeg{]}>}\end{tabular} & \begin{tabular}[c]{@{}l@{}}\texttt{select-sample(X,CS) $\equiv$} \\
\texttt{provide(X:input.raw\_data,CS:{[}input.raw\_data{]})}\\$\leftarrow$\texttt{select(X,CS)}\end{tabular} \\ \hline

\multicolumn{1}{|l|}{\texttt{B6}} & \begin{tabular}[c]{@{}l@{}}\texttt{<model$\rightarrow$user,\textbf{req-sample\_class(X,Y)},}\\\texttt{{[}X:playSegment;Y:isTargetSound,labelBtn{]}>}\end{tabular} & \begin{tabular}[c]{@{}l@{}}\texttt{req-sample\_class(X,Y) $\equiv$}\\\texttt{request(Y:output.label,X:input.raw\_data)}\\$\leftarrow$ \texttt{map(X,Y)}\end{tabular} \\ \hline

\multicolumn{1}{|l|}{\texttt{B7}} & \begin{tabular}[c]{@{}l@{}}\texttt{<user$\rightarrow$model,\textbf{annotate-sample(X,Y)},}\\\texttt{{[}X:selSegment; Y:isTargetSound,labelBtn{]}>}\end{tabular} & \begin{tabular}[c]{@{}l@{}}\texttt{annotate-sample(X,Y)$\equiv$}\\\texttt{provide(Y:output.label,X:input.raw\_data)}\\$\leftarrow$\texttt{map(X,Y)}\end{tabular} \\ \hline

\multicolumn{1}{|l|}{\texttt{B8}} & \begin{tabular}[c]{@{}l@{}}\texttt{<model$\rightarrow$user,\textbf{req-modified-sample(X,M)},}\\\texttt{{[}X:selSegment;M:modifySegment{]}>}\end{tabular} & \begin{tabular}[c]{@{}l@{}}
\texttt{req-modified-sample(X,M)$\equiv$}\\\texttt{request(M:input.raw\_data,X:input.raw\_data)}\\$\leftarrow$\texttt{modify(X,M)}\end{tabular} \\ \hline

\multicolumn{1}{|l|}{\texttt{B9}} & \begin{tabular}[c]{@{}l@{}}\texttt{<model$\rightarrow$user,\textbf{modify-sample(X,M)},}\\\texttt{{[}X:selSegment;M:modifySegment{]}>}\end{tabular} & \begin{tabular}[c]{@{}l@{}}
\texttt{modify-sample(X,M)$\equiv$}\\\texttt{provide(M:input.raw\_data,X:input.raw\_data)}\\$\leftarrow$\texttt{modify(X,M)}\end{tabular} \\ \hline

\end{tabular}
}
\caption{Messages and action definitions for the interactive sound annotation system interactions}
\label{tab:actions2}
\end{table}

\begin{table}[h]
\resizebox{0.95\columnwidth}{!}{
\begin{tabular}{|c|c|c|}
\hline
\textbf{pattern} & \textbf{actions/messages} & \textbf{description} \\ \hline
\multirow{2}{*}{new\_class\_sample} & \texttt{req-new-class\_sample(S,L)} & model asks user for a (positive) class sample \\ \cline{2-3} 
 & \texttt{generate-class\_sample(S,L)} & user provides a (positive) class sample \\ \hline
\multirow{3}{*}{candidate\_samples} & \texttt{req-candidate\_samples(CS,S)} & user asks for candidate (similar) samples \\ \cline{2-3} 
 & \texttt{show-candidate\_samples(CS,S)} & model provides a set of candidate samples \\ \cline{2-3} 
 & \texttt{select-sample(X,CS)} & user selects a candidate sample \\ \hline
\multirow{2}{*}{sample-annotation} & \texttt{req-sample\_class(X,Y)} & model asks user to provide a label for the input \\ \cline{2-3} 
 & \texttt{annotate-sample(X,Y)} & user provides the correct label for the input \\ \hline
\multirow{2}{*}{sample-modification} & \texttt{req-modified\_sample(X,M)} & model asks user to modify the sample (if needed) \\ \cline{2-3} 
 & \texttt{modify-sample(X,M)} & user modifies the selected sample \\ \hline
\end{tabular}
}
\caption{Interaction patterns for the interactive sound annotation system}
\label{tab:interaction2}
\end{table}

Based on these definitions, we define the following patterns (Table \ref{tab:interaction2}): \texttt{[B1-B2] new\_class\_sample}, \texttt{[B3-B5] candidate\_samples}, \texttt{[B5-B7] sample-annotation}, and \texttt{[B8-B9] sample-modification}. User and model interact through input (segments) and output (labels) in an interactive manner in order to annotate all candidate samples. More specifically, the first pattern is required to set the target class which represents samples which include a target sound. This is achieved by asking the user to provide a segment which includes the required sound (positive sample). During the second pattern, the model uses the positive sample to identify and visualize a list of candidate inputs for the user to label. The third pattern describes how user provides feedback to the model by listening and annotating the selected segment. The fourth pattern describes the modification of a sample (input) by adjusting the boundaries of the selected segment.
    
\begin{figure}[h]
\centering
\includegraphics[width=\columnwidth]{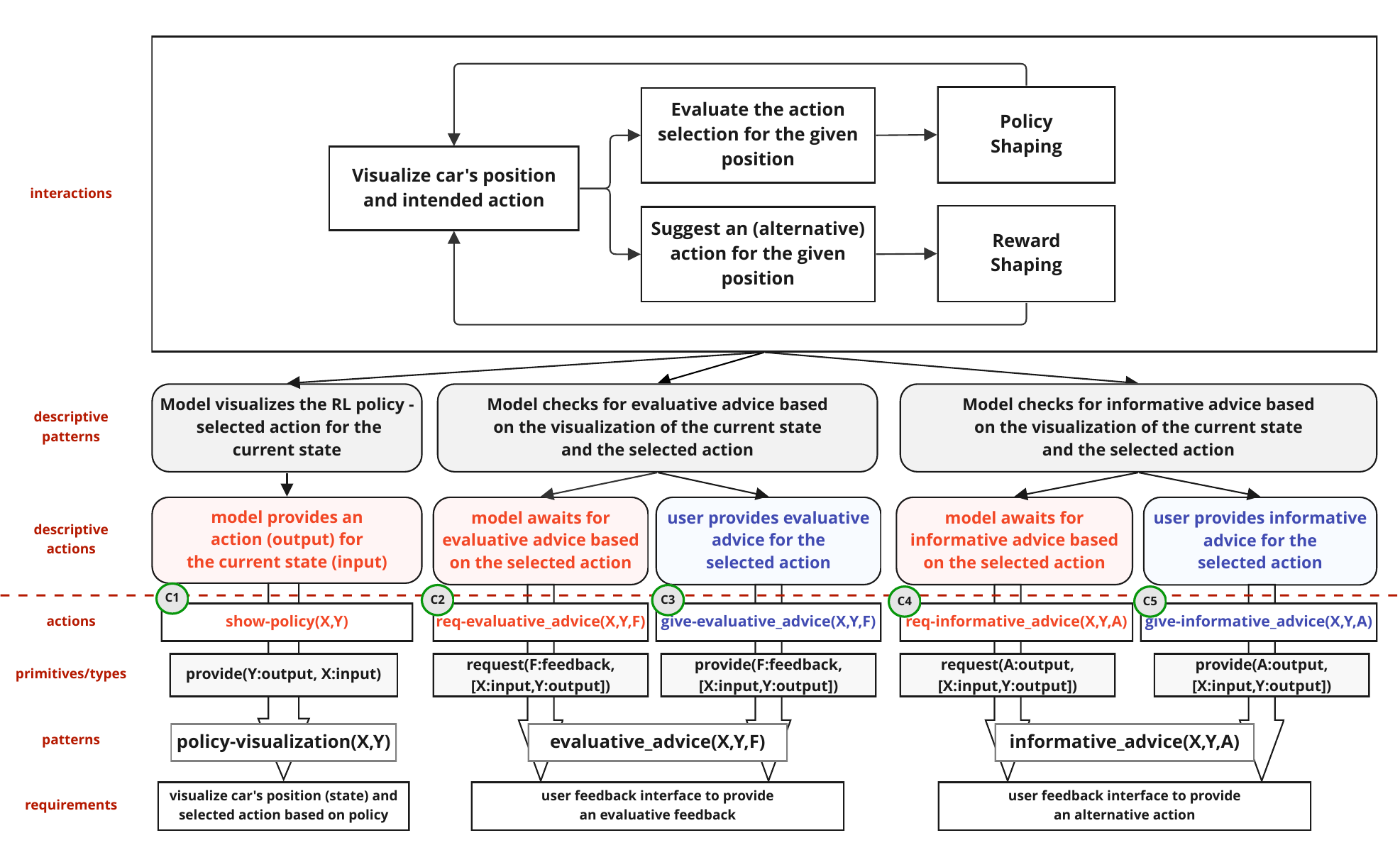}
\caption{Unpacking an interactive RL interaction into interaction patterns. Image adapted from \cite{bignold2022human} }
\label{fig:unpacking3}
\end{figure}

\subsubsection{Interactive RL and human trainer engagement (Figure \ref{fig:unpacking3}).} The proposed system integrates human-provided feedback to an RL agent to improve its performance while executing the Mountain Car task. More specifically, the RL agent visualizes the current state and the selected action based on the model's policy (input-output pair) and allows the user to provide two types of feedback (evaluative/informative) in order to complete the task. Evaluative advice assesses the past performance of an agent and it is provided in the form of a reward, while informative advice supplements future decision-making and it is provided as an intervention - modified action. The goal of the interaction is to utilize human advice and maximize model's performance.  
    
We identify the following interactions: \textbf{\textit{(a) policy visualization:}} model visualizes its policy as a selected action for the current state (input-output pair), \textbf{\textit{(b) informative advice:}} model queries the user for informative advice -- user provides an alternate action (output) for the current state-action pair, and \textbf{\textit{(c)} evaluative advice:} model queries the user for evaluative advice -- user provides a reward (feedback) for the current state-action pair. Interactions take place in the form of a transparent interactive RL policy (visualization) and human-feedback for advice or evaluation (through keyboard input). The goal of the user study is to measure human engagement and model performance for the two types of feedback. In terms of the interaction design, human users are autonomous in terms of when they can provide feedback. According to the type of feedback, the model integrates it to the learning mechanism in different ways. If the user provides informative advice (learning from guidance), feedback is integrated to the learning mechanism through policy shaping, while for evaluative advice (learning from feedback), the RL agent uses reward shaping. The system uses an interface to visualize the RL policy and the task execution. Both types of advice are provided through a keyboard input, specific for each type. The design of the interactions plays an essential role in both model performance and human engagement. An important aspect to consider is the human feedback quality and consistency. The design challenge is to maintain user's engagement and performance considering user's perception. According to the analysis, users who provided informative advice were more engaged and accurate, which may be linked to how users perceive the different advice methods. Table \ref{tab:actions3} shows the messages and action definitions. 

\begin{table}[h]
\resizebox{\columnwidth}{!}{
\begin{tabular}{|ll|l|}
\hline
\multicolumn{2}{|c|}{\textbf{Message}} & \multicolumn{1}{c|}{\textbf{Action definition}} \\ \hline

\multicolumn{1}{|l|}{\texttt{C1}} & \begin{tabular}[c]{@{}l@{}}\texttt{<model$\rightarrow$user,\textbf{show-policy(X,Y)},}\\\texttt{{[}X:CarPosition;Y:selectedAction{]}>}\end{tabular} & \begin{tabular}[c]{@{}l@{}}
\texttt{show-policy(X,Y)$\equiv$}\\\texttt{provide({[}X:input.state,Y:output.action{]}))}\\$\leftarrow$ \texttt{select(Y),map(X,Y)}\end{tabular} \\ \hline

\multicolumn{1}{|l|}{\texttt{C2}} & \begin{tabular}[c]{@{}l@{}}\texttt{<model$\rightarrow$user,\textbf{req-informative\_advice(X,Y,A)},}\\\texttt{{[}X:CarPosition;Y:selectedAction;A:keyboardAction{]}>}\end{tabular} & \begin{tabular}[c]{@{}l@{}}\texttt{req-informative\_advice(X,Y,A)$\equiv$} \\ \texttt{request(A:output.action,{[}X:input.state,Y:output.action{]})}\\$\leftarrow$\texttt{modify(Y,A), map(X,A)}\end{tabular} \\ \hline

\multicolumn{1}{|l|}{\texttt{C3}} & \begin{tabular}[c]{@{}l@{}}\texttt{<user$\rightarrow$model,\textbf{give-informative\_advice(X,Y,A)},}\\\texttt{{[}X:CarPosition;Y:selectedAction;A:keyboardAction{]}>}\end{tabular} & \begin{tabular}[c]{@{}l@{}}\texttt{give-evaluative\_advice(X,Y,A)$\equiv$}\\
\texttt{provide(A:output.action,{[}X:input.state,Y:output.action{]})}\\
$\leftarrow$\texttt{modify(Y,A), map(X,A)}\end{tabular} \\ \hline

\multicolumn{1}{|l|}{\texttt{C4}} & \begin{tabular}[c]{@{}l@{}}\texttt{<model$\rightarrow$user,\textbf{req-evaluative\_advice(X,Y,F)},}\\\texttt{{[}X:CarPosition;Y:selectedAction;F:keyboardFeedback{]}>}\end{tabular} & \begin{tabular}[c]{@{}l@{}}\texttt{req-evaluative\_advice(X,Y,F)$\equiv$} \\ \texttt{request(F:feedback.eval,{[}X:input.state,Y:output.action{]})}\\$\leftarrow$\texttt{select(F), map(X,A)}\end{tabular} \\ \hline

\multicolumn{1}{|l|}{\texttt{C5}} & \begin{tabular}[c]{@{}l@{}}\texttt{<user$\rightarrow$model,\textbf{give-evaluative\_advice(X,Y,A)},}\\\texttt{{[}X:CarPosition;Y:selectedAction;A:keyboardAction{]}>}\end{tabular} & \begin{tabular}[c]{@{}l@{}}\texttt{give-evaluative\_advice(X,Y,A)$\equiv$}\\
\texttt{provide(F:feedback.eval,{[}X:input.state,Y:output.action{]})}\\
$\leftarrow$\texttt{select(F),map(X,A)}\end{tabular} \\ \hline

\end{tabular}
}
\caption{Messages and action definitions for the interactive RL interactions}
\label{tab:actions3}
\end{table}

Based on these definitions, we define the following patterns (Table \ref{tab:interaction3}): \texttt{[C1] policy-visualization}, \texttt{[C2-C3] informative\_advice}, and \texttt{evaluative\_advice}. The interaction starts with the policy visualization and a human teaching method. For both patterns, the RL agent communicates its policy by visualizing the current state and the selected action. Based on the teaching method, there are two different types of interactions between the human trainer and the model: informative advice in the form of an alternate action/guidance and evaluative advice as a feedback/reward. Each patterns requires an appropriate model update mechanism for online learning. For the evaluation pattern, the user can evaluate the policy by providing evaluative feedback (reward shaping), while for the informative pattern, the user can suggest an action as informative advice (policy shaping). 

\begin{table}[h]
\begin{tabular}{|c|c|c|}
\hline
\textbf{pattern} & \textbf{actions/messages} & \textbf{description} \\ \hline
policy-visualization & \texttt{show-policy(X,Y)} & model visualizes action for current state \\ \hline
\multirow{2}{*}{informative\_advice} & \texttt{req-informative\_advice(X,Y,A)} & model checks if user provided advice \\ \cline{2-3} 
 & \texttt{give-informative\_advice(X,Y,A)} & user provides informative advice (action) \\ \hline
\multirow{2}{*}{evaluative\_advice} & \texttt{req-evaluative\_advice(X,Y,F)} & model checks if user provided feedback \\ \cline{2-3} 
 & \texttt{give-evaluative\_advice(X,Y,F)} & user provides evaluative feedback (reward) \\ \hline
\end{tabular}
\caption{Interaction Patterns and actions for interactive RL.}
\label{tab:interaction3}
\end{table}

\begin{figure}[h]
\centering
\includegraphics[width=\columnwidth]{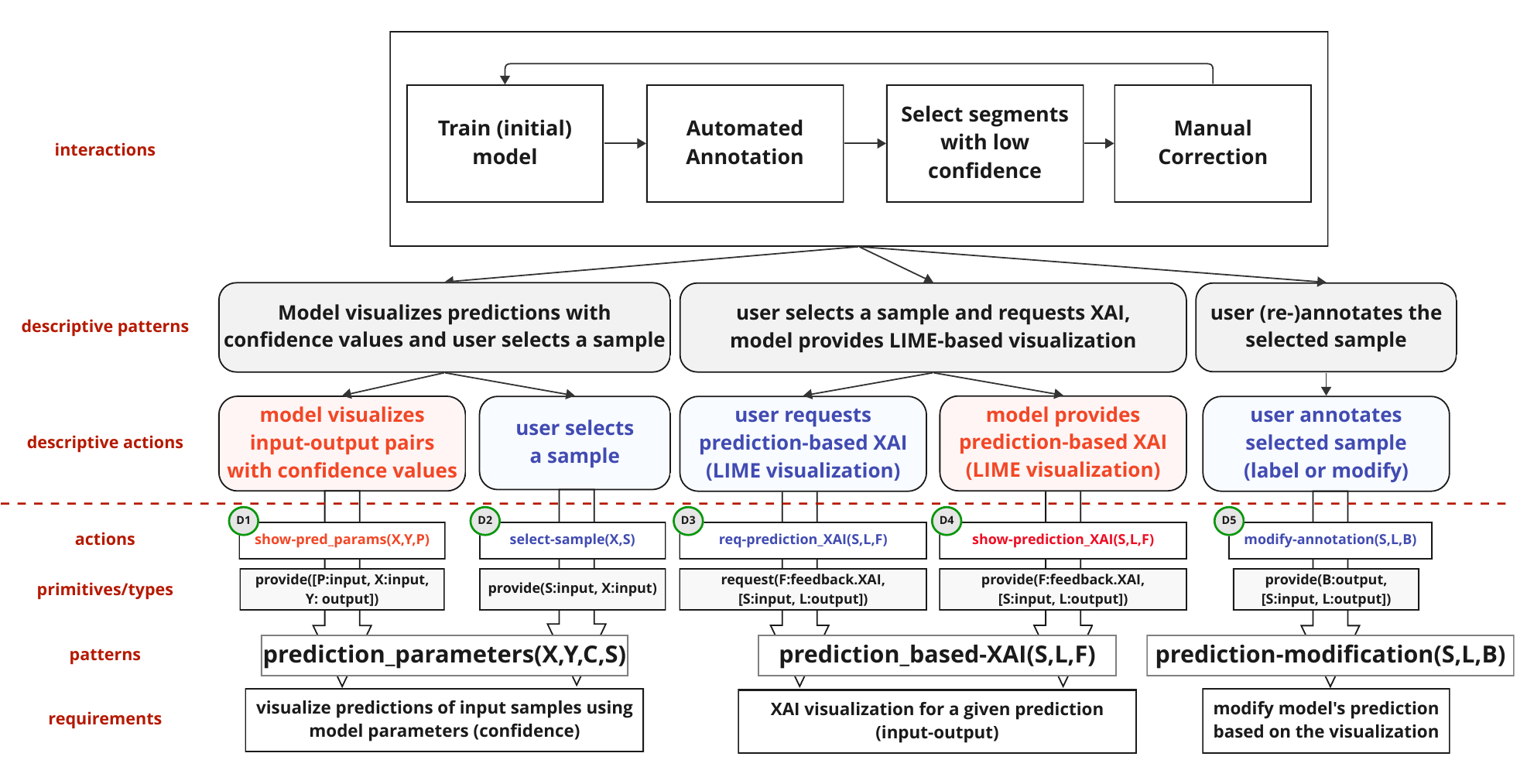}
\caption{Unpacking an explainable active learning interaction into patterns. Image adapted from \cite{heimerl2019nova} }
\label{fig:unpacking4}
\end{figure}

\subsubsection{Explainable active learning for collaborative emotion labeling (Figure \ref{fig:unpacking4}).} This use case is based on an application of a multimodal annotation tool, called NOVA. The use case describes a collaborative annotation task for emotion recognition. The proposed tool includes XAI functionalities (transparency and visualizations) which aim to enhance user's decision making and trust to the system. More specifically, the proposed annotation system follows an active learning approach to select which samples should be labeled by the user by visualizing the model's predictions confidence for these samples. The user can choose any sample and re-label it. Moreover, an XAI method (LIME visualization) is used to support user's decision making (emotion detection) through saliency maps; visualizations of important visual features for the selected frame classification. 
    
We identify the following interaction patterns: (a) model provides input-output pairs with confidence values and visualizations, and (b) user selects and visualizes an input-output pair and updates it. These patterns can be further described as: (a1) model provides input, (a2) model provides output and (a3) model provides XAI-based feedback (confidence-based visualization), and (b1) user provides input (selection), (b2) user provides output (selection), (b3) user provides output (edit). The proposed system follows an explainable semi-supervised active learning approach. One of the main challenges of active learning is to identify the appropriate queries (data points) to ask for user labeling. The proposed approach aims to improve model performance through interactive labeling. Considering both design and implementation aspects, the system utilizes a user interface for model transparency and visual explanations, and integrates the human-in-the-loop to facilitate the active learning process, by identifying which data should be (re-)labeled. Both design and implementation choices are made to satisfy such requirements. Transparency and explainability are used to support user's decision making to refine a model through design features. The model guides the user to improve its performance for low-confidence predictions (active learning), while additional explanations (LIME) can be requested to further support user's annotation task. Based on these, we define the following actions (Table \ref{tab:actions4}): 

\begin{table}[h]
\resizebox{\columnwidth}{!}{
\begin{tabular}{|ll|l|}
\hline
\multicolumn{2}{|c|}{\textbf{Message}} & \multicolumn{1}{c|}{\textbf{Action definition}} \\ \hline

\multicolumn{1}{|l|}{\texttt{D1}} & \begin{tabular}[c]{@{}l@{}}\texttt{<model$\rightarrow$user,\textbf{show-prediction\_params(X,Y,P)},}\\\texttt{{[}X:frames;Y:emotionLabels;P:confValues{]}>}\end{tabular} & \begin{tabular}[c]{@{}l@{}}
\texttt{show-prediction\_params(X,Y,P)$\equiv$}\\\texttt{provide({[}X:{[}input.raw\_data{]},{[}Y:output.label{]},{[}P:input.}\\\texttt{model\_params{]}{]}))$\leftarrow$ \texttt{map(X,Y,P)}}\end{tabular} \\ \hline

\multicolumn{1}{|l|}{\texttt{D2}} & \begin{tabular}[c]{@{}l@{}}\texttt{<user$\rightarrow$model,\textbf{select-sample(S,X)},}\\\texttt{{[}X:frames;S:selFrame{]}>}\end{tabular} & \begin{tabular}[c]{@{}l@{}}\texttt{select-sample(S,X)$\equiv$} \\ \texttt{provide(S:input.raw\_data,X:{[}input.raw\_data{]})}\\$\leftarrow$\texttt{select(S,X)}\end{tabular} \\ \hline

\multicolumn{1}{|l|}{\texttt{D3}} & \begin{tabular}[c]{@{}l@{}}\texttt{<user$\rightarrow$model,\textbf{modify-prediction(S,L,A)},}\\\texttt{{[}S:selFrame;L:emotionLabel;A:newLabel{]}>}\end{tabular} & \begin{tabular}[c]{@{}l@{}}\texttt{modify-prediction(S,L,A)$\equiv$}\\
\texttt{provide(A:output.label,{[}S:input.raw\_data,L:output.label{]})}\\
$\leftarrow$\texttt{modify(L,A), map(X,A)}\end{tabular} \\ \hline

\multicolumn{1}{|l|}{\texttt{D4}} & \begin{tabular}[c]{@{}l@{}}\texttt{<model$\rightarrow$user,\textbf{req-prediction\_XAI(S,L,F)},}\\\texttt{{[}S:selFrame;L:emotionLabel;F:LIMEVisualization{]}>}\end{tabular} & \begin{tabular}[c]{@{}l@{}}\texttt{req-prediction\_XAI(S,L,F)$\equiv$} \\ \texttt{request(F:feedback.XAI,{[}S:input.raw\_data,L:output.label{]})}\\$\leftarrow$\texttt{map(F,S,L)}\end{tabular} \\ \hline

\multicolumn{1}{|l|}{\texttt{D5}} & \begin{tabular}[c]{@{}l@{}}\texttt{<model$\rightarrow$user,\textbf{show-prediction\_XAI(S,L,F)},}\\\texttt{{[}S:selFrame;L:emotionLabel;F:LIMEVisualization{]}>}\end{tabular} & \begin{tabular}[c]{@{}l@{}}\texttt{show-prediction\_XAI(S,L,F)$\equiv$}\\
\texttt{provide(F:feedback.XAI,{[}S:input.raw\_data,L:output.label{]})}\\$\leftarrow$\texttt{map(F,S,L)}\end{tabular} \\ \hline

\end{tabular}
}
\caption{Messages and action definitions for the active learning emotion recognition}
\label{tab:actions4}
\end{table}   

We define the following patterns (Table \ref{tab:interaction4}): \texttt{[D1-D2] prediction\_parameters}, \texttt{[D3] prediction-modification}, and  \texttt{[D4-D5] prediction-based\_XAI}. The first pattern describes the sample selection process, where the model supports the user to select the appropriate samples to annotate through visualizing the model confidence for its predictions. This approach is part of the active learning process based on which a set of candidate samples is selected for annotation. In this case, model instances with low confidence are presented to the user. The second pattern describes human labeling through a prediction modification approach. The third pattern describes a user request for XAI of a selected sample, where the model provides a LIME-based visualization for the user to understand the current prediction and modify it if needed. Local explanations are provided to the user to support their decision making through local interpretability. The output of LIME is a visualization of explanations representing the contribution of each feature to the prediction of the current frame. These patterns (and their actions) can be combined to design the interactions, e.g., the user can explore and select a sample based on the visualizations and either annotate it or request sample-based explanations.  

\begin{table}[h]
\begin{tabular}{|c|c|c|}
\hline
\textbf{pattern} & \textbf{actions/messages} & \textbf{description} \\ \hline
\multirow{2}{*}{prediction\_parameters} & \texttt{show-prediction\_params(X,Y,P)} & parameter-based sample visualization\\ \cline{2-3} 
& \texttt{select-sample(S,X)} & user selects a sample from list \\ \hline
prediction-modification & \texttt{modify-prediction(S,L,A)} & user annotates/modifies an annotation \\ \hline
\multirow{2}{*}{prediction-based\_XAI} & \texttt{req-prediction\_XAI(S,L,F)} & user request XAI for selected sample \\ \cline{2-3} 
& \texttt{show-prediction\_XAI(S,L,F)} & model provides sample-based XAI \\ \hline
\end{tabular}
\caption{Interaction patterns and actions for the active learning emotion recognition}
\label{tab:interaction4}
\end{table}

\begin{figure}[h]
\centering
\includegraphics[width=0.65\columnwidth]{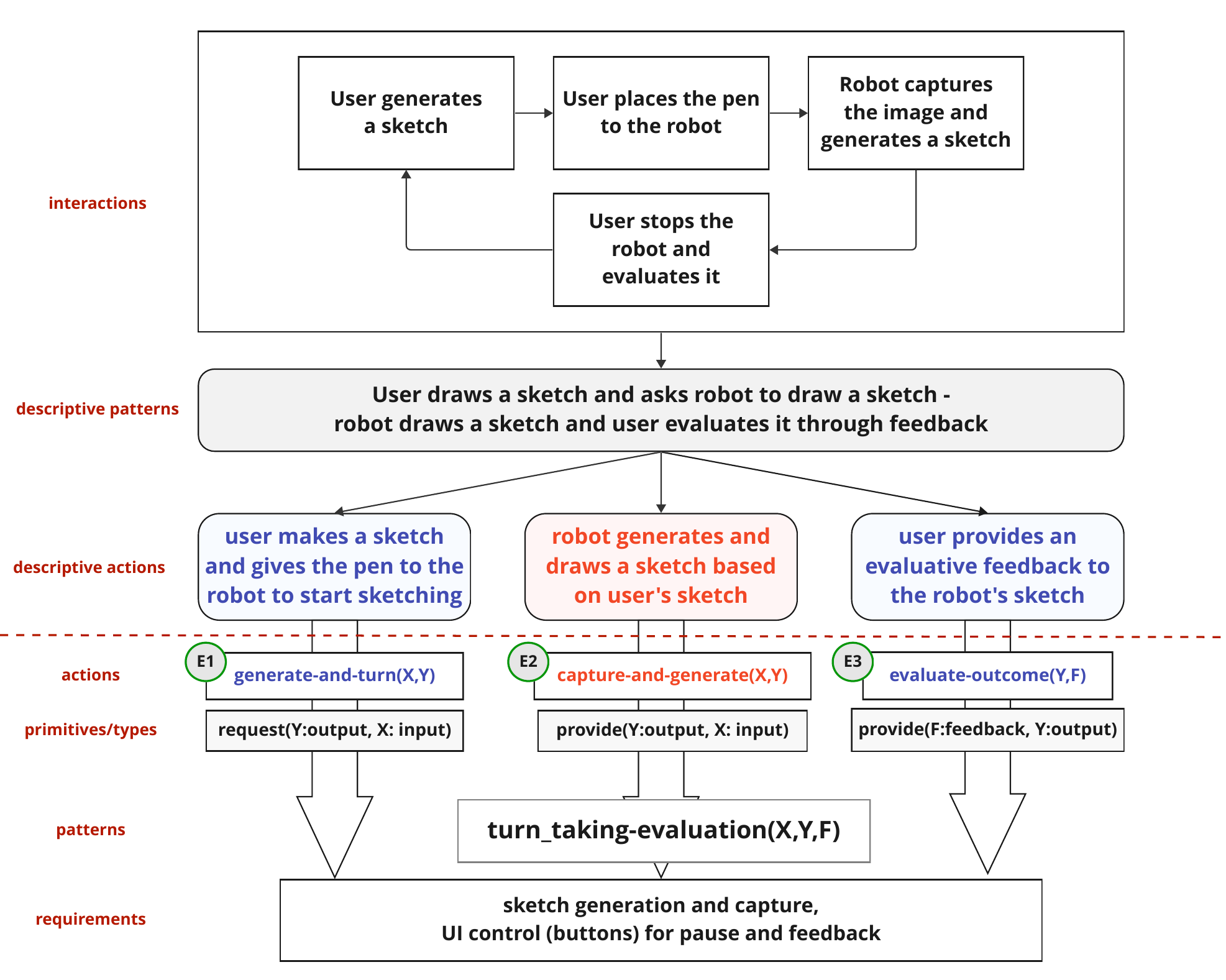}
\caption{Unpacking a human-robot collaborative sketching interaction into patterns. Image adapted from \cite{lin2020your} }
\label{fig:unpacking5}
\end{figure}

\subsubsection{Human-robot collaborative sketching (Figure \ref{fig:unpacking5}).} 
The proposed system describes a collaborative sketching approach between a user an a robot (Cobbie). More specifically, the proposed system utilizes the mechanism of conceptual shift to support human-AI co-creation and collaborative sketch ideation. Based on this approach, the user initiates the interaction by sketching an image on paper. Once finished, the user gives the pen to the robot, which captures and analyzes the user's sketch. Based on its analysis, it generates a new sketch on paper. The user can pause the robot and provide an evaluative feedback for the robot's drawing. If feedback is negative, the robot starts drawing a new sketch until new feedback is received. If feedback is positive, the user draws a new sketch by combining the two sketches. The robot utilizes the provided feedback to adjust its model in order to provide more useful ideas to the user. 
    
\begin{table}[h]
\resizebox{\columnwidth}{!}{
\begin{tabular}{|ll|l|}
\hline
\multicolumn{2}{|c|}{\textbf{Message}} & \multicolumn{1}{c|}{\textbf{Action definition}} \\ \hline

\multicolumn{1}{|l|}{\texttt{E1}} & \begin{tabular}[c]{@{}l@{}}\texttt{<model$\rightarrow$user,\textbf{generate-and-turn(X,Y)},}\\\texttt{{[}X:userSketch;Y:onPenClipper,robotSketch{]}>}\end{tabular} & \begin{tabular}[c]{@{}l@{}}
\texttt{generate-and-turn(X,Y)$\equiv$}\\\texttt{request(Y:input.raw\_data,C:output.raw\_data))}\\$\leftarrow$ \texttt{create(X), create(Y), map(X,Y)}\end{tabular} \\ \hline

\multicolumn{1}{|l|}{\texttt{E2}} & \begin{tabular}[c]{@{}l@{}}\texttt{<user$\rightarrow$model,\textbf{capture-and-generate(X,Y)},}\\\texttt{{[}X:userSketch; Y:robotSketch{]}>}\end{tabular} & \begin{tabular}[c]{@{}l@{}}\texttt{capture-and-generate(X,Y)$\equiv$} \\ \texttt{provide(Y:input.raw\_data,C:output.raw\_data))}\\$\leftarrow$\texttt{create(Y),map(X,Y)}\end{tabular} \\ \hline

\multicolumn{1}{|l|}{\texttt{E3}} & \begin{tabular}[c]{@{}l@{}}\texttt{<user$\rightarrow$model,\textbf{evaluate-outcome(Y,F)},}\\\texttt{{[}Y:robotSketch;F:pauseBtn,feedbackBtn{]}>}\end{tabular} & \begin{tabular}[c]{@{}l@{}}\texttt{evaluate-outcome(Y,F)$\equiv$}\\
\texttt{provide(F:feedback.eval,Y:output.raw\_data)}\\
$\leftarrow$\texttt{select(F), map(Y,F)}\end{tabular} \\ \hline
\end{tabular}
}
\caption{Messages and action definitions for the collaborative sketching interactions.}
\label{tab:actions5}
\end{table}

The interaction takes place as a turn taking sketching-based interaction, where user and model sketch a drawing considering previous drawing (co-ideation). In terms of the types of communicated information, user and robot communicate through drawing sketches, giving the pen to the robot and providing feedback to the robot through buttons. In terms of implementation aspects, the robot deploys an RNN-based recognizer to capture and classify the user's input and an adapted version of the RNN-sketch model to generate a new image based on user's input. User's feedback is used to update the network weights, and thus In terms of interaction design, the user is the dominant member of the co-creation session. The user can determine when Cobbie should start drawing an image, by placing the pen to the robot's clipper. The user can also select the robot's position (what to capture and where to draw) as well as the pen strokes. Users can use the on-platform buttons to pause and resume robot's drawing and provide feedback. Robot expressive movements and sounds indicate that the robot has successfully received a command (button pressed). In terms or model performance, the deployed AI models (image classification and rnn-sketch) are well-performing models. The goal of the interaction is to support user's creative thinking and ideation processes. The model provides appropriate outputs (not the most accurate) in order to facilitate this co-ideation process. Table \ref{tab:actions5} shows the message and actions definitions for this interaction.

\begin{table}[h]
\begin{tabular}{|c|c|c|}
\hline
\textbf{pattern} & \textbf{actions} & \textbf{description} \\ \hline
\multirow{3}{*}{turn\_taking-evaluation} & \texttt{generate-and-turn(X,Y)} & user asks robot to sketch based on the drawing \\ \cline{2-3} 
 & \texttt{capture-and-generate(X,Y)} & robot captures and generates new sketch \\ \cline{2-3} 
 & \texttt{evaluate-outcome(Y,F)} & user pauses robot and provides feedback \\ \hline
\end{tabular}
\caption{Interaction patterns and actions for the collaborative robot sketching.}
\label{tab:interaction5}
\end{table}

Based on this interaction, we can define \texttt{[E]:turn\_taking-evaluation} as a user-driven control and evaluation pattern, where the user can control and evaluate the collaboration with the model (Table \ref{tab:interaction5}). Based on this pattern, the user initiates the interaction by drawing a sketch and giving the pen to the robot for idea generation. The goal of the robot is to support user's ideation process by generating creative and diverse sketches. This is achieved through object detection (image classification) and conceptual shift (sketc-rnn). The user evaluates the robot's  sketches (model output) using the feedback buttons. This interaction pattern can describe an explorative collaborative learning process, where both user and system aim to explore and identify new insights through their interaction.    

\begin{figure}[h]
\centering
\includegraphics[width=\columnwidth]{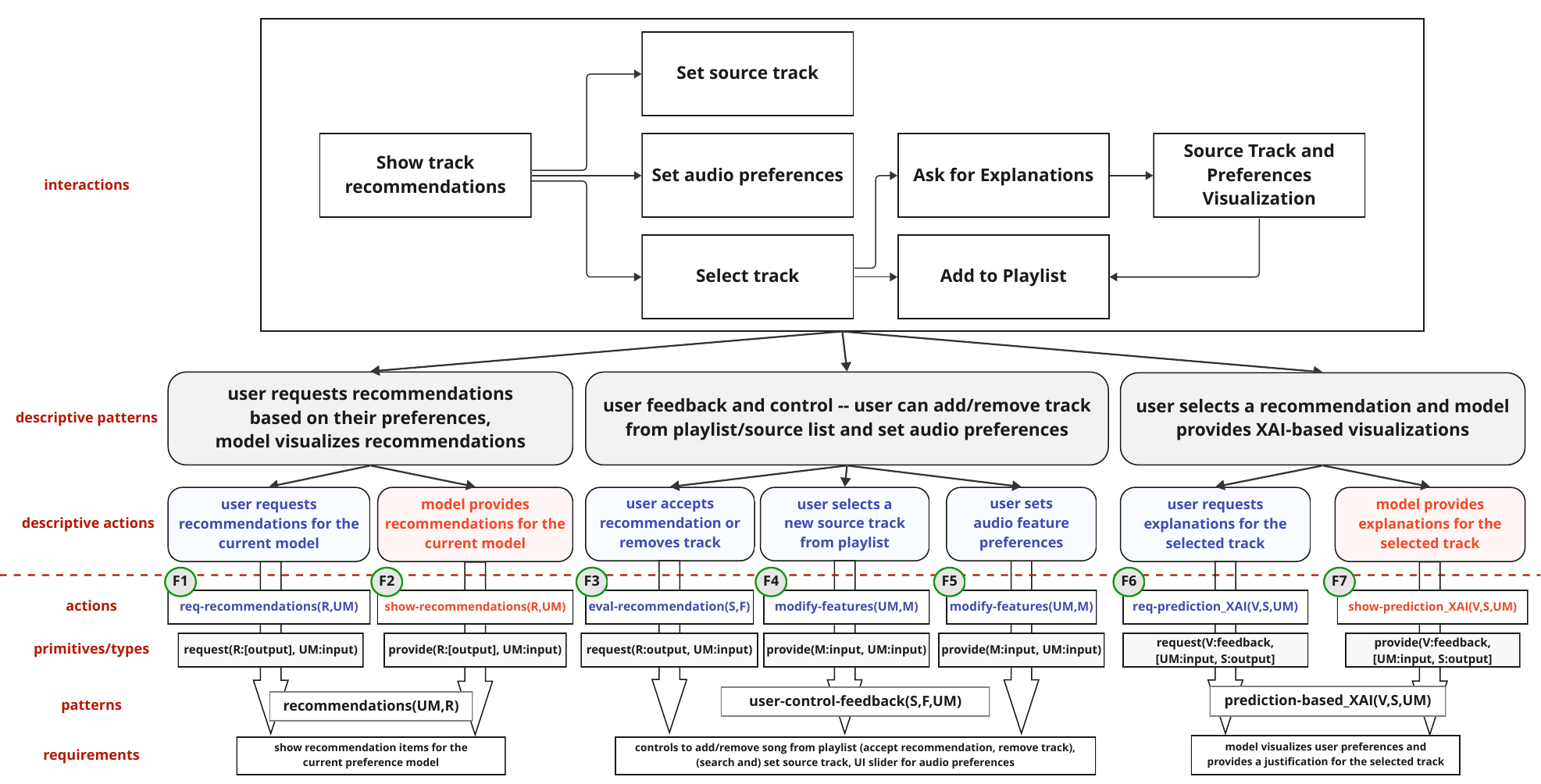}
\caption{Unpacking an explainable music recommendation system into patterns. Image adapted from \cite{martijn2022knowing}}
\label{fig:unpacking6}
\end{figure}

\subsubsection{Explainable Music Recommendation System (Figure \ref{fig:unpacking6}).} The proposed system aims to explore the design of explanations in a music recommender system in order to fit the user's preferences (selected songs, audio preferences) and personal characteristics (i.e., need for cognition, musical sophistication and openness). The system provides recommendations to the user based on a source song (i.e., a playlist song) and user preferences for audio features (danceability, energy, happiness, and popularity). User can search and add recommended tracks to the playlist, remove existing track from the playlist, select a playlist song as a source song for recommendations, set their preferences through the audio features, and request and explore explanations. Explanations could be requested (and provided) both for a selected recommendation as well as for all songs at once.

We identify the following types of interaction: (a) model visualizes playlist and recommendations with control options -- user adds (or removes) a song from the playlist/source list songs, as well as through the audio feature preferences, and (b) user can request further explanations for a given track or all songs --  model provides visual and textual explanations to justify recommendations. In terms of the communicated information, the model and the user interact through showing and selecting (explainable) recommendations and audio feature preferences. In terms of implementation, the model utilizes user's actions (adding/removing songs, setting preferences, asking for XAI), in order to provide personalized recommendations and explanations. After certain interactions with the user, i.e., add/remove/search song or change audio preferences, the model updates its recommendations (updated feature vector). In terms of design aspects, the model provides different types of explanations in order to match the individual's preference and characteristics. Based on the outcomes from a set of user studies, the authors provide a set of design suggestions towards selecting appropriate explanation styles and levels of transparency based on user's personal characteristics. For example, users with low musical sophistication may prefer brief explanations that do not require domain knowledge. we can define the following actions (Table \ref{tab:actions6}): 

\begin{table}[h]
\resizebox{\columnwidth}{!}{
\begin{tabular}{|ll|l|}
\hline
\multicolumn{2}{|c|}{\textbf{Message}} & \multicolumn{1}{c|}{\textbf{Action definition}} \\ \hline

\multicolumn{1}{|l|}{\texttt{F1}} & \begin{tabular}[c]{@{}l@{}}\texttt{<model$\rightarrow$user,\textbf{req-recommendations(R,UM)},}\\\texttt{{[}[UM:audioPrefs,srcTrack;R:reccTracks{]}>}\end{tabular} & \begin{tabular}[c]{@{}l@{}}\texttt{req-recommendations(R,UM)$\equiv$}\\\texttt{request(R:{[}output.item{]},UM:input.fvector))}\\$\leftarrow$ \texttt{select(R),map(UM,R)}\end{tabular} \\ \hline

\multicolumn{1}{|l|}{\texttt{F2}} & \begin{tabular}[c]{@{}l@{}}\texttt{<model$\rightarrow$user,\textbf{show-recommendations(R,UM)},}\\\texttt{{[}[UM:audioPrefs,srcTrack;R:reccTracks{]}>}\end{tabular} & \begin{tabular}[c]{@{}l@{}}\texttt{show-recommendations(R,UM)$\equiv$}\\\texttt{provide(R:{[}output.item{]},UM:input.fvector))}\\$\leftarrow$ \texttt{select(R),map(UM,R)}\end{tabular} \\ \hline

\multicolumn{1}{|l|}{\texttt{F3}} & \begin{tabular}[c]{@{}l@{}}\texttt{<user$\rightarrow$model,\textbf{modify-features(UM,M)},}\\\texttt{{[}UM:audioPrefs;M:modifyPrefs,UIslider{]}>}\end{tabular} & \begin{tabular}[c]{@{}l@{}}\texttt{modify-features(UM,M)$\equiv$}\\
\texttt{provide(M:input.fvector,UM:input.fvector)}\\$\leftarrow$\texttt{modify(UM,M)}\end{tabular} \\ \hline

\multicolumn{1}{|l|}{\texttt{F4}} & \begin{tabular}[c]{@{}l@{}}\texttt{<model$\rightarrow$user,\textbf{evaluate-recommendation(F,S)},}\\\texttt{{[}F:addPlaylistTrack;S:selTrack{]}>}\end{tabular} & \begin{tabular}[c]{@{}l@{}}\texttt{evaluate-recommendation(F,R)$\equiv$}\\
\texttt{provide(F:feedback.eval,S:output.item)}\\$\leftarrow$\texttt{select(S),map(S,F)}\end{tabular} \\ \hline

\multicolumn{1}{|l|}{\texttt{F5}} & \begin{tabular}[c]{@{}l@{}}\texttt{<user$\rightarrow$model,\textbf{modify-features(UM,M)},}\\\texttt{{[}UM:srcTrack;M:newSrcTrack,click{]}>}\end{tabular} & \begin{tabular}[c]{@{}l@{}}\texttt{modify-features(UM,M)$\equiv$}\\
\texttt{provide(M:input.fvector,UM:input.fvector)}\\$\leftarrow$\texttt{modify(UM,M)}\end{tabular} \\ \hline

\multicolumn{1}{|l|}{\texttt{F6}} & \begin{tabular}[c]{@{}l@{}}\texttt{<model$\rightarrow$user,\textbf{req-prediction\_XAI(V,S,UM)},}\\\texttt{{[}V:clickXAIBtn;S:selTrack;UM:audioPrefs,srcTrack{]}>}\end{tabular} & \begin{tabular}[c]{@{}l@{}}\texttt{req-prediction\_XAI(UM,S,V) $\equiv$}\\\texttt{request(V:feedback.XAI;{[}UM:input.fvector,}\\\texttt{S:output.item{]})}\\$\leftarrow$ \texttt{select(S), map(V,S,UM)}\end{tabular} \\ \hline

\multicolumn{1}{|l|}{\texttt{F7}} & \begin{tabular}[c]{@{}l@{}}\texttt{<model$\rightarrow$user,\textbf{show-prediction\_XAI(V,S,UM)},}\\\texttt{{[}V:openUserModel;S:selTrack;UM:audioPrefs,srcTrack{]}>}\end{tabular} & \begin{tabular}[c]{@{}l@{}}\texttt{show-prediction\_XAI(UM,S,V) $\equiv$}\\\texttt{provide(V:feedback.XAI;{[}UM:input.fvector,}\\\texttt{S:output.item{]})}\\$\leftarrow$ \texttt{select(S), map(V,S,UM)}\end{tabular} \\ \hline

\end{tabular}
}
\caption{Messages and action definitions for the interactive sound annotation system interactions}
\label{tab:actions6}
\end{table}

\begin{table}[h]
\resizebox{0.95\columnwidth}{!}{
\begin{tabular}{|c|c|c|}
\hline
\textbf{pattern} & \textbf{actions} & \textbf{description} \\ \hline
\multirow{2}{*}{recommendations} & \texttt{req-recommendations(UM,TR)} & user requests recommendations for preferences \\ \cline{2-3} 
 & \texttt{show-recommendations(UM,TR)} & model visualizes recommendations for preferences \\ \hline
\multirow{2}{*}{user-control-feedback} & \texttt{modify-features(UM,M)} & user sets preferences and updates feature vector \\ \cline{2-3} 
& \texttt{evaluate-recommendation(S,PL)} & user adds or removes playlist song \\  \hline
\multirow{2}{*}{prediction-based\_XAI} & \texttt{req-prediction\_XAI(S,V,UM)} & user requests XAI for recommendations \\ \cline{2-3} 
& \texttt{show-prediction\_XAI(S,V,UM)} & model visualizes user model \\ \hline
\end{tabular}
}
\caption{Interaction patterns for the explainable music recommendation system}
\label{tab:interaction6}
\end{table}

We define the following patterns (Table \ref{tab:interaction6}): \texttt{[F1] recommendations}, \texttt{[F2-F3] user-control-feedback}, and \texttt{[F4-F5] prediction-based\_XAI}.
Based on the first pattern, the user is provided with a list of recommended tracks for the current preferences. The second pattern described the user's interactions with the recommendations and their preferences. The model updates its decisions based on the these user actions. The third pattern is part of an XAI-interaction where the system visualizes the preference model and the similarity to justify a given recommendation. The user makes the final decision about a recommended track (output) and the feature values (input). 
    
\begin{figure}[h]
\centering
\includegraphics[width=\columnwidth]{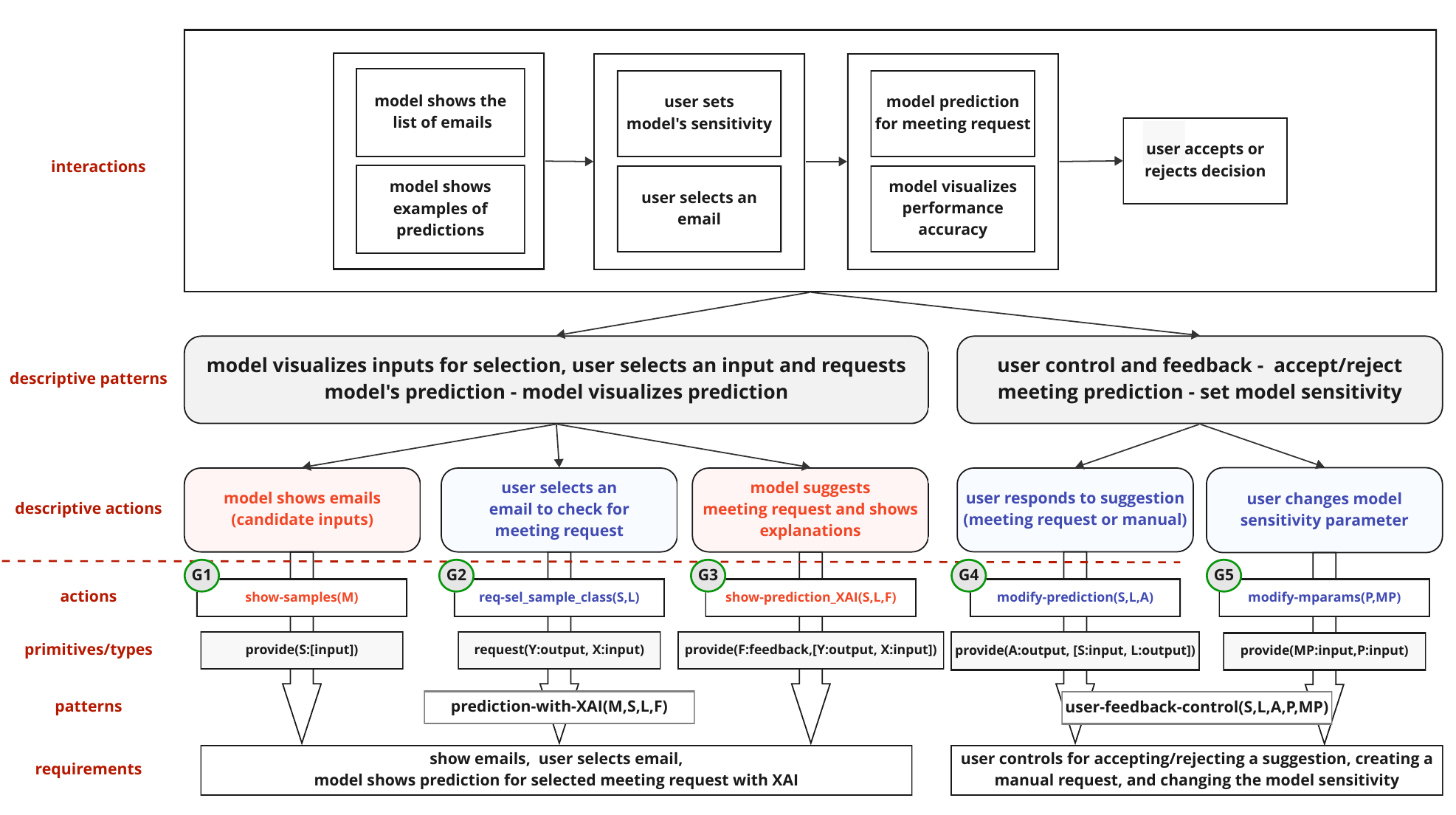}
\caption{Unpacking the meeting scheduling assistant  \cite{kocielnik2019will} into patterns and primitives.}
\label{fig:unpacking7}
\end{figure}

\subsubsection{Transparent Meeting Scheduling Assistant (Figure \ref{fig:unpacking7}).} The proposed system uses an AI model to automatically detect meeting requests from free-text emails. The scheduling assistant provides the user with additional information (XAI), including an accuracy indicator component and a textual description for example-based explanations. The accuracy indicator visualizes the model's accuracy for the predictions in the form of a chart. The example-based explanations aim to enhance user's understanding about the underlying AI model and include a set of example sentences (inputs) and the model prediction (output) for each sentence, which can vary from "very unlikely" to "very likely" to describe the model's confidence, and depend on the model's sensitivity. The user is able to control model's sensitivity and see the updated results, through a UI slider which provides information about how sensitivity affects model's decisions.  
    
We can identify the following interactions: (a) model visualizes the inputs (e-mails) and user selects an item to check model's prediction - model provides prediction with XAI, (b) user provides feedback to the model through accepting or rejecting suggestions (predictions) and/or setting the sensitivity value through the slider. During these interactions, user and model communicate messages through visualization and selection of model inputs, textual and visual explanations and control interfaces (slider), as well as accepting or rejecting model's decisions. IN order to enhance user's decision making, model provides explanations and transparency in terms of performance. Interactions with low-performance models (low confidence) can be effective for the user, if they provide an appropriate level of transparency and explainability. Such interaction can enhance user's decision making. In terms of implementation aspects, the different types of user feedback (e.g.,accepting/rejecting suggestions) can be used to update the model in an interactive way. Users can set the model sensitivity parameter value based on their observations of how it affects model's decisions. we define the following actions (Table \ref{actions7}): 

\begin{table}[h]
\resizebox{\columnwidth}{!}{
\begin{tabular}{|ll|l|}
\hline
\multicolumn{2}{|c|}{\textbf{Message}} & \multicolumn{1}{c|}{\textbf{Action definition}} \\ \hline

\multicolumn{1}{|l|}{\texttt{G1}} & \begin{tabular}[c]{@{}l@{}}\texttt{<model$\rightarrow$user,\textbf{show-samples(M)},}\\\texttt{{[}M:emailList,freeText{]}>}\end{tabular} & \begin{tabular}[c]{@{}l@{}}\texttt{show-samples(M)$\equiv$}\\\texttt{provide(M:{[}input.raw\_data{]}))}\\$\leftarrow$ \texttt{select(M)}\end{tabular} \\ \hline

\multicolumn{1}{|l|}{\texttt{G2}} & \begin{tabular}[c]{@{}l@{}}\texttt{<user$\rightarrow$model,
\textbf{req-sel\_sample\_class(S,L)},}\\\texttt{{[}S:selectedEmail;L:L:isMeeting{]}>}\end{tabular} & \begin{tabular}[c]{@{}l@{}}\texttt{req-sel\_sample\_class(S,L)$\equiv$}\\\texttt{request(L:output.label,S:input.raw\_data)}\\$\leftarrow$ \texttt{select(S),map(S,L)}\end{tabular} \\ \hline

\multicolumn{1}{|l|}{\texttt{G3}} & \begin{tabular}[c]{@{}l@{}}\texttt{<model$\rightarrow$user,
\textbf{show-prediction-XAI(S,L,F)},}\\\texttt{{[}S:selectedEmail;L:L:isMeeting;F:XAIexamples{]}>}\end{tabular} & \begin{tabular}[c]{@{}l@{}}\texttt{show-prediction-XAI(S,L,F)$\equiv$}\\\texttt{provide(F:feedback.XAI;{[}S:input.raw\_data,L:output.label{]})}\\$\leftarrow$ \texttt{map(S,L), map(F,S,L)}\end{tabular} \\ \hline

\multicolumn{1}{|l|}{\texttt{G4}} & \begin{tabular}[c]{@{}l@{}}\texttt{<model$\rightarrow$user,\textbf{modify-prediction(S,L,A)},}\\\texttt{{[}S:selectedEmail;
L:isMeeting;A:acceptBtn,createBtn{]}>}\end{tabular} & \begin{tabular}[c]{@{}l@{}}\texttt{modify-prediction(S,L,A)$\equiv$}\\\texttt{provide(A:output.label,{[}S:input.raw\_data,}\\\texttt{L:output.label{]})}$\leftarrow$ \texttt{select(S),modify(L,A),map(S,A)}\end{tabular} \\ \hline

\multicolumn{1}{|l|}{\texttt{G5}} & 
\begin{tabular}[c]{@{}l@{}}\texttt{<model$\rightarrow$user,\textbf{modify-mparams(P,MP)},}\\\texttt{{[}P:sensitivityValue; MP:modifiedValue,UIslider{]}>}\end{tabular} & \begin{tabular}[c]{@{}l@{}}\texttt{modify-mparams(P,MP)$\equiv$}\\\texttt{provide(MP:input.model\_params,P:model\_params))}\\$\leftarrow$ \texttt{modify(P,MP)}\end{tabular} \\ \hline
\end{tabular}
}
\caption{Messages and action definitions for the meeting scheduling assistant}
\label{tab:actions7}
\end{table}

We identify two patterns: \texttt{[G1-G3] prediction-with-XAI}, and \texttt{[G4-G5] user-feedback-control} (Table \ref{interaction7}). The first pattern describes an XAI-based interaction for input selection and prediction, where the model provides textual explanations of prediction examples to the user to enhance their understanding about the model's prediction. The second pattern describes a feedback control pattern; the user provides feedback to set the model sensitivity and interacts with the model decisions. Depending on the prediction, the user can agree with the model and accept a predicted request (true positive) or a predicted non-request (true negative). If the model does not predict a meeting request (false negative) the user can manually create a meeting request. If the model predicts a false meeting request (false positive), the user can ignore the predicted request.

\begin{table}[h]
\resizebox{0.9\columnwidth}{!}{
\begin{tabular}{|c|c|c|}
\hline
\textbf{pattern} & \textbf{actions} & \textbf{description} \\ \hline
\multirow{3}{*}{prediction-with-XAI} & \texttt{show-samples(M)} & model visualizes all inputs (emails) \\ \cline{2-3} 
& \texttt{req-sel\_sample\_class(S,L)} & user selects an email and asks for suggestion \\ \cline{2-3} 
& \texttt{show-prediction\_XAI(S,L,F)} & model visualizes prediction with XAI \\ \hline
\multirow{2}{*}{user-feedback-control} & \texttt{modify-prediction(S,L,A)} & user accepts or modifies prediction  \\ \cline{2-3} 
& \texttt{modify-mparams(P,MP)} & user sets (a new) model sensitivity parameter\\ \hline
\end{tabular}
}
\caption{Interaction patterns for the transparent meeting scheduling assistant}
\label{tab:interaction7}
\end{table}
   
\begin{figure}[h]
\centering
\includegraphics[width=0.6\columnwidth]{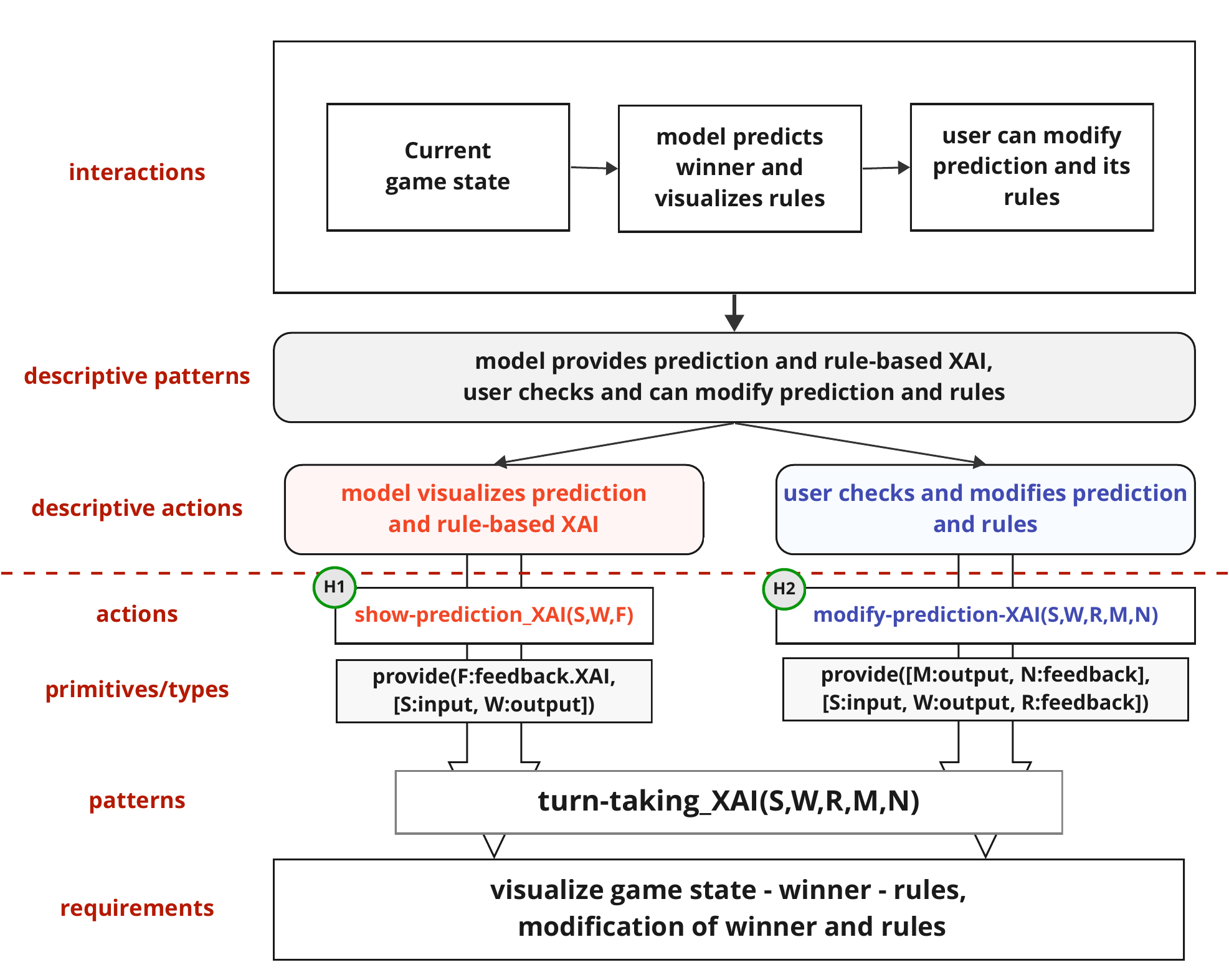}
\caption{Unpacking an interactive ML system for game rules into patterns and primitives. Image adapted from \cite{guo2022building}}
\label{fig:unpacking8}
\end{figure}

\subsubsection{Explainable-driven Interactive Machine Learning for game outcome prediction (Figure \ref{fig:unpacking8}).} The proposed system integrates an explanation-driven interactive machine learning (XIML) mechanism to improve user's trust and satisfaction during the interaction with the system. The use case is the Tic-Tac-Toe game, where user and model make predictions about the winner of the game for a given state (game instance), in a turn-taking interaction. Both user and model can justify their predictions using rule-based explanations.
    
\begin{table}[h]
\resizebox{\columnwidth}{!}{
\begin{tabular}{|ll|l|}
\hline
\multicolumn{2}{|c|}{\textbf{Message}} & \multicolumn{1}{c|}{\textbf{Action definition}} \\ \hline

\multicolumn{1}{|l|}{\texttt{G1}} & \begin{tabular}[c]{@{}l@{}}\texttt{<model$\rightarrow$user,
\textbf{show-prediction\_XAI(S,W,R)},}\\\texttt{{[}S:gameState;W:predWinner;R:XAIrules{]}>}\end{tabular} & \begin{tabular}[c]{@{}l@{}}\texttt{show-prediction-and-XAI(S,W,R)$\equiv$}\\\texttt{provide(R:feedback.XAI;{[}S:input.raw\_data,W:output.label{]})}\\$\leftarrow$ \texttt{map(S,L), map(F,S,L)}\end{tabular} \\ \hline

\multicolumn{1}{|l|}{\texttt{G2}} & \begin{tabular}[c]{@{}l@{}}\texttt{<model$\rightarrow$user,\textbf{modify-prediction-and-XAI(S,W,R,M,N)},}\\\texttt{{[}S:gameState;W:predWinner;R:XAIrules;}\\\texttt{M:modifiedPred;N:modifiedRules{]}>}\end{tabular} & \begin{tabular}[c]{@{}l@{}}\texttt{modify-prediction-and-XAI(S,W,R,M,N)$\equiv$}\\\texttt{provide({[}M:output.label,N:feedback.XAI{]},}\\\texttt{{[}S:input.raw\_data,W:output.label,R:feedback.XAI{]})}\\$\leftarrow$ \texttt{modify(W,M),modify(R,N),map(S,M,N)}

\end{tabular} \\ \hline

\end{tabular}
}
\caption{Messages and action definitions for the XAI-based interactive ML for game rules}
\label{tab:actions8}
\end{table}
    
During this turn-taking interaction, the model visualizes a game instance and its prediction for the winner. In order to justify its prediction, it visualizes the rule based on which the decision was made. The user can accept or modify both prediction and rules. The rules are a set of Boolean rules in disjunctive normal form (DNF). The authors conducted a user study to evaluate the effects of interactivity and visualization on user's trust and satisfaction. The outcomes of their analysis indicate that both aspects can have an effect on users’ perception of control over the different types of visualization. The model visualizes its prediction and reasoning to enhance user's trust in model performance and does not update its prediction based on user 's input. However, similar interactions can take place in hybrid intelligence systems, where both users and models support their own decisions/predictions in order to augment each other's perception. Table \ref{tab:actions8} shows the action definitions of the pattern (Table \ref{tab:interaction8}). \texttt{turn-taking\_XAI} a turn-taking pattern where both user and model can exchange the same information in an interactive manner.    

\begin{table}[h]
\resizebox{\columnwidth}{!}{
\begin{tabular}{|c|c|c|}
\hline
\textbf{pattern} & \textbf{actions} & \textbf{description} \\ \hline
\multirow{2}{*}{turn-taking\_XAI} & show-prediction\_XAI(S,W,R) & model provides rule-based explanations for prediction \\ \cline{2-3} 
& modify\_prediction\_rules(S,W,MW,MR) & user accepts or modifies prediction and rules \\ \hline
\end{tabular}
}
\caption{Interaction patterns for the explainable/interactive game outcome predictions}
\label{tab:interaction8}
\end{table}

\end{document}